%% file: main.tex
\documentclass{article}

\usepackage{microtype}
\usepackage{graphicx}
\usepackage{subcaption}
\usepackage{booktabs}
\usepackage{hyperref}

\usepackage[arxiv]{icml2024}

\usepackage{amsmath}
\usepackage{amssymb}
\usepackage{mathtools}
\usepackage{amsthm}
\usepackage{multicol}
\usepackage{multirow}
\usepackage{enumitem}
\usepackage{placeins}
\usepackage{wrapfig}
\usepackage{bm}
\usepackage{bbm}
\usepackage{multirow}
\usepackage[normalem]{ulem}
\usepackage[export]{adjustbox}
\usepackage{tikz}

\definecolor{linkColor}{rgb}{0.18,0.39,0.62} 
\usepackage{pifont}
\definecolor{BrickRed}{rgb}{0.8, 0.25, 0.33}

\usepackage[capitalize,noabbrev]{cleveref}

\theoremstyle{plain}

\theoremstyle{definition}

\theoremstyle{remark}

\usepackage[textsize=tiny]{todonotes}

\newcommand{\eg}{\textit{e.g.}}
\newcommand{\ie}{\textit{i.e.}}
\newcommand{\xx}{\bm{x}}
\usepackage{bbm, dsfont}

\input{macro}
\input{math_command}

\icmltitlerunning{Rethinking RGB Color Representation for Image Restoration Models}

\begin{document}

\twocolumn[
\icmltitle{Rethinking RGB Color Representation for
Image Restoration Models
}

\icmlsetsymbol{equal}{*}

\begin{icmlauthorlist}
\icmlauthor{Jaerin Lee}{yyy}
\icmlauthor{JoonKyu Park}{yyy}
\icmlauthor{Sungyong Baik}{zzz}
\icmlauthor{Kyoung Mu Lee}{yyy,www}
\end{icmlauthorlist}

\icmlaffiliation{yyy}{ASRI, Department of ECE, Seoul National University, Seoul, Korea}
\icmlaffiliation{zzz}{Department of Data Science, Hanyang University, Seoul, Korea}
\icmlaffiliation{www}{Interdisciplinary Program in Artificial Intelligence, Seoul National University, Seoul, Korea}

\icmlcorrespondingauthor{Kyoung Mu Lee}{kyoungmu@snu.ac.kr}

\icmlkeywords{representation learning, image restoration, representation space, loss function, image super-resolution, image deblurring, image denoising, interpretability, Machine Learning, ICML}

\vskip 0.3in
]

\printAffiliationsAndNotice{}

\input{sections/0_abstract}

\input{sections/1_introduction}
\input{sections/2_related_works}
\input{sections/3_propsed_methods}
\input{sections/4_experiments}
\input{sections/5_discussion}
\input{sections/6_conclusion}

\clearpage

\section*{Broader Impact}
The goal of this paper is to advance the image restoration through representation learning.
Better image restoration techniques imply both beneficial and potentially harmful societal consequences.
For example, restoring old, compressed, or degraded images to high-quality images not only can help archiving old historical records or personal memories but also can strengthen surveillance systems.
 
\bibliography{reference}
\bibliographystyle{icml2024}

\newpage
\appendix
\onecolumn

\section{Theoretical study on the structure embedded in $a$RGB space}
\label{sec:a_math}
\input{sections/A_math}

\section{Implementation detail}
\label{sec:b_impl_detail}
\input{sections/B_impl_detail}

\section{More quantitative results on perceptual image super-resolution}
\label{sec:d_esrgan}
\input{sections/C_esrgan}

\section{More qualitative results}
\label{sec:c_more_result}
\input{sections/D_more_result}

\section{Understanding the $a$RGB representation space}
\label{sec:f_decomposition}
\input{sections/E_decomposition}

\end{document}

%% file: macro.tex
\newcommand{\xhat}{\hat{\bm{x}}}

\newcommand{\yy}{\bm{y}}
\newcommand{\zz}{\bm{z}}

\newcommand{\xxi}{\bm{\xi}}

\newcommand{\cmark}{\ding{51}}%
\newcommand{\xmark}{\ding{55}}%

\usepackage{arydshln}

\makeatletter
\def\adl@drawiv#1#2#3{%
        \hskip.5\tabcolsep
        \xleaders#3{#2.5\@tempdimb #1{1}#2.5\@tempdimb}%
                #2\z@ plus1fil minus1fil\relax
        \hskip.5\tabcolsep}
\newcommand{\cdashlinelr}[1]{%
  \noalign{\vskip\aboverulesep
           \global\let\@dashdrawstore\adl@draw
           \global\let\adl@draw\adl@drawiv}
  \cdashline{#1}
  \noalign{\global\let\adl@draw\@dashdrawstore
           \vskip\belowrulesep}}
\makeatother

%% file: math_command.tex
\usepackage{amsmath,amsfonts,bm}

\def\eqref#1{equation~\ref{#1}}

\def\1{\bm{1}}

\def\vb{{\bm{b}}}

\def\vk{{\bm{k}}}

\def\vv{{\bm{v}}}

\def\mA{{\bm{A}}}

\def\mI{{\bm{I}}}

\def\mW{{\bm{W}}}

\DeclareMathAlphabet{\mathsfit}{\encodingdefault}{\sfdefault}{m}{sl}
\SetMathAlphabet{\mathsfit}{bold}{\encodingdefault}{\sfdefault}{bx}{n}

\newcommand{\R}{\mathbb{R}}

\DeclareMathOperator*{\argmax}{arg\,max}
\DeclareMathOperator*{\argmin}{arg\,min}

%% file: sections/0_abstract.tex
\begin{abstract}
Image restoration models are typically trained with a pixel-wise distance loss defined over the RGB color representation space, which is well known to be a source of blurry and unrealistic textures in the restored images.
The reason, we believe, is that the three-channel RGB space is insufficient for supervising the restoration models.
To this end, we augment the representation to hold structural information of local neighborhoods at each pixel while keeping the color information and pixel-grainedness unharmed.
The result is a new representation space, dubbed augmented RGB~($a$RGB) space.
Substituting the underlying representation space for the per-pixel losses facilitates the training of image restoration models, thereby improving the performance without affecting the evaluation phase.
Notably, when combined with auxiliary objectives such as adversarial or perceptual losses, our $a$RGB space consistently improves overall metrics by reconstructing both color and local structures, overcoming the conventional perception-distortion trade-off.
\end{abstract}

%% file: sections/1_introduction.tex
\section{Introduction}
\label{sec:1_intro}

Since VDSR~\cite{cv:sr:kim16-vdsr} and EDSR~\cite{cv:sr:lim17-edsr}, training an image-to-image deep neural network has been a promising practice for dealing with image restoration tasks.
This has led to much interest in the learning objectives for better supervision of image restoration networks.
Most of the works have focused on exploiting semantic prior knowledge~\cite{cv:sr:wang18-esrgan,cv:sr:zhang19-ranksrgan,park2023content}, typically in the form of adversarial \cite{cv:gan:goodfellow14-gan} and VGG perceptual losses \cite{cv:obj:johnson16-perceptualloss}.
These loss functions are regarded as add-ons to a per-pixel RGB distance, which has been a unanimous choice regardless of the underlying restoration problem.

\input{sections/figures/concept}

Unfortunately, as highlighted in~\cite{cv:sr:ledig17-srgan}, this per-pixel loss defined over the RGB color representation is the primary cause of blurriness commonly found in the restoration results.
We attribute these well-known shortcomings to the lack of local structural information within the three-dimensional feature at each pixel.
Since per-pixel distance metrics are applied to each pixel independently, the network is guided towards a mean RGB value estimator for each pixel.
Hence, the result appears to be blurry in a global sense.
Nevertheless, the per-pixel RGB color difference has been considered a necessary evil due to its \emph{pixel-grained} supervision.
The image restoration model should be trained to preserve the extremely dense, pixel-grained correspondence between the low-quality inputs, the reconstructions, and the ground truth images.
This is the reason previous solutions utilizing auxiliary loss functions such as perceptual loss~\cite{cv:obj:johnson16-perceptualloss} and adversarial loss~\cite{cv:sr:ledig17-srgan,cv:sr:blind:zhang21-bsrgan} still rely on the per-pixel loss in the RGB space.
Perceptual losses using classifier backbones reduce the spatial dimension of the image, and thus fail to provide pixel-perfect structural information even if they are used along with a per-pixel RGB difference loss.
Adversarial losses, on the other hand, do not rely on the pairwise distances but only on the distributional shift of each unit patch received by the discriminator network, inevitably leading to inaccurate restoration.
Consequently, the common practice of mixing the per-pixel RGB distance with those additional losses cannot provide accurate, pixel-grained supervision of color and local structure.

Instead, we seek our solution by directly changing the per-pixel loss.
Focusing on the deficiency of information in the underlying RGB representation space, we propose to \emph{augment} this representation with local structural information.
This leads to our \emph{augmented RGB}~($a$RGB) \emph{representation space}, serving as a substitute for the traditional RGB space over which per-pixel losses are defined.
Our solution relies on a translating autoencoder consisting of a nonlinear, mixture-of-experts~\cite{ml:moe:jacobs91-mixture_of_experts} encoder and a linear decoder that translates images between the RGB and the $a$RGB spaces.
This architecture ensures almost perfect preservation of the color information ($> 60\,\mathrm{dB}$ PSNR) while embedding diverse, multimodal distributions of local image structure.

Overall, our contributions can be summarized as follows:
\begin{itemize}[leftmargin=1em]
    \item We present a novel approach to solve the perception-distortion trade-off by training a network over alternative representation space—our $a$RGB space.
    This involves creating a novel autoencoder to obtain the $a$RGB space.

    \item With only a few lines of additional code, our method is seamlessly applied to any existing restoration models.

    \item The pixel-wise loss over our $a$RGB space not only enhances distortion metrics but also consistently improves perceptual metrics when combined with traditional perceptual objectives, all without affecting the testing phase.

    \item We provide comprehensive analysis on our $a$RGB space for interpretable low-level image representation.
\end{itemize}

%% file: sections/figures/concept.tex
\begin{figure}[t]
\centering
\includegraphics[width=.98\linewidth]{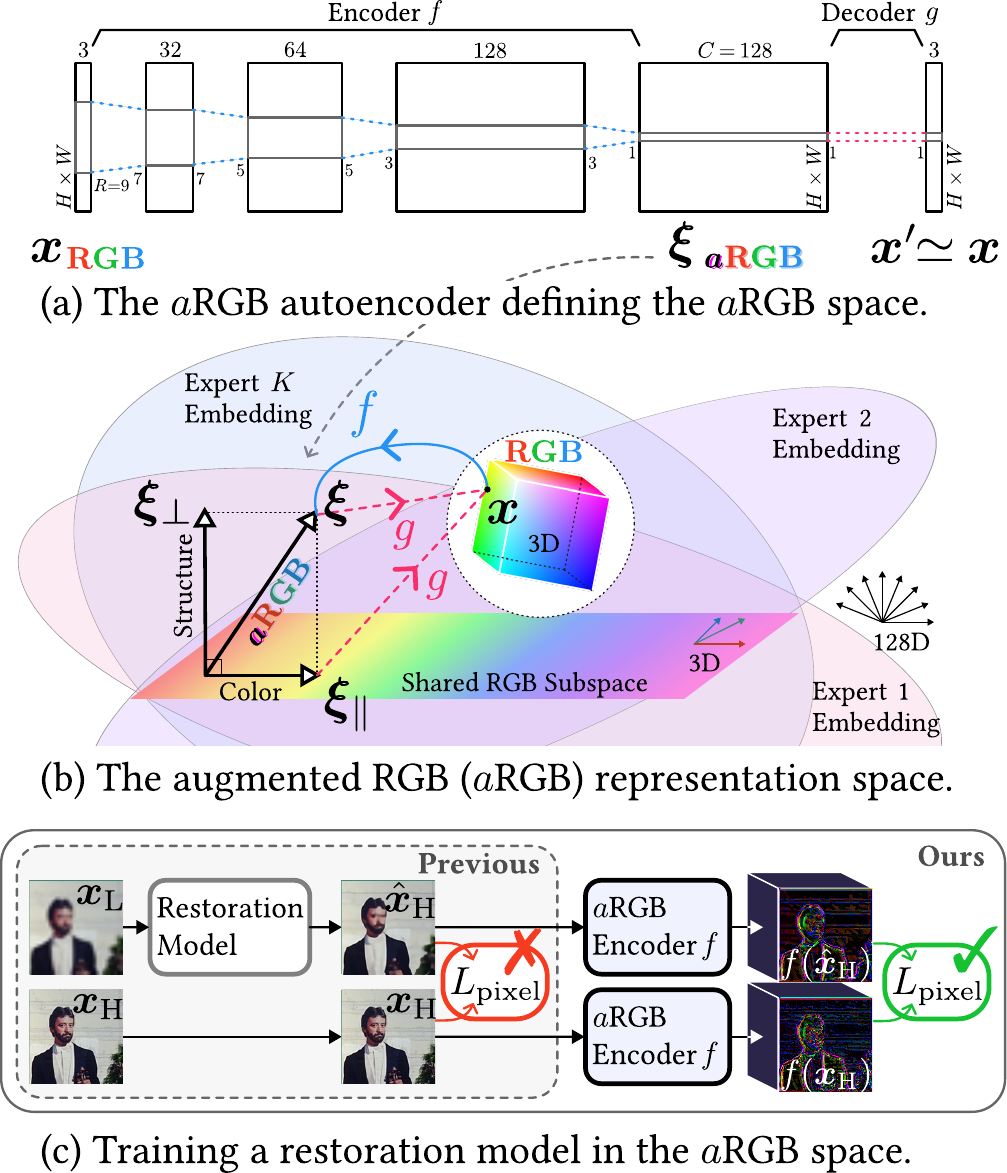}
\vspace{-2mm}
\caption{
\small
\textbf{The $a$RGB representation space.}
Our augmented RGB~($a$RGB) space is designed to replace the RGB space for calculating per-pixel losses to train image restoration models.
Unlike conventional per-pixel distances over the RGB space, the same distances defined over our $a$RGB space convey pixel-grained structural information, which is crucial for high-fidelity image reconstruction.
Our $a$RGB space also enjoys interpretability, for any embedding $\xxi$ is orthogonally decomposable into the color encoding $\xxi_{\parallel}$ and the structure encoding parts $\xxi_{\perp}\,$.
}
\label{fig:concept}
\vspace{-1.5em}
\end{figure}

%% file: sections/2_related_works.tex
\section{Related Work}
\label{sec:6_related_work}

\paragraph{Pairwise loss in image restoration.}
Training a deep neural network that translates low-quality images into high-quality estimates has undoubtedly become the standard way of solving image restoration.
While most of the advancements have been made in the network architecture \cite{cv:sr:kim16-vdsr,cv:sr:lim17-edsr,cv:deblur:nah17-deepdeblur,cv:sr:tong17-srdensenet,cv:sr:wang18-esrgan,cv:sr:zhang18-rcan,cv:deblur:zamir21-mprnet,cv:res:liang21-swinir,cv:res:waqas_zamir22-restormer,cv:res:chen22-nafnet}, the importance of loss functions is also widely acknowledged.
Since SRCNN \cite{cv:sr:dong16-srcnn}, the first pioneer, employed the MSE loss, the first image restoration models had been trained with the MSE loss \cite{cv:sr:kim16-drcn,cv:sr:kim16-vdsr,cv:deblur:nah17-deepdeblur,cv:denoise:zhang17-dncnn}.
However, after EDSR \cite{cv:sr:lim17-edsr} reported that better convergence can be achieved with $L_{1}$ loss, various pairwise loss functions are explored.
LapSRN \cite{cv:sr:lai17-lapsrn} rediscovers Charbonnier loss \cite{cv:obj:bruhn05-charbonnierloss}, a type of smooth $L_{1}$ loss, for image super-resolution, which is also employed in image deraining \cite{cv:derain:jiang20-mspfn} with a new edge loss, defined as a Charbonnier loss between Laplacians, which is then employed in general restoration by MPRNet \cite{cv:deblur:zamir21-mprnet}.
NAFNet \cite{cv:res:chen22-nafnet}, on the other hand, uses the PSNR score directly as a loss function.
In accordance with these approaches, we attempt a more general approach to design a representation space over which those loss functions can be redefined.

\paragraph{Structural prior of natural images.}
It is generally recognized that a convolutional neural network, either trained \cite{cv:cls:simonyan15-vgg} or even untrained \cite{cv:misc:prior:ulyanov18-dip}, contains structural prior that resonates with the internal structure of natural images.
Attempts to exploit this information include the perceptual loss \cite{cv:obj:johnson16-perceptualloss,cv:obj:zhang18-lpips,cv:obj:Ding20-dists}.
Adversarial losses \cite{cv:gan:goodfellow14-gan} can also be seen as utilization of structural priors \cite{cv:sr:wang18-esrgan,cv:sr:zhang19-ranksrgan,park2023content}, as they rely on the gradients calculated from the structural differences between real and restored images.
On the other hand, dual domain-based losses \cite{cui2023selective,cho21-mimounet} seek a way to provide supervision regarding nonlocal structures by calculating difference in the Fourier domain.
However, all of those losses are auxiliary, and thus cannot eliminate the strong averaging effect of the pixel-wise loss over the RGB space.
Instead, we directly \emph{replace} the RGB space with our $a$RGB space that is both color-preserving, pixel-perfect, and contains pixel-grained structural information.
Our approach, therefore, can be orthogonally used with the auxiliary losses, such as perceptual loss, for better performance.

\input{sections/figures/model}

\paragraph{Mixture of Experts.}
Instead of relying on a single model to handle complex large-scale data, a more effective approach is to distribute the workload among multiple workers--the \emph{experts}.
Mixture of Experts~(MoE)~\cite{ml:moe:jacobs91-mixture_of_experts}, now a classic paradigm in machine learning, consists of a routing strategy~\cite{shazeer2017outrageously} and multiple expert models, each of which processes a subset of the training data partitioned and given by the router.
Recent studies~\cite{zhou2022mixture,nlp:fedus21-switch_transformer} have shown the advantages of MoE in deep learning.
Two main challenges arise when working with MoE in deep learning: limited computational resources and training stability \cite{2021arXiv210103961F,he2021fastmoe}.
In response to these challenges, we employ a balancing loss~\cite{nlp:fedus21-switch_transformer} to ensure the stable training of expert networks and incorporate MoE exclusively during the training phase, leaving the testing phase unaffected.

%% file: sections/figures/model.tex
\begin{figure*}[t]
\centering
\includegraphics[width=\linewidth]{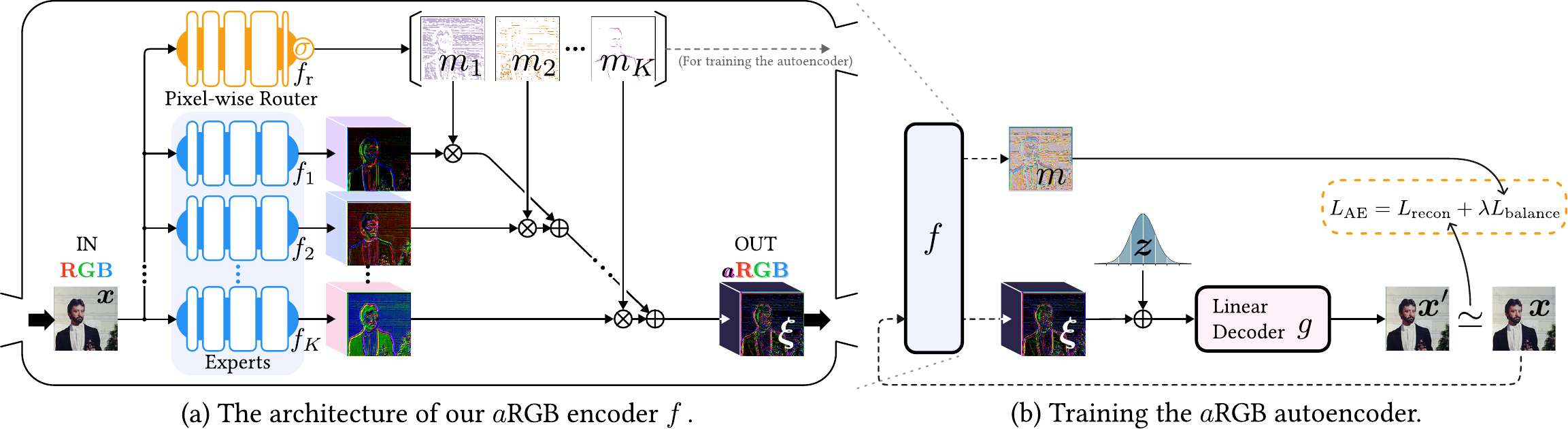}
\vspace{-1em}
\caption{
\small
\textbf{The design and the training of the $a$RGB autoencoder.}
Defined with a mixture-of-experts encoder and a linear decoder, the $a$RGB autoencoder translates an RGB image into $a$RGB and back.
This design allows us to imbue gradient-based supervision from any per-pixel distance loss with rich pixel-grained structural information, while preserving color information.
After training the $a$RGB encoder as an autoencoder fashion, it remains frozen during the training of restoration models.
}
\label{fig:model}
\vspace{-3mm}
\end{figure*}

%% file: sections/3_propsed_methods.tex
\section{Lifting the RGB to $a$RGB}
\label{sec:3_method}

\subsection{The $a$RGB Autoencoder}
\label{sec:3_method:autoencoder}
Our primary goal is to design a representation space for low-level vision tasks in order to facilitate training of image restoration networks.
Designing a representation space is achieved by defining the encoder and the decoder to translate images back and forth between the RGB space and the target space.
We can split our goal into two parts: (1) the feature at each \emph{pixel} in our space is required to encode its neighboring structure, and (2) the integrity of the color information should be preserved.
To fulfill the first requirement, our encoder is a size-preserving ConvNet with nonlinearities to capture the structure among adjacent pixels.
For the latter, we employ a per-pixel linear decoder, \ie, a $1 \times 1$ convolution, to strongly constrain the embedding of a pixel to include its RGB color information.
The overall architecture is illustrated in Figure~\ref{fig:model}.

\input{sections/tables/deblur}
\input{sections/figures/result_denoise}
\input{sections/figures/result_deblur}

We start from an RGB image $\xx \in \R^{3 \times H \times W}\,$.
Our convolutional encoder $f$ transforms image $\xx$ into a feature $\xxi \in \R^{C \times H \times W}$ of a new representation space.
Unlike typical undercomplete autoencoders, which remove information from their inputs, we aim to \textit{add more} information regarding local structures for each pixel $[\xxi]_{ij}$ at coordinate $(i, j)\,$.
Therefore, $C$ must be greater than 3, and the receptive field size $R$ should be greater than unity.
Our decoder $g: \xxi \mapsto \xx$ is effectively a single $1 \times 1$ convolution.
That is, we can express $g([\xxi]_{ij})$ as a per-pixel linear operation: $g([\xxi]_{ij}) = \mA [\xxi]_{i j} + \vb\,$, where $\mA \in \R^{3 \times C}$ and $\vb \in \R^{3}\,$.
This ensures that each feature $[\xxi]_{ij}$ in our representation space extends the color information presented in $[\xx]_{ij}\,$, hence the name of our new representation, \emph{augmented} RGB.
Additionally, using a linear decoder~$g$ offers an interpretability: we can regard the nullspace of $\mA\,$, \ie, the set of undecoded information, as a reservoir of any extra information captured by the encoder $f$ other than local colors.

What is crucial at this juncture is to define our $a$RGB space to effectively capture the highly varying, complex mixture of information from the color and the neighboring structure at each pixel.
To this end, we employ a mixture-of-experts~(MoE) architecture \citep{ml:moe:jacobs91-mixture_of_experts,ml:moe:shazeer17-sparsely_gated_moe,nlp:fedus21-switch_transformer} within our encoder.
We choose this design based on our conjecture that the topology of the space of image patches is disconnected and, therefore, can be more efficiently modeled with a MoE architecture than a single ConvNet.
For the set of the smallest images, \ie, a set of pixels, we can argue that their domain is a connected set under absence of quantization, since a pixel can take arbitrary color value.
This does not hold in general if the size of the patches becomes large enough to contain semantic structures.
In fact, we cannot interpolate between two images of semantically distinct objects \emph{in the natural image domain}, \eg, there is no such thing as a half-cat half-airplane object \emph{in nature}.
This implies that topological disconnectedness emerges from the domain of patches as the size of its patches increases.
Since a single-module encoder is a continuous function, learning a mapping over a disconnected set may require deeper architecture with a lot of parameters.
An MoE encoder, per contra, can model a discontinuous map more effectively through its discrete routing strategy between small, specialized experts.
We will revisit our conjecture in Section~\ref{sec:5_discussion}.

In practice, an RGB image $\xx \in \R^{3 \times H \times W}$ is fed into the router $f_{\text{r}}$ as well as $K$ encoders $f_{1}, \ldots, f_{K}\,$.
The router $f_{\text{r}}$ is a five-layer ConvNet classifier with a softmax at the end.
The output of the router $\yy = f_{\text{r}}(\xx) \in [0, 1]^{K \times H \times W}$ partitions each pixel of $\xx$ into $K$ different bins with top-$1$ policy.
This is equivalent to generating mutually exclusive and jointly exhaustive $K$ masks $m_{1}, \ldots, m_{K}$ of size $H \times W\,$.
The features $\xxi_{1} = f_{1}(\xx), \ldots, \xxi_{K} = f_{K}(\xxi)$ are aggregated into a single feature $\xxi = f(\xx) \in \R^{C \times H \times W}\,$:
\begin{align}
\label{eq:mixture_of_experts}
    \xxi &= \sum_{k = 1}^{K} m_{k} \odot f_{k} (\xx) \\
\label{eq:mixture_of_experts2}
    &= \sum_{k = 1}^{K} \mathbbm{1}(k = \argmax_{k'} [f_{\text{r}}(\xx)]_{k'}) \odot f_{k} (\xx)\,,
\end{align}
where $\odot$ is an element-wise multiplication and $\mathbbm{1}$ is the indicator function.
We ensure that $(g \circ f)(\xx) = \xx' \simeq \xx$ by training $f$ and $g$ jointly in an autoencoder scheme.
After the training, the decoder $g$ is discarded and the encoder $f$ is used to generate $a$RGB representations from RGB images.

\subsection{Training the Autoencoder}
\label{sec:3_method:training}
Our objective is to ensure that the $a$RGB encoder~$f$ effectively learns accurate low-level features from clean~(or sharp) and natural images.
To achieve this goal, we make use of a dataset~$D$, consisting of clean image patches.
With this dataset, the $a$RGB autoencoder is trained to minimize the $L_{1}$ distance between a patch $\xx \in D$ and its reconstruction $(g \circ f)(\xx)\,$.
In addition, likewise in Switch Transformer \citep{nlp:fedus21-switch_transformer}, a load-balancing loss $L_{\text{balance}}$ is applied to encourage the router $f_{\text{r}}$ to distribute pixels evenly across the $K$ experts during training:
\begin{equation}
\label{eq:balancing_loss}
L_{\text{balance}} = K^{2} \sum_{i = 1}^{H} \sum_{j = 1}^{W} \left[ \max_{k} [f_{\text{r}}(\xx)]_{k} \right]_{ij}\,,
\end{equation}
which is minimized when the distribution is uniform with the value of unity.
Furthermore, to increase the sensitivity of the encoder $f\,$, we simply add an isotropic Gaussian noise at the output of the encoder only during the training of the $a$RGB autoencoder.
Therefore, the reconstruction loss for the autoencoder is:
\begin{equation}
\label{eq:reconstruction_loss}
L_{\text{recon}} = \| g(f(\xx) + \zz) - \xx \|_{1}\,,
\end{equation}
where $\zz \sim \mathcal{N}(\bm{0}, \bm{I})\,$.
Although the decoder is only informed with three color channels of each pixel during the training, we observe that the latent space does not degenerate into trivial solutions.
See Appendix~\ref{sec:a_math} for more information.
Overall, the training loss for the $a$RGB autoencoder is:
\begin{equation}
\label{eq:autoencoder_loss}
L_{\text{AE}} = L_{\text{recon}} + \lambda L_{\text{balance}}\,.
\end{equation}
In practice, we choose $\lambda = 0.01\,$.
The final autoencoder achieves 67.21~$\mathrm{dB}$ in the reconstruction of the Set5 benchmark \citep{cv:data:bevilacqua12-set5}.
In other words, the average RGB color difference is below a tenth of the quantization step.
Henceforth, we will consider our $a$RGB autoencoder lossless in the analysis in Section~\ref{sec:5_discussion}.
More implementation details are provided in Appendix~\ref{sec:b_impl_detail}.

\subsection{Integration into Existing Restoration Frameworks}
Figure~\ref{fig:concept}c illustrates the overall pipeline for integrating our $a$RGB into the existing restoration framework.
The interface is similar to the famous perceptual loss~\cite{cv:obj:johnson16-perceptualloss}, but unlike perceptual loss, the resulting $a$RGB losses are pixel-grained and color-preserving, and completely replaces the RGB per-pixel loss.
Training image restoration models with respect to the $a$RGB space only requires a few lines of code modification.
Typically, an image restoration model is trained to minimize a per-pixel distance $L_{\text{pixel}}\,$, optionally combined with some auxiliary losses $L_{\text{aux}}\,$, such as a perceptual loss \citep{cv:obj:johnson16-perceptualloss} or an adversarial loss \citep{cv:sr:ledig17-srgan}.
The overall loss function can be represented as:
\begin{equation}
\label{eq:restoration_loss}
L_{\text{RGB}} (\xx, \xhat) = L_{\text{pixel}} (\xx, \xhat) + L_{\text{aux}} (\xx, \xhat)\,,
\end{equation}
where $\xx$ is the ground-truth image and  $\xhat$ is the restoration result.
To train the model in the $a$RGB space, we are only required to modify the input to the per-pixel loss $L_{\text{pixel}}$.
That is, the per-pixel distances are now computed between the images in the $a$RGB space, namely, $f(\xx)$ and $f(\xhat)\,$.
\begin{equation}
\label{eq:restoration_loss_aRGB}
L_{\text{$a$RGB}} (\xx, \xhat) = L_{\text{pixel}} (f(\xx), f(\xhat)) + L_{\text{aux}} (\xx, \xhat)\,.
\end{equation}
Since what we present is not a specific loss function but the underlying \emph{space} itself, our method can be seamlessly integrated with any existing restoration framework regardless of the type of per-pixel loss it uses.

%% file: sections/tables/deblur.tex
\begin{table*}[t]
\begin{minipage}{.426\linewidth}
    \caption{\small
    \textbf{Results on real image denoising.}
    }
    \label{tab:result_denoise}
    \centering
    \vspace{.5em}
    \resizebox{\linewidth}{!}{%
    \begin{tabular}{llcc}
        \toprule
        & & \multicolumn{2}{c}{{SIDD}} \\
        \cmidrule(lr){3-4} 
        Model & Objective & PSNR$\uparrow$ & SSIM$\uparrow$ \\
        \midrule
        NAFNet-width32 & $L_{\text{PSNR}}$ & 39.9672 & 0.9599 \\
        NAFNet-width32 & {\color{BrickRed} $L_{\text{PSNR}, \text{$a$RGB}}$ } & \textbf{39.9864} & \textbf{0.9601} \\
        NAFNet-width32 & {\color{BrickRed} $L_{1, \text{$a$RGB}}$} & \textbf{40.0106} & \textbf{0.9602} \\
        \cdashlinelr{1-4}
        NAFNet-width64 & $L_{\text{PSNR}}$ & 40.3045 & 0.9614 \\
        NAFNet-width64 & {\color{BrickRed} $L_{1, \text{$a$RGB}}$} & \textbf{40.3364} & \textbf{0.9620} \\
        \bottomrule
    \end{tabular}%
    }%
\end{minipage}
\hfill
\begin{minipage}{.56\linewidth}
    \caption{\small
    \textbf{Results on motion blur deblurring.}
    }
    \label{tab:result_deblur}
    \centering
    \vspace{.5em}
    \resizebox{\linewidth}{!}{%
    \begin{tabular}{llcccc}
        \toprule
        & & \multicolumn{2}{c}{{GoPro}} & \multicolumn{2}{c}{{HIDE}} \\
        \cmidrule(lr){3-4} \cmidrule(lr){5-6}
        Model & Objective & PSNR$\uparrow$ & SSIM$\uparrow$ & PSNR$\uparrow$ & SSIM$\uparrow$ \\
        \midrule
        DeepDeblur & $L_{1}$ & 27.8333 & 0.8707 & 25.7313 & 0.8400 \\
        DeepDeblur & ${\color{BrickRed} L_{1, \text{$a$RGB}}}$ & \textbf{27.8409} & \textbf{0.8710} & \textbf{25.8146} & \textbf{0.8424} \\
        \cdashlinelr{1-6}
        MPRNet & $L_{\text{Char}} \qquad\, + 0.05 L_{\text{Edge}}$ & 32.6581 & 0.9589 & 30.9622 & 0.9394 \\
        MPRNet & ${\color{BrickRed} L_{\text{Char}, \text{$a$RGB}}} + 0.05 L_{\text{Edge}}$ & \textbf{32.7118} & \textbf{0.9594} & \textbf{31.0248} & \textbf{0.9398} \\
        \cdashlinelr{1-6}
        MPRNet-TLC & $L_{\text{Char}} \qquad\, + 0.05 L_{\text{Edge}}$ & 33.3137 & 0.9637 & 31.1868 & 0.9418 \\
        MPRNet-TLC & ${\color{BrickRed} L_{\text{Char}, \text{$a$RGB}}} + 0.05 L_{\text{Edge}}$ & \textbf{33.3886} & \textbf{0.9642} & \textbf{31.2082} & \textbf{0.9421} \\
        \bottomrule
    \end{tabular}%
    }%
\end{minipage}
\vspace{-.1em}
\end{table*}

%% file: sections/figures/result_denoise.tex
\begin{figure*}
\vspace{-1mm}
\newcommand{\figwidth}{0.1418\linewidth}
\newcommand{\varA}{LQ}%
\newcommand{\varB}{N32}%
\newcommand{\varC}{N32+LPaRGB}%
\newcommand{\varD}{N32+L1aRGB}%
\newcommand{\varE}{N64}%
\newcommand{\varF}{N64+L1aRGB}%
\newcommand{\varG}{GT}%
\centering
\newcommand{\seed}{0564}%
    \subfloat{\includegraphics[width=\figwidth]{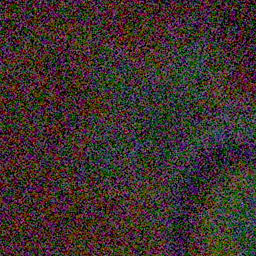}}
    \hfill
    \subfloat{\includegraphics[width=\figwidth]{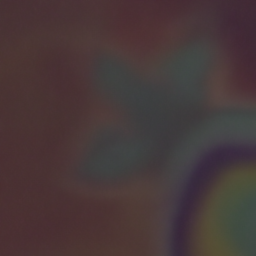}}
    \hfill
    \subfloat{\includegraphics[width=\figwidth]{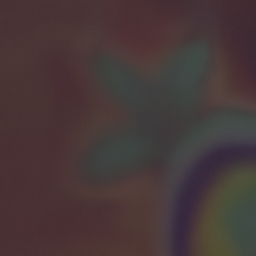}}
    \hfill
    \subfloat{\includegraphics[width=\figwidth]{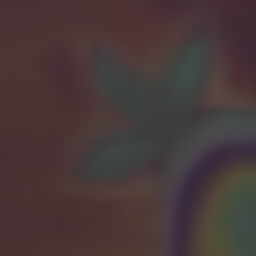}}
    \hfill
    \subfloat{\includegraphics[width=\figwidth]{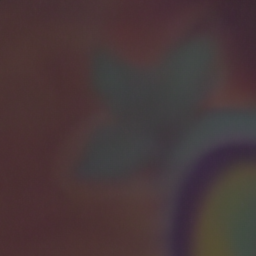}}
    \hfill
    \subfloat{\includegraphics[width=\figwidth]{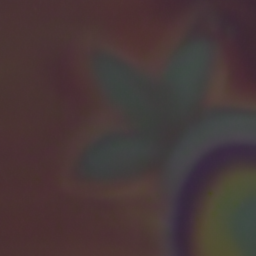}}
    \hfill
    \subfloat{\includegraphics[width=\figwidth]{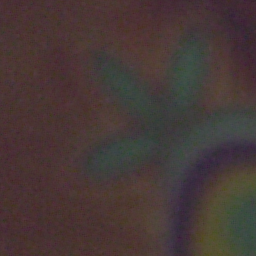}}
    \\
    \addtocounter{subfigure}{-7}
    \subfloat[Noisy\\ \hspace*{1.3em}\resizebox{3.5em}{!}{17.1647 $\mathrm{dB}$}]{\includegraphics[width=\figwidth]{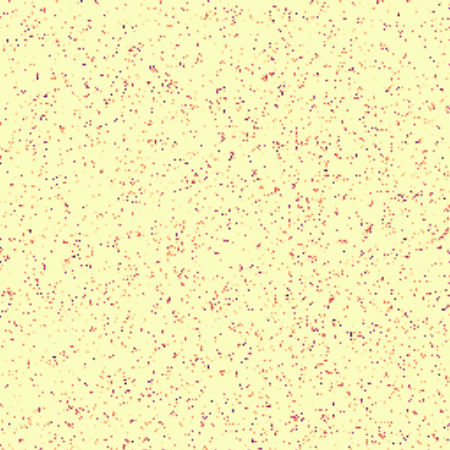}}
    \hfill
    \subfloat[N32 $L_{\text{PSNR}}$\\ \hspace*{1.3em}\resizebox{3.5em}{!}{36.0312 $\mathrm{dB}$}]{\includegraphics[width=\figwidth]{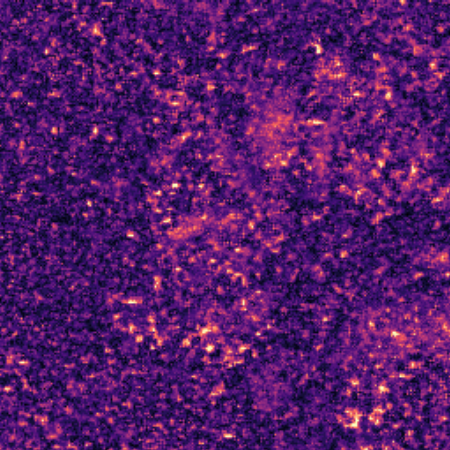}}
    \hfill
    \subfloat[{\scriptsize N32 {\color{BrickRed}$L_{\text{PSNR}, \text{$a$RGB}}$}}\\ \hspace*{1.3em}\resizebox{3.5em}{!}{\textbf{36.3540} $\mathrm{dB}$}]{\includegraphics[width=\figwidth]{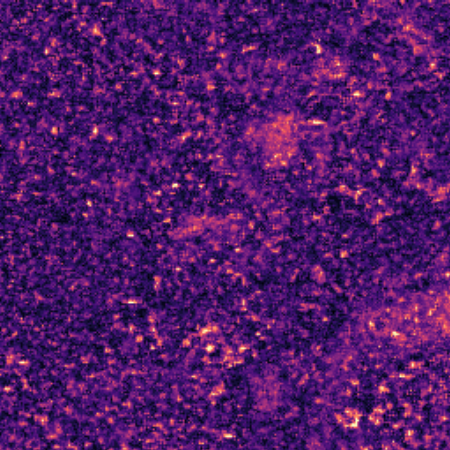}}
    \hfill
    \subfloat[\resizebox{4.9em}{!}{N32 {\color{BrickRed}$L_{1, \text{$a$RGB}}$}}\\ \hspace*{1.3em}\resizebox{3.5em}{!}{\textbf{36.3747} $\mathrm{dB}$}]{\includegraphics[width=\figwidth]{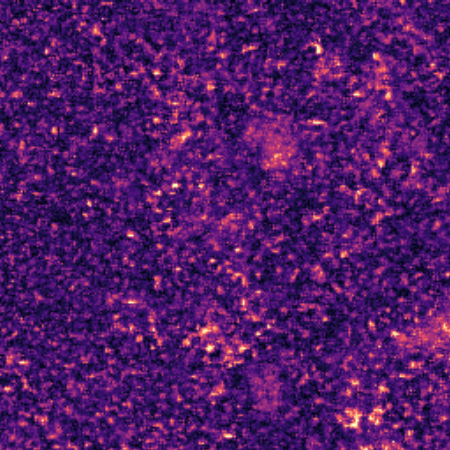}}
    \hfill
    \subfloat[N64 $L_{\text{PSNR}}$\\ \hspace*{1.3em}\resizebox{3.5em}{!}{35.5460 $\mathrm{dB}$}]{\includegraphics[width=\figwidth]{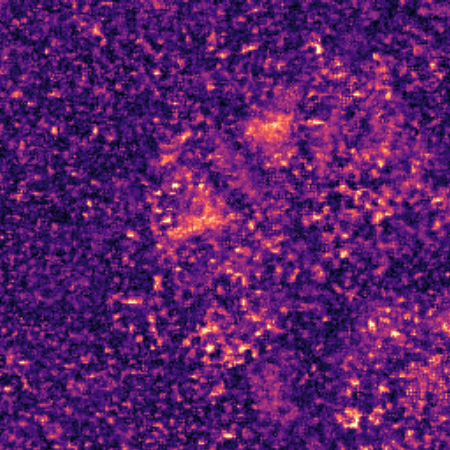}}
    \hfill
    \subfloat[\resizebox{4.9em}{!}{N64 {\color{BrickRed}$L_{1, \text{$a$RGB}}$}}\\ \hspace*{1.3em}\resizebox{3.5em}{!}{\textbf{36.4639} $\mathrm{dB}$}]{\includegraphics[width=\figwidth]{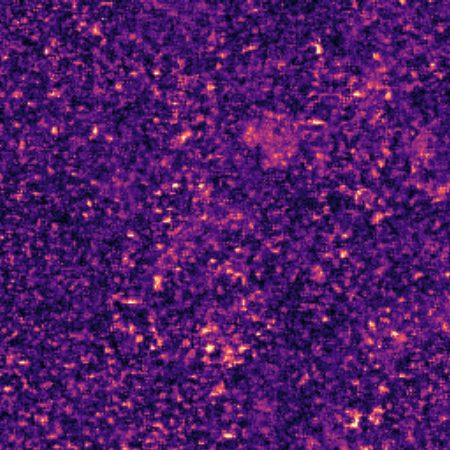}}
    \hfill
    \subfloat[Clean GT\\ \hspace*{1.4em}\resizebox{3.5em}{!}{PSNR ($\mathrm{dB}$)}]{\vspace*{-0.72mm}\includegraphics[width=\figwidth]{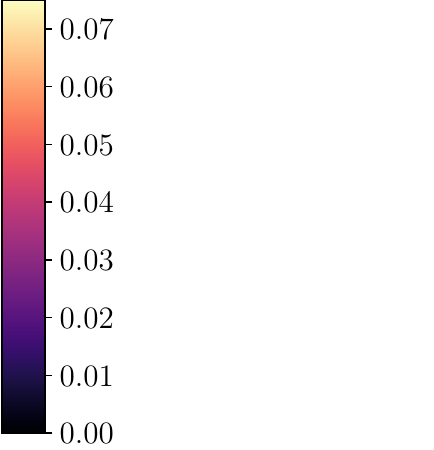}}
\\
\vspace{-1mm}
\caption{\small
    \textbf{Qualitative comparison of real image denoising models trained with different loss functions.}
    Each column corresponds to each row in Table~\ref{tab:result_denoise}.
    N32 corresponds to NAFNet-width32 and N64 corresponds to NAFNet-width64.
    The bottom row shows the maximum absolute difference in color with a range of $[0, 1]\,$.
}
\label{fig:result_denoise}
\vspace{-3mm}
\end{figure*}

%% file: sections/figures/result_deblur.tex
\begin{figure*}[t]
\newcommand{\figwidth}{0.165\linewidth}
\newcommand{\patchsize}{192_256}
\newcommand{\varA}{LQ}%
\newcommand{\varB}{Lchar}%
\newcommand{\varC}{LcharaRGB}%
\newcommand{\varD}{Lchar+TLC}%
\newcommand{\varE}{LcharaRGB+TLC}%
\newcommand{\varF}{GT}%
\centering
\newcommand{\seed}{HIDE_34fromGOPR1089.MP4_255_575}%
    \subfloat{\includegraphics[width=\figwidth]{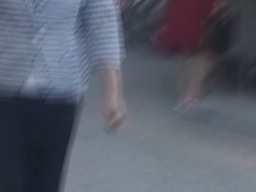}}
    \hfill
    \subfloat{\includegraphics[width=\figwidth]{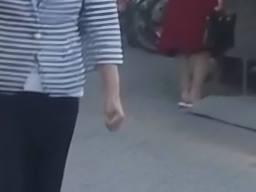}}
    \hfill
    \subfloat{\includegraphics[width=\figwidth]{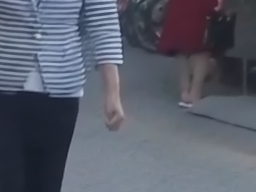}}
    \hfill
    \subfloat{\includegraphics[width=\figwidth]{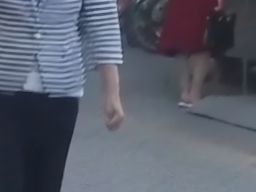}}
    \hfill
    \subfloat{\includegraphics[width=\figwidth]{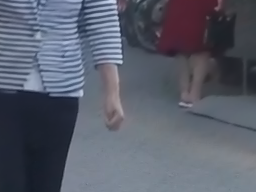}}
    \hfill
    \subfloat{\includegraphics[width=\figwidth]{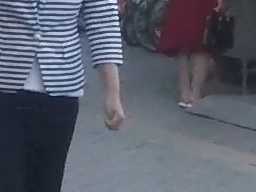}}
    \\[-0.14em]
    \addtocounter{subfigure}{-6}
    \subfloat[Blurry\\ \hspace*{1.3em}\resizebox{3.5em}{!}{21.1887 $\mathrm{dB}$}]{\includegraphics[width=\figwidth]{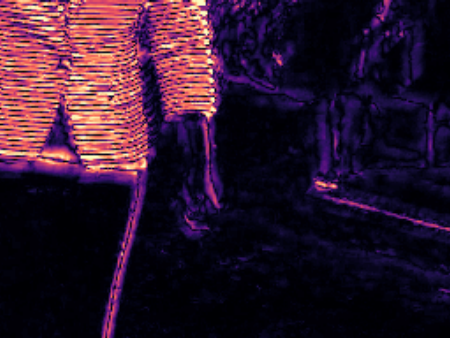}}
    \hfill
    \subfloat[$L_{\text{Char}}$\\ \hspace*{1.3em}\resizebox{3.5em}{!}{18.9074 $\mathrm{dB}$}]{\includegraphics[width=\figwidth]{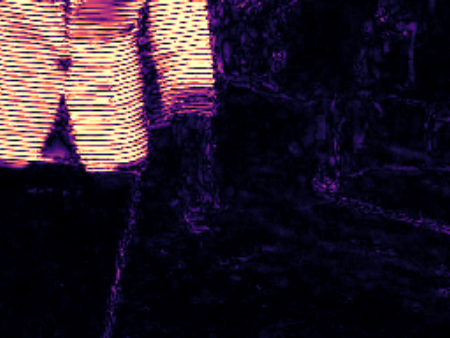}}
    \hfill
    \subfloat[{\color{BrickRed}$L_{\text{Char}, \text{$a$RGB}}$}\\ \hspace*{1.3em}\resizebox{3.5em}{!}{\textbf{19.2360} $\mathrm{dB}$}]{\includegraphics[width=\figwidth]{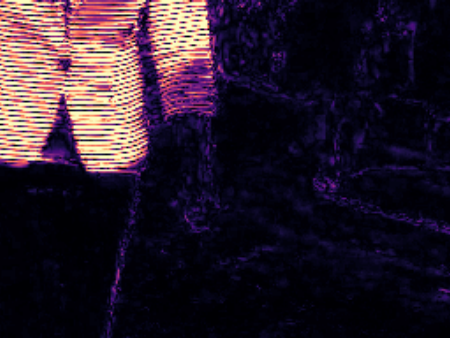}}
    \hfill
    \subfloat[$L_{\text{Char}}\,$,TLC\\ \hspace*{1.3em}\resizebox{3.5em}{!}{19.3418 $\mathrm{dB}$}]{\includegraphics[width=\figwidth]{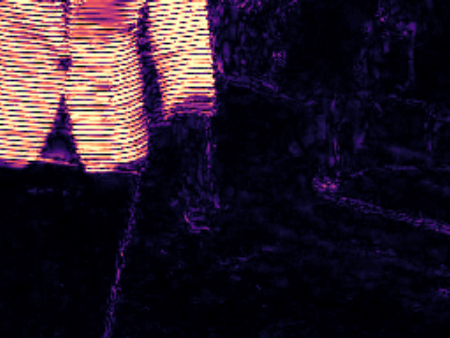}}
    \hfill
    \subfloat[\resizebox{6em}{!}{{\color{BrickRed}$L_{\text{Char}, \text{$a$RGB}}\,$},TLC}\\ \hspace*{1.3em}\resizebox{3.5em}{!}{\textbf{23.3990} $\mathrm{dB}$}]{\includegraphics[width=\figwidth]{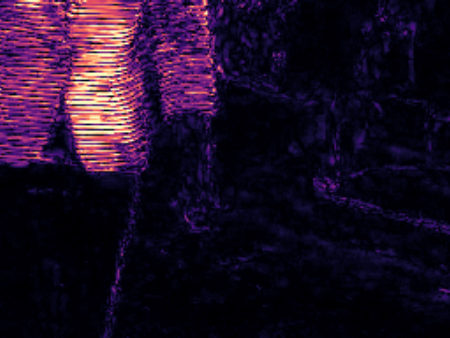}}
    \hfill
    \subfloat[Clean GT\\ \hspace*{1.4em}\resizebox{3.5em}{!}{PSNR ($\mathrm{dB}$)}]{\vspace*{-1.02mm}\includegraphics[width=\figwidth]{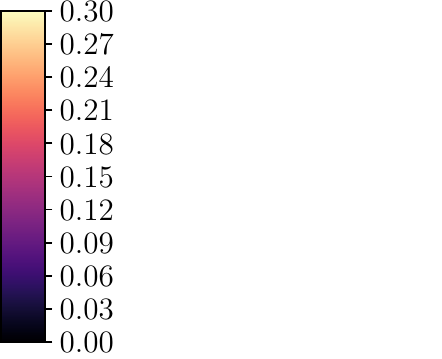}}
\\
\vspace{-1mm}
\caption{\small
    \textbf{Qualitative comparison of motion blur deblurring models trained with different loss functions.}
    Each column corresponds to each row in Table~\ref{tab:result_deblur}.
    The bottom row is the maximum absolute RGB difference.
}
\label{fig:result_deblur}
\vspace{-3mm}
\end{figure*}

%% file: sections/4_experiments.tex
\section{Experiments}
\label{sec:4_experiment}

\input{sections/tables/sr}
\input{sections/figures/result_gan}

\subsection{Comparison with Previous Methods}
To demonstrate our proposed $a$RGB space, we conduct experiments across various image restoration tasks, including denoising, deblurring, and super-resolution.
We also show the versatility of our representation space-based approach by applying diverse types of pixel-wise losses over the $a$RGB space, \eg, $L_{1}\,$, PSNR and Charbonnier loss.
Furthermore, we empirically show the benefit of combining our per-pixel $a$RGB loss function with conventional perceptual quality-oriented objectives, including VGG and adversarial losses.

\paragraph{Distortion-oriented:}
Tables~\ref{tab:result_denoise} and \ref{tab:result_deblur} demonstrate the efficacy of our $a$RGB representation across various pixel-wise loss objectives.
In all comparisons, we substitute the RGB space with our $a$RGB space in the target domain of pixel-wise loss, as illustrated in Equation~\ref{eq:reconstruction_loss}.
Specifically, in Table~\ref{tab:result_denoise}, using NAFNet~\citep{cv:res:chen22-nafnet} with $L_{2}$ loss and $L_{\text{PSNR}}$, our $a$RGB representation yields better PSNR and SSIM scores than the model trained directly with the original PSNR metric.
Here, a per-pixel PSNR loss $L_{\text{PSNR}}\,$ represents a mathematically equivalent form of the $L_{2}$ loss.
The table indicates that our $a$RGB representation preserves RGB information and incorporates additional local structural information, enhancing denoising supervision.
Furthermore, experiments with different metrics highlight the impact of changing the representation space on training dynamics, offering unexpected benefits, as seen with NAFNets trained for the $L_{1}$ metric in our $a$RGB space~(last rows of Table~\ref{fig:result_denoise}).

Moreover, in Table~\ref{tab:result_deblur}, employing a Charbonnier loss~\cite{cv:obj:bruhn05-charbonnierloss}, we observe significant improvements in training MPRNet~\citep{cv:deblur:zamir21-mprnet} for motion blur deblurring when the loss is applied in our $a$RGB space, compared to the RGB space.
Furthermore, Table~\ref{tab:result_deblur} and Figure~\ref{fig:result_deblur} demonstrate significant improvements, orthogonal to existing enhancement techniques such as test-time local converter (TLC)~\citep{cv:deblur:chu22-tlc}.
Our experiments indicate that our $a$RGB representation enhances the training of image restoration models across various tasks, architectures, and loss functions, synergizing with other techniques like edge loss and test-time local converter.

\paragraph{Perceptual quality-oriented:}
To demonstrate the strong synergy of our proposed pixel-wise loss with perceptual losses (\eg, VGG loss~\cite{cv:obj:johnson16-perceptualloss} and adversarial loss~\cite{cv:gan:goodfellow14-gan}), we compare our approach on photo-realistic super-resolution.
To elaborate, Table~\ref{tab:result_gan} shows a comparison using the ESRGAN \citep{cv:sr:wang18-esrgan} as a baseline.
The consistent improvement in both distortion metrics~(\eg, PSNR and SSIM) and perceptual metrics~(\eg, LPIPS, NIQE, and FID), deviating from the conventional trade-off between them, underscores the effectiveness of our modified $L_{1, \text{$a$RGB}}$ objective in stabilizing the adversarial training of a super-resolution model.
This indicates that the local structural information within our $a$RGB representation complements supervision from a pre-trained classifier, resulting in superior performance in both distortion-based and perceptual metrics compared to the original ESRGAN.
The improvements in PSNR and SSIM scores align with our design philosophy, emphasizing that RGB colors are included as a subspace in our $a$RGB representation.

\input{sections/tables/computation}

\paragraph{Computational overhead:}
Since our method incorporates an additional $a$RGB encoder model~$f$ during the training of the image restoration model, there is an associated increase in computational overhead.
Table~\ref{tab:computation_comparison} provides a comparison of the computational burden with other conventional objectives that utilize additional modules during the training process.
Notably, our $a$RGB encoder~$f$ requires fewer parameters compared to the alternatives, demonstrating its efficiency.
Specifically, for the adversarial loss $L_{\text{Adv}}$, we quantify the number of parameters based on the discriminator of ESRGAN~\cite{cv:sr:wang18-esrgan}.
In comparison to the super-resolution model itself, our overhead is not substantial, accounting for only 9\% of the entire model.
Also, please note that our method exclusively applies to the training phase of the restoration model, incurring no additional computational overhead during test time.

\input{sections/figures/discussion}

%% file: sections/tables/sr.tex
\begin{table*}[t]
\caption{\small
\textbf{Quantitative results on training $4 \times$ super-resolution ESRGAN in the $a$RGB space.}
In our methods using $a$RGB representation, we modify only the $L_{1}$ loss by exchanging it with the $L_{1, \text{$a$RGB}}$ loss.
All the other training hyperparameters are left untouched.
Better scores in each block are shown in \textbf{boldface} text.
More results are in Table~\ref{tab:appx:result_gan} in Appendix~\ref{sec:d_esrgan}.
}
\vspace{1mm}
\label{tab:result_gan}
\resizebox{\linewidth}{!}{%
\begin{tabular}{lcccccccccc}
    \toprule
    & \multicolumn{5}{c}{{DIV2K-Val}} & \multicolumn{5}{c}{{Urban100}} \\
    \cmidrule(lr){2-6} \cmidrule(lr){7-11}
    Objective & PSNR$\uparrow$ & SSIM$\uparrow$ & LPIPS$\downarrow$ & NIQE$\downarrow$ & FID$\downarrow$ & PSNR$\uparrow$ & SSIM$\uparrow$ & LPIPS$\downarrow$ & NIQE$\downarrow$ & FID$\downarrow$ \\
    \midrule
    $0.01 L_{1} \qquad \hspace{.07em} + L_{\text{VGG}} + 0.005 L_{\text{Adv}}$\textsuperscript{$\dagger$}
        & 26.627 & 0.7033 & 0.1154 & 3.0913 & 13.557 & 22.776 & 0.7033 & 0.1232 & 4.2067 & 20.616 \\
    \textcolor{BrickRed}{${ 0.01 L_{1, \text{$a$RGB}}} + L_{\text{VGG}} + 0.005 L_{\text{Adv}}$}
        & \textbf{26.845} & \textbf{0.7500} & \textbf{0.1110} & \textbf{2.9615} & \textbf{12.799} & \textbf{23.270} & \textbf{0.7196} & \textbf{0.1183} & \textbf{3.8982} & \textbf{17.739} \\
    \bottomrule
    \multicolumn{11}{l}{\textsuperscript{$\dagger$}\footnotesize{The official ESRGAN model \citep{cv:sr:wang18-esrgan}.}}
\end{tabular}%
}%
\vspace*{-.7em}
\end{table*}

%% file: sections/figures/result_gan.tex
\begin{figure*}[t]
\newcommand{\figwidth}{.199\linewidth}
\newcommand{\varA}{LR}%
\newcommand{\varB}{RRDBNet}%
\newcommand{\varC}{L1+Adv}%
\newcommand{\varD}{L1aRGB+Adv}%
\newcommand{\varE}{ESRGAN}%
\newcommand{\varF}{L1aRGB+VGG+Adv}%
\newcommand{\varG}{GT}%
\centering
\newcommand{\seed}{DIV2K100_0830_a8ce}
    \subfloat{\includegraphics[width=\figwidth]{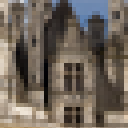}}
    \hfill
    \subfloat{\includegraphics[width=\figwidth]{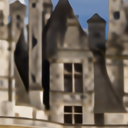}}
    \hfill
    \subfloat{\includegraphics[width=\figwidth]{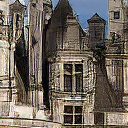}}
    \hfill
    \subfloat{\includegraphics[width=\figwidth]{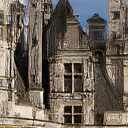}}
    \hfill
    \subfloat{\includegraphics[width=\figwidth]{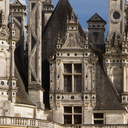}}
    \\[-0.05em]
\renewcommand{\seed}{DIV2K100_img076_390b}
    \addtocounter{subfigure}{-5}
    \subfloat[LR]{\includegraphics[width=\figwidth]{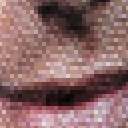}}
    \hfill
    \subfloat[RRDBNet]{\includegraphics[width=\figwidth]{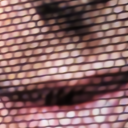}}
    \hfill
    \subfloat[ESRGAN]{\includegraphics[width=\figwidth]{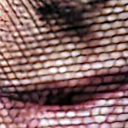}}
    \hfill
    \subfloat[\resizebox{7em}{!}{\scriptsize\color{BrickRed} $L_{1, \text{$a$RGB}} + L_{\text{VGG}} + L_{\text{Adv}}$}]{\includegraphics[width=\figwidth]{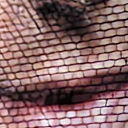}}
    \hfill
    \subfloat[HR]{\includegraphics[width=\figwidth]{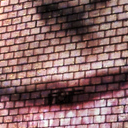}}
    \\[-.5em]
\caption{\small
\textbf{Qualitative comparison of ESRGAN models trained with different loss functions.}
Each column corresponds to each row in Table~\ref{tab:result_gan}.
The loss weights are omitted for brevity, 
ESRGAN corresponds to the $0.01 L_{1} + L_{\text{VGG}} + 0.005 L_{\text{Adv}}$ in Table~\ref{tab:result_gan}.
}
\label{fig:result_gan}
\vspace{-3mm}
\end{figure*}

%% file: sections/tables/computation.tex
\begin{table}[t]
\vspace{0em}
    \caption{\textbf{The number of parameters of the losses.}}
    \label{tab:computation_comparison}
    \vspace{.5em}
  \centering
\resizebox{0.48\linewidth}{!}{
    \begin{tabular}{cc}
        \toprule
        Objective & \# params~(M)\\
        \midrule
        $L_{\text{VGG}}$ & 20.2\\
        $L_{\text{Adv}}$ & 80.2\\
        \textcolor{BrickRed}{$L_{\text{$a$RGB}}$} & \textbf{5.3} \\ 
        \bottomrule
    \end{tabular}}
    \vspace{-1em}
\end{table}

%% file: sections/figures/discussion.tex
\begin{figure*}[t]
\newcommand{\h}{2.4mm}
\newcommand{\hh}{2.5mm}
\newcommand{\height}{3.88cm}
\newcommand{\heightfig}{1.8cm}
\centering
\subfloat[\scriptsize Inverting orthogonal mixture of two $a$RGB embeddings. \label{fig:discussion:inversion}]{%
\begin{minipage}{.21\textwidth}
    \includegraphics[height=\heightfig]{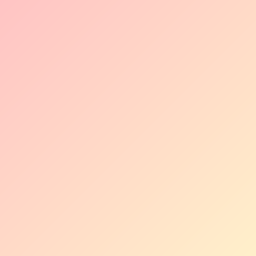}\hspace*{0.2mm}%
    \includegraphics[height=\heightfig]{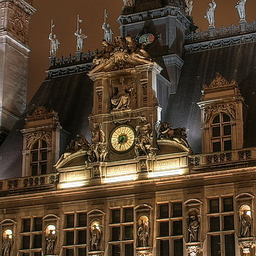}\\[-7.5mm]
    \raisebox{-3mm}[0mm][0mm]{%
        \hspace*{7.6mm}{\makebox[\h][c]{\pgfsetfillopacity{1}\scriptsize $\xx_{1}$}}%
        \hspace*{15.02mm}{\pgfsetfillopacity{0.0}\colorbox{white!20}{\color{white}\makebox[\h][c]{\pgfsetfillopacity{1}\scriptsize $\xx_{2}$}}}%
        \pgfsetfillopacity{1}
    }\\[3.5mm]
    \includegraphics[height=\heightfig]{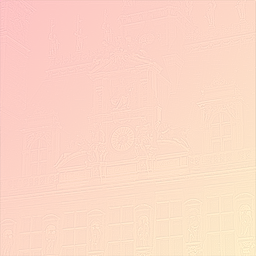}\hspace*{0.2mm}%
    \includegraphics[height=\heightfig]{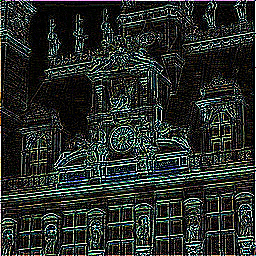}\\[-7.5mm]%
    \raisebox{-3mm}[0mm][0mm]{%
        \hspace*{3.6mm}{\makebox[11mm][c]{\scriptsize $f^{-1} (\xxi_{\text{mix}})$}}%
        \hspace*{6.72mm}{\pgfsetfillopacity{0.0}\colorbox{white!20}{\color{white}\makebox[11mm][c]{\pgfsetfillopacity{1}\scriptsize $\nabla^{2} f^{-1} (\xxi_{\text{mix}})$}}}%
        \pgfsetfillopacity{1}
    }\\[0.6mm]
\end{minipage}%
}
\hfill
\subfloat[\scriptsize Expert selection map of the MoE router $f_{\text{r}}\,$. \label{fig:discussion:segm}]{\includegraphics[height=\height]{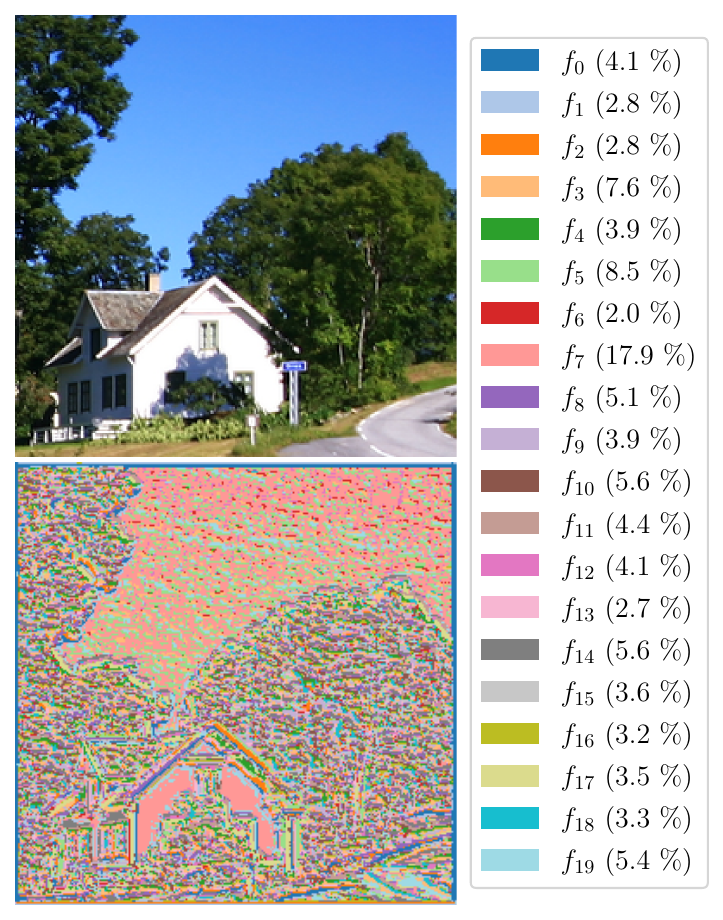}}
\hfill
\subfloat[\scriptsize t-SNE plot of the $a$RGB embedding $\xxi$ of pixels in image \ref{fig:discussion:segm}. \label{fig:discussion:tsne}]{\includegraphics[height=\height]{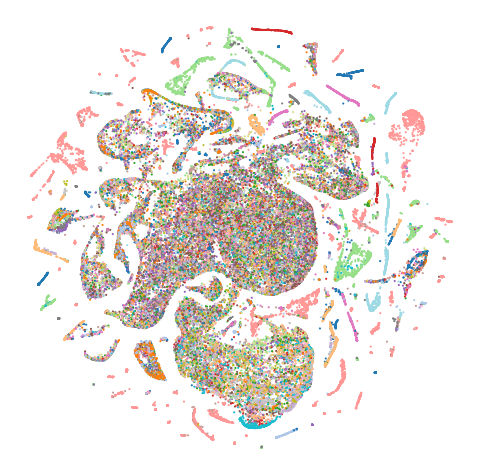}}
\hfill
\subfloat[\scriptsize Change of $L_{2}$ metrics in the $a$RGB space relative to the $L_{2}$ metrics in the RGB space. \label{fig:discussion:metric}]{\includegraphics[height=\height]{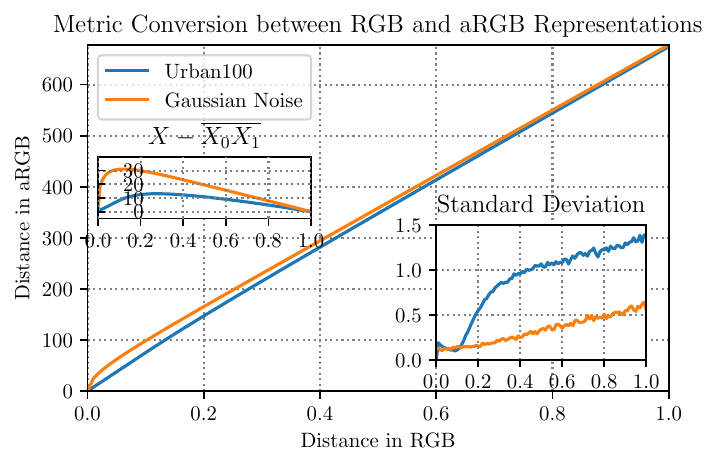}}
\\
\caption{\small
    \textbf{Understanding the learned $a$RGB representation.}
    Figure~\ref{fig:discussion:inversion} show a visual example of $a$RGB embedding inversion.
    Figure~\ref{fig:discussion:segm} and \ref{fig:discussion:tsne} reveal clear evidence that the experts of our $a$RGB encoder $f$ are specialized for a particular type of input structures, and that even the embedding vectors within a single patch are clustered in a complicated manner, justifying our usage of MoE architecture.
    Figure~\ref{fig:discussion:metric} shows how the distance metric changes in the $a$RGB space relative to the distance in the RGB space.
    Mean distances and their standard deviations are measured by MSE losses between an image and the same image with 100 AWGNs with the same standard deviation.
    Note that the $a$RGB space slightly exaggerates the distance more outside natural image domain, \eg, Gaussian noise, and the metric's variance is negligibly small.
    More examples are in Appendix~\ref{sec:f_decomposition}.
}
\vspace{-3mm}
\label{fig:discussion}
\end{figure*}

%% file: sections/5_discussion.tex
\section{Discussion}
\label{sec:5_discussion}

\subsection{Nullspace of the Decoder}
\label{sec:5_discussion:nullspace}
In addition to the design simplicity, our pixel-wise \emph{linear} decoder enjoys an additional benefit: decomposability.
Since our autoencoder is almost lossless, we will consider that the RGB $\xx \in \R^{3}$ and the $a$RGB $\xxi = f(\xx) \in \R^{C}$ representations of any given image equivalent.
That is, $\xx' = g(\xxi) = \mA \xxi + \vb = \xx\,$.
As a result of the linearity of our decoder $g\,$, the $a$RGB representation $\xxi$ can be decomposed into the sum of two orthogonal components:
\begin{align}
\label{eq:ortho_decomposition}
 \xxi &\,= \xxi_{\parallel} + \xxi_{\perp}\,, \\
\label{eq:ortho_decomposition2}
 \text{s.t.} \quad f_{\parallel} (\xx) &\coloneqq \xxi_{\parallel} = \mA^{\dagger} \mA \xxi \\
\label{eq:ortho_decomposition3}
 \text{and} \quad  f_{\perp} (\xx) &\coloneqq \xxi_{\perp} = (\mI - \mA^{\dagger} \mA) \xxi\,,
\end{align}
where $\mA^{\dagger}$ is the Moore-Penrose pseudoinverse of $\mA\,$.
The parallel component $\xxi_{\parallel}$ of the $a$RGB representation lies in the three-dimensional subspace of $\R^{C}$ that is projected onto the RGB colors by the decoder $g\,$, \ie, $\mA \xxi_{\parallel} = \mA \mA^{\dagger} \mA \xxi = \mA \xxi\,$.
The remaining perpendicular part $\xxi_{\perp}$ can be regarded as the information the $a$RGB space encodes in addition to the RGB colors.
The contribution of the two components can be visualized by inverting the encoder $f$ with respect to a mixed embedding:
\begin{align}
\label{eq:optimize_mixture}
&\,\, f^{-1} (\xxi_{\text{mix}}) = \argmin_{\zz}\,\, \| f(\zz) - \xxi_{\text{mix}} \|_{2}^{2}\,, \\
\label{eq:optimize_mixture2}
\text{s.t.} \,\, &\xxi_{\text{mix}} = \xxi_{1 \parallel} + \xxi_{2 \perp} = \mA^{\dagger} \mA \xxi_{1} + (\mI - \mA^{\dagger} \mA) \xxi_{2}\,.
\end{align}
We use a SGD optimizer with a learning rate of 0.1 for 50 iterations.
As shown in Figure~\ref{fig:discussion:inversion} and Appendix~\ref{sec:f_decomposition}, the inversion of the mixed embedding inherits color information from the parallel embedding $\xxi_{1 \parallel}\,$, while the perpendicular part $\xxi_{2 \perp}$ contributes to the high-frequency edge information.

\subsection{Expert Specialization and Learned Structures}
\label{sec:5_discussion:moe}
Figure~\ref{fig:discussion:segm} visualizes how individual pixels of a natural image are distributed to $K=20$ experts.
Unlike in semantic segmentation, where segmentation maps are chunked into large blocks of semantically correlated pixels, our pixel-wise router $f_{\text{r}}$ generates fine-grained distributions of pixels.
That is, multiple experts are jointly involved in encoding the same texture, such as the blue sky and the leafy trees.
Another salient feature we can observe in the figure is that edges of different orientations are dealt with by different experts, implying their specialization.
Visualizing the $a$RGB embedding space using t-SNE \citep{ml:vis:vdmaaten08-tsne} provides us with additional insights into the topology of the space.
Figure~\ref{fig:discussion:tsne} reveals that the $a$RGB embeddings cluster into multiple disconnected groups in two different types: \emph{common} groups where multiple experts are involved in the encoding process and \emph{specialized} groups where a single expert is exclusively allocated for the embeddings.
These observations align well with our initial design principles in Section~\ref{sec:3_method:autoencoder}, where the feature embeddings occupy a highly complicated, disconnected set, and an MoE architecture effectively deals with this structure by specializing each expert to a subset of the embedding space.

\input{sections/tables/disc_ablations}

\subsection{The $a$RGB Metric Space and Produced Gradients}
\label{sec:5_discussion:metricspace}
The main purpose of our the $a$RGB space is to provide alternative supervision to the existing image restoration framework.
This supervision is realized with a metric defined over the space and its gradients generated from pairs of images.
To this end, we first visualize the correlation between $L_{2}$ distances defined in the RGB and our $a$RGB spaces in Figure~\ref{fig:discussion:metric}.
We plotted additional figure with title $X - \overline{X_{0} X_{1}}$ to show the deviation of the graph over the straight line, showing clear convexity of the graph.
This implies that the metrics within $a$RGB spaces are inflated when the given two images are similar.
Figure~\ref{fig:gradient_vis} shows the gradients from two per-pixel $L_{1}$ losses between a restored image and its high-quality counterpart defined over both spaces.
Unlike RGB $L_{1}$ loss which exhibits a highly off-centered, discrete distribution, the $L_{1, \text{$a$RGB}}$ loss shows smooth and centered distribution of gradients.
We believe that this allows for the stable training of the image restoration models despite its huge scale of the generated gradients from the $L_{1, \text{$a$RGB}}$ loss, which is more than a hundredfold as shown in the x axis of Figure~\ref{fig:gradient_vis}b.
In the RGB domain, the same scale of gradient is achievable only through increasing the learning rate, which leads to destabilization of the training.
Overall, the analyses show how our $a$RGB encoder helps the training of image restoration models.

\subsection{Ablation Study}
\label{sec:5_discussion:ablation}
Lastly, we provide ablation studies to determine the best hyperparameters for our $a$RGB autoencoder.
We compare the models by the results of training an RRDBNet \citep{cv:sr:wang18-esrgan} only on DIV2K dataset.
The results are summarized in Table~\ref{tab:ablation}.
More information is elaborated in Appendix~\ref{sec:b_impl_detail}.

\paragraph{Number of experts:}
As our encoder employs a Mixture-of-Experts~(MoE) architecture, the overall performance of forming the $a$RGB space can be influenced by the number of experts. 
To find the optimal number of experts, we conduct an ablation study in Table~\ref{tab:ablation}.
The first four rows of Table~\ref{tab:ablation} show the effect of the number of experts of the $a$RGB encoder $f$ on its supervision quality.
While applying more experts yields improved results by letting encoder to jointly learn input textures in multiple spaces, as shown in the t-SNE visualization in Figure~\ref{fig:discussion:tsne}, using 30 experts shows slightly worse performance compared to 20 experts.
This is attributed to the larger number of experts causing input textures to be learned in a too-specified distribution, hindering the encoder's ability to learn the distribution of input samples properly.
From the results, we fix the number of experts to 20 throughout our experiments.

\paragraph{Dataset dependence:}
As the second part of Table~\ref{tab:ablation} presents, the training data for the $a$RGB autoencoder decides the quality of supervision the model gives.
This implies that our $a$RGB autoencoder utilizes structural priors of its training data.
Appendix~\ref{fig:appx_math_imperfect} provides additional theoretical and empirical evidence that our $a$RGB autoencoder learns image structures to reconstruct given images.

\paragraph{Regularizers:}
In the last row of Table~\ref{tab:ablation}, we observe that the regularizing noise $\zz$ added at the end of the encoder during training helps the $a$RGB encoder to produce stronger supervision for image restoration models.
In practice, we observe more than tenfold reduction in the scale of produced gradients when the $a$RGB autoencoder trained without the regularizing noise is applied.
This correlates to our discussion in Section~\ref{sec:5_discussion:metricspace}, that our $a$RGB encoder helps training image restoration models by stably increasing the scale of gradients.

%% file: sections/tables/disc_ablations.tex
\begin{table*}[t]
\begin{minipage}{.28\linewidth}
\vspace{-5mm}
    \newcommand{\h}{2.4mm}
    \newcommand{\hh}{2.5mm}
    \newcommand{\height}{3.15cm}
    \newcommand{\heightfig}{.97cm}
    \centering
    \subfloat[\scriptsize $L_{1}$ loss gradients visualized. \label{fig:gradient_vis:images}]{%
    \hspace{4mm}
    \begin{minipage}{\linewidth}
        \vspace{8mm}
        \includegraphics[height=\heightfig]{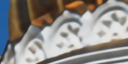}\hspace*{0.2mm}%
        \includegraphics[height=\heightfig]{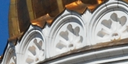}\\[-7.5mm]
        \raisebox{-3mm}[0mm][0mm]{%
            \hspace*{15.mm}{\pgfsetfillopacity{0.9}\colorbox{white!20}{\makebox[\h][c]{\pgfsetfillopacity{1}\scriptsize $\xx_{\text{SR}}$}}}%
            \hspace*{15.mm}{\pgfsetfillopacity{0.9}\colorbox{white!20}{\makebox[\h][c]{\pgfsetfillopacity{1}\scriptsize $\xx_{\text{GT}}$}}}%
            \pgfsetfillopacity{1}
        }\\[3.5mm]
        \includegraphics[height=\heightfig]{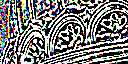}\hspace*{0.2mm}%
        \includegraphics[height=\heightfig]{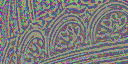}\\[-7.5mm]%
        \raisebox{-3mm}[0mm][0mm]{%
            \hspace*{9.2mm}{\pgfsetfillopacity{0.9}\colorbox{white!20}{\makebox[8.2mm][c]{\pgfsetfillopacity{1}\scriptsize $\nabla L_{1, \text{RGB}}$}}}%
            \hspace*{8.4mm}{\pgfsetfillopacity{0.9}\colorbox{white!20}{\makebox[9.2mm][c]{\pgfsetfillopacity{1}\color{BrickRed}\scriptsize $\nabla L_{1, \text{$a$RGB}}$}}}%
            \pgfsetfillopacity{1}
        }\\[0.6mm]
    \vspace*{-0.7em}
    \end{minipage}%
    }
    \\
    \subfloat[\scriptsize Histogram of gradients of 6a. \label{fig:gradient_vis:histogram}]{\includegraphics[width=\linewidth]{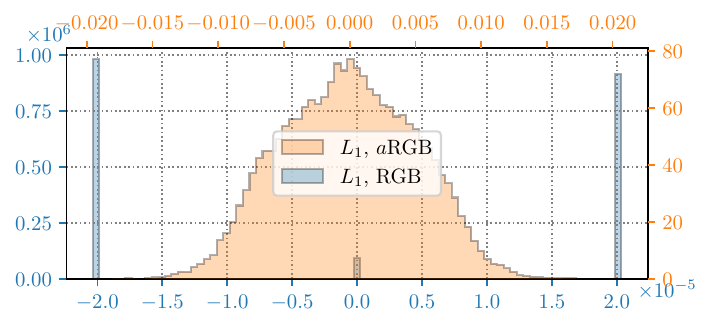}\vspace*{-0.8em}}
    \\
    [-.4em]
    \captionof{figure}{\small
    \textbf{Gradients from $L_{1, \text{RGB}}$ and $L_{1, \text{$a$RGB}}$ losses.}
    }
    \label{fig:gradient_vis}
\end{minipage}
\hfill
\begin{minipage}{.7\linewidth}
    \addtocounter{table}{-1}
    \caption{\small
    \textbf{Ablation studies on the $a$RGB autoencoder.}
    RRDBNets \citep{cv:sr:wang18-esrgan} are trained with DIV2K \citep{cv:data:agustsson17-div2k} for 300k iterations for $4 \times$ SISR tasks with only the $L_{1}$ loss between the $a$RGB embeddings.
    }
    \label{tab:ablation}
    \vspace*{.5em}
    \centering
    \resizebox{\linewidth}{!}{%
    \begin{tabular}{cccccccccc}
        \toprule
        \multicolumn{4}{c}{RRDBNet in $4 \times$ SISR} & \multicolumn{2}{c}{{Set14}} & \multicolumn{2}{c}{{Urban100}} & \multicolumn{2}{c}{{DIV2K-Val}} \\
        \cmidrule(lr){1-4} \cmidrule(lr){5-6} \cmidrule(lr){7-8} \cmidrule(lr){9-10}
        {\small {\# experts}} & {\small Routing} & {\small $a$RGB train set} & {\small Reg. noise} & PSNR$\uparrow$ & SSIM$\uparrow$ & PSNR$\uparrow$ & SSIM$\uparrow$ & PSNR$\uparrow$ & SSIM$\uparrow$ \\
        \midrule
         1 & -   & DIV2K & \cmark & 26.87 & 0.7467 & 24.75 & 0.7735 & 29.08 & 0.8222 \\
         5 & MoE & DIV2K & \cmark & 26.87 & 0.7477 & 24.83 & 0.7745 & 29.12 & 0.8231 \\
        10 & MoE & DIV2K & \cmark & 26.89 & 0.7474 & 24.84 & 0.7750 & 29.11 & 0.8231 \\
        \textbf{20} & MoE & DIV2K & \cmark & 26.91 & 0.7471 & 24.87 & 0.7745 & 29.14 & 0.8227 \\
        30 & MoE & DIV2K & \cmark & 26.89 & 0.7476 & 24.84 & 0.7750 & 29.11 & 0.8231 \\
        \cdashlinelr{1-10}
        20 & MoE & GoPro & \cmark & 26.89 & 0.7459 & 24.83 & 0.7728 & 29.12 & 0.8220 \\
        20 & MoE & SIDD  & \cmark & 26.86 & 0.7420 & 24.80 & 0.7691 & 29.08 & 0.8186 \\
        20 & MoE & Noise & \cmark & 26.65 & 0.7461 & 24.64 & 0.7729 & 28.87 & 0.8212 \\
        \cdashlinelr{1-10}
        20 & MoE & DIV2K & \xmark & 26.91 & 0.7469 & 24.85 & 0.7722 & 29.13 & 0.8223 \\
        \bottomrule
    \end{tabular}%
    }%
\end{minipage}
\vspace{-3mm}
\end{table*}

%% file: sections/6_conclusion.tex
\section{Conclusion}
\label{sec:7_conclusion}
We hypothesized that blurry restoration due to per-pixel losses can be alleviated if the underlying representation space contains enough structural information for each pixel.
To prove this, we proposed augmented RGB~($a$RGB) representation space to supply pixel-accurate local structural information and to preserve color information while training the restoration models.
Improved performance could be observed across diverse set of restoration tasks such as denoising, deblurring, and perceptual super-resolution by only substituting the underlying representation space for the per-pixel losses to our $a$RGB space.
This suggests that the RGB color space is not the optimal representation space for low-level computer vision tasks.
We hope our work spurs more interests in the optimal representation spaces.

%% file: sections/A_math.tex
The goal of this section is to provide a simple theoretical analysis on how the structure is learned in our $a$RGB encoder $f\,$.
This section comprises of two parts.
The first part shows that our $a$RGB encoder is a piecewise linear function over the connected neighborhood regions in the RGB pixel domain.
From this, we can equivalently transform our $a$RGB autoencoder to a coordinate-wise function, which is useful in the upcoming analyses.
In the second part, we will show that an autoencoder with a neural network encoder and a linear decoder like ours does not learn structural priors if it is perfectly lossless.
In other words, we claim that imperfect, yet almost perfect ($> 60 \mathrm{dB}$ PSNR), reconstruction capability of our $a$RGB autoencoder helps its encoder to learn local structures within natural images.
Based on the mathematical analysis, we provide another method to measure how much image structure is captured in the $a$RGB autoencoder.

\subsection{The $a$RGB encoder is a piecewise linear function over the local neighborhood structures}
\label{sec:a_math:piecewise_linear}
It is known that a multi-layer perceptron (MLP) with a continuous piecewise linear activation is also a continuous piecewise linear (CPWL) function \citep{ml:rahaman19-spectral_bias}.
That is, given an input $\xx \in \R^{C}\,$, the network $f$ has an output $\yy = f(\xx) \in \R^{C'}\,$, which can be explicitly written as:
\begin{equation}
\label{eq:cpwl_fn}
f(\xx) = \sum_{\epsilon} \mathbbm{1}_{P_{\epsilon} (\xx)} (\mW_{\epsilon} \xx + \vb_{\epsilon})\,,
\end{equation}
where $\epsilon$ is an index of a (connected) region $P_{\epsilon} \subset \R^{C}$ and $\mathbbm{1}_{P_{\epsilon}}\,$, the region's indicator function.
$\mW_{\epsilon} \in \R^{C' \times C}$ and $\vb_{\epsilon} \in \R^{C'}$ are the effective weight and bias at the region $P_{\epsilon}\,$, respectively.
This interpretation of neural networks is straightforward for multilayer perceptrons.
An MLP is a composition of linear layers and rectifiers, which are indeed continuous and piecewise linear, and a composition of a continuous piecewise linear function is also continuous and piecewise linear.

Convolutional neural networks with rectifiers are no different.
A linear convolution operation
\begin{equation}
\label{eq:linear_convolution2}
[\vk \star \xx]_{h, w} \coloneqq \sum_{i = -\lfloor k_{h} / 2 \rfloor}^{\lfloor k_{h} / 2 \rfloor} \sum_{j = -\lfloor k_{w} / 2 \rfloor}^{\lfloor k_{w} / 2 \rfloor} [\vk]_{i, j} [\xx]_{h + i, w + j}\,,
\end{equation}
where $k_{h}$ and $k_{w}$ define the height and width of the discrete kernel, can be viewed as a coordinate-wise linear layer with a flattened weight operating on a concatenation of translations of the input $\xx\,$, \ie,
\begin{align}
\label{eq:linear_convolution}
\vk \star \xx &= \mW_{\vk} \tilde{\xx}\,, \quad \text{where} \\
\label{eq:linear_convolution_weight}
\mW_{\vk} &= \mathrm{flatten} (\vk)\,, \quad \text{and} \\
\label{eq:linear_convolution_input}
\tilde{\xx} &= \underset{i, j}{\mathrm{concat}} [\mathrm{translate} [\xx, (i, j)]]\,.
\end{align}
In other words, with a receptive field of size $R \times R\,$, an image $x \in \R^{C \times H \times W}$ is (linearly) transformed into the extended space $\tilde{x} \in \R^{CR^{2} \times H \times W}\,$, and the ConvNet $f : \R^{C \times H \times W} \rightarrow \R^{C' \times H \times W}$ is equivalent to the coordinate-wise function $\tilde{f}: \R^{CR^{2}} \rightarrow \R^{C'}\,$, which is continuous and piecewise linear.

Since our transform from $f$ to $\tilde{f}$ can be applied to all the $K$ experts $f_{1}, \ldots f_{K}$ of our $a$RGB autoencoder, we can abstract away the coordinates for simplicity.
Another virtue of this reform is that we do not need to care about the router $f_{\text{r}}$ in our analysis hereafter.
From \eqref{eq:mixture_of_experts}, for each coordinate $c \in [H] \times [W]\,$, we have:
\begin{equation}
\label{eq:mixture_of_experts_coordinatewise}
[\xxi]_{c} = [f(\xx)]_{c} = \sum_{k = 1}^{K} \left[m_{k} \odot f_{k} (\xx)\right]_{c} = \sum_{k = 1}^{K} [m_{k}]_{c} [f_{k} (\xx)]_{c} = \sum_{k = 1}^{K} [m_{k}]_{c} \tilde{f}_{k} ([\tilde{\xx}]_{c}) \eqqcolon \tilde{f}([\tilde{\xx}]_{c}) \,,
\end{equation}
and since the coordinate-wise equivalent of each expert $\tilde{f}_{k}$ is continuous and piecewise linear, its weighted summation is also piecewise linear, yet it may not be continuous.
Moreover, the function's argument $[\tilde{\xx}]_{c} \in \R^{CR^{2}}$ is a reshaping of the receptive field of size $R \times R$ at the locus $c\,$.
As a result, our mixture of experts encoder $f$ is equivalent to a piecewise linear, yet not generally continuous, function over $R \times R$ neighborhood of each pixel in the RGB representation $\xx\,$.
No matter how each pixel is distributed by the router, each feature vector $[\xxi]_{c}$ at an arbitrary coordinate $c \in [H] \times [W]$ is a piecewise linear function of a local $R \times R$ neighborhood of image $\xx$ at the locus $c\,$.

\subsection{A flawless autoencoder is a bad structure encoder}
\label{sec:a_math:flawless_ae}
From the previous analysis, we may omit the spatial coordinate $c$ and assume $\xx \in \R^{C}$ be an image in the RGB color space, $\xxi \in \R^{C'}$ be the same image in the $a$RGB representation.
Let $\tilde{\xx} = \begin{bmatrix} \xx & B(\xx) \end{bmatrix}^{\top} \in \R^{CR^{2}}$ be the flattened $R \times R$ neighborhood patch in the RGB domain.
$B: \R^{C} \rightarrow \R^{C(R^{2} - 1)}$ is a structure function that maps a center pixel $\xx$ to its peripheral pixels.
For the simplicity, we discard the function notation and regard $B$ as a vector in $\R^{C(R^{2} - 1)}$ from now on.
The $a$RGB autoencoder is equivalently reformed coordinate-wise.
Let our $a$RGB encoder be $f: \R^{CR^{2}} \rightarrow \R^{C'}$ and the $a$RGB decoder be $g: \R^{C'} \rightarrow \R^{C}\,$.
Note that we discard tilde from the encoder equivalent $\tilde{f}$ for the sake of simplicity.
Again, since the decoder is linear, we can also write $g(\xxi) = \mA \xxi + \vb\,$, where $\mA \in \R^{C \times C'}$ and $\vb \in \R^{C}\,$.

The piecewise linear characteristic of the encoder $f$ lets us rewrite the autoencoder into a form:
\begin{align}
\label{eq:cpwl_fn_single_encoder}
f(\tilde{\xx}) &= \sum_{\epsilon} \mathbbm{1}_{P_{\epsilon} (\tilde{\xx})} (\mW_{\epsilon} \tilde{\xx} + \vb_{\epsilon}) = \xxi\,, \\
g(\xxi) &= \mA \xxi + \vb = \xx' \simeq \xx\,,
\label{eq:linear_decoder}
\end{align}
following \eqref{eq:cpwl_fn}.
The autoencoder $h = g \circ f$ can be written as:
\begin{align}
\label{eq:cpwl_fn_autoencoder}
h(\tilde{\xx}) = g(f(\tilde{\xx})) &= \mA \sum_{\epsilon} \mathbbm{1}_{P_{\epsilon} (\tilde{\xx})} (\mW_{\epsilon} \tilde{\xx} + \vb_{\epsilon}) + \vb\,, \\
&= \sum_{\epsilon} \mathbbm{1}_{P_{\epsilon} (\tilde{\xx})} \left( \mA \mW_{\epsilon} \tilde{\xx} + \mA \vb_{\epsilon} + \vb \right)\,.
\label{eq:cpwl_fn_autoencoder2}
\end{align}
Let $\epsilon'$ is a subscript for a connected region in the partition $P$ that contains the coordinate $c\,$.
That is, $\mathbbm{1}_{P_{\epsilon'} (\tilde{\xx})} = 1$ and $\mathbbm{1}_{P_{\epsilon} (\tilde{\xx})} = 0$ for $\epsilon \neq \epsilon'\,$.
The summation from \eqref{eq:cpwl_fn_autoencoder2}, then, can be simplified:
\begin{equation}
\label{eq:cpwl_fn_autoencoder_simple}
h(\tilde{\xx}) = \mA \mW_{\epsilon'} \tilde{\xx} + \mA \vb_{\epsilon'} + \vb = \xx' \simeq \xx\,.
\end{equation}
In other words, we only care about the specific region of the partition generated by the mixture of experts $a$RGB encoder that includes our pixel of interest.
The flattened region $\tilde{\xx}$ is decomposed into the center pixel $\xx$ and the peripherals $B\,$.
\begin{equation}
\label{eq:cpwl_fn_autoencoder_decomposed}
h(\tilde{\xx}) = h\left( \begin{bmatrix} \xx \\ B \end{bmatrix} \right) = \mA \mW_{\epsilon'} \begin{bmatrix} \xx \\ B \end{bmatrix} + \mA \vb_{\epsilon'} + \vb \simeq \xx\,.
\end{equation}
This can be further decomposed if we decompose $\mW_{\epsilon'}$ into two matrices,
\begin{align}
\label{eq:cpwl_fn_autoencoder_decomposed2}
h(\tilde{\xx}) &= \mA \begin{bmatrix} \mW_{\text{cen}, \epsilon'} & \mW_{\text{per}, \epsilon'} \end{bmatrix} \begin{bmatrix} \xx \\ B \end{bmatrix} + \mA \vb_{\epsilon'} + \vb\,, \\
\label{eq:cpwl_fn_autoencoder_decomposed3}
&= \mA \mW_{\text{cen}, \epsilon'} \xx + \mA \mW_{\text{per}, \epsilon'} B + \mA \vb_{\epsilon'} + \vb \simeq \xx\,,
\end{align}
where $\mW_{\text{cen}, \epsilon'} \in \R^{C' \times C}$ is the effective weight for the center pixel $\xx\,$, and $\mW_{\text{per}, \epsilon'} \in \R^{C' \times C(R^{2} - 1)}$ is the effective weight for the peripheral pixels $B\,$.

Let us now assume a perfect lossless autoencoder $h^{\star}(\tilde{\xx}) = \xx$ for every combination of $\xx$ and $B\,$.
That is, the autoencoder $h^{\star}$ always returns the exact same pixel $\xx$ without requiring $B$ to be a function of $\xx\,$.
In this case, \eqref{eq:cpwl_fn_autoencoder_decomposed3} can be further decomposed into three equations:
\begin{align}
\label{eq:cpwl_fn_autoencoder_decomposed_fin1}
\mA \mW_{\text{cen}, \epsilon'} \xx &= \xx \in \R^{C}\,, \\
\label{eq:cpwl_fn_autoencoder_decomposed_fin2}
\mA \mW_{\text{per}, \epsilon'} B &= \bm{0} \in \R^{C(R^{2} - 1)}\,, \quad \text{and} \\
\label{eq:cpwl_fn_autoencoder_decomposed_fin3}
\mA \vb_{\epsilon'} + \vb &= \bm{0} \in \R^{C}\,.
\end{align}
In particular, \eqref{eq:cpwl_fn_autoencoder_decomposed_fin1} should be satisfied for every combinations of RGB colors $\xx\,$, and therefore we can conclude that $\mA \mW_{\text{cen}, \epsilon'} = \mI$ for the perfect autoencoder.
Furthermore, \eqref{eq:cpwl_fn_autoencoder_decomposed_fin2} should be satisfied for every possible $B\,$, signifying that the encoder $f^{\star}$ of the perfect autoencoder $h^{\star}$ should project peripheral pixels to the nullspace of the decoder's weight $\mA\,$, nullifying the information from the peripherals to propagate into estimating the pixel $\xx$ at the center.
In other words, the encoder $f$ does not learn to infer either $\xx$ from $B$ or $B$ from $\xx$ but simply acts as a separation of information between the center pixel and the peripheral pixels.

In practice, however, this does not happen.
In fact, the matrices $\mW_{\text{cen}, \epsilon'}$ and $\mW_{\text{per}, \epsilon'}$ are not just mathematical tools to represent the mechanism how the complicated nonlinear encoder $f$ acts upon the given input $\xx\,$.
From \eqref{eq:cpwl_fn_autoencoder_decomposed2}, we can express only the encoder $f$ as:
\begin{equation}
\label{eq:cpwl_fn_encoder_decomposed}
f(\tilde{\xx}) = f\left( \begin{bmatrix} \xx \\ B \end{bmatrix} \right) = \begin{bmatrix} \mW_{\text{cen}, \epsilon'} & \mW_{\text{per}, \epsilon'} \end{bmatrix} \begin{bmatrix} \xx \\ B \end{bmatrix} + \vb_{\epsilon'} = \mW_{\text{cen}, \epsilon'} \xx + \mW_{\text{per}, \epsilon'} B + \vb_{\epsilon'}\,.
\end{equation}
Taking the derivatives reveals interesting relationship between the gradients of $f$ and the effective weight at a particular pixel.
\begin{equation}
\label{eq:cpwl_fn_encoder_derivatives}
\frac{\partial f}{\partial \xx} = \mW_{\text{cen}, \epsilon'} \quad \text{and} \quad \frac{\partial f}{\partial B} = \mW_{\text{per}, \epsilon'}\,.
\end{equation}
Therefore, for the lossless autoencoder, the relationship between the encoder's gradient and the decoder's weights are:
\begin{equation}
\label{eq:cpwl_fn_encoder_derivatives2}
\mA \frac{\partial f}{\partial \xx} = \mI \quad \text{and} \quad \mA \frac{\partial f}{\partial B} B = \bm{0}\,.
\end{equation}
We can calculate the matrix multiplication $\mA\partial f / \partial \xx$ and see how this differs from the identity to determine how the encoder mixes information between $\xx$ and $B\,$.
Note that $\partial f / \partial \xx \in \R^{C' \times C}$ is a gradient only at the locus $c\,$.
Figure~\ref{fig:appx_math_imperfect} shows a simple experiment to check if the value of $\mA\partial f / \partial \xx$ deviates from unity.
As the result presents, our $a$RGB encoder relies strongly on the neighboring structure to reconstruct a pixel.

\input{figures/appx_math_imperfect/main}

%% file: figures/appx_math_imperfect/main.tex
\begin{figure}[t]
\newcommand{\height}{5cm}
\centering
\subfloat[\small Sample image patch. \label{fig:appx_math_imperfect:image}]{\includegraphics[height=\height]{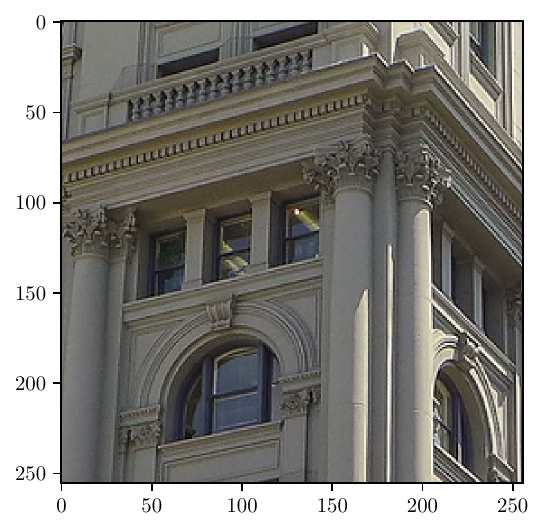}}
\hfill
\subfloat[\small $\mA \partial f / \partial \xx\,$. \label{fig:appx_math_imperfect:heatmap}]{\includegraphics[height=\height]{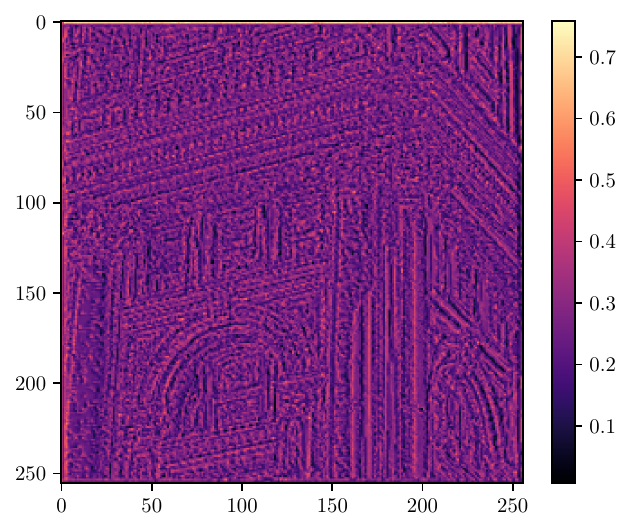}}
\hfill
\subfloat[\small Histogram of $\mA \partial f / \partial \xx\,$. \label{fig:appx_math_imperfect:histogram}]{\includegraphics[height=\height]{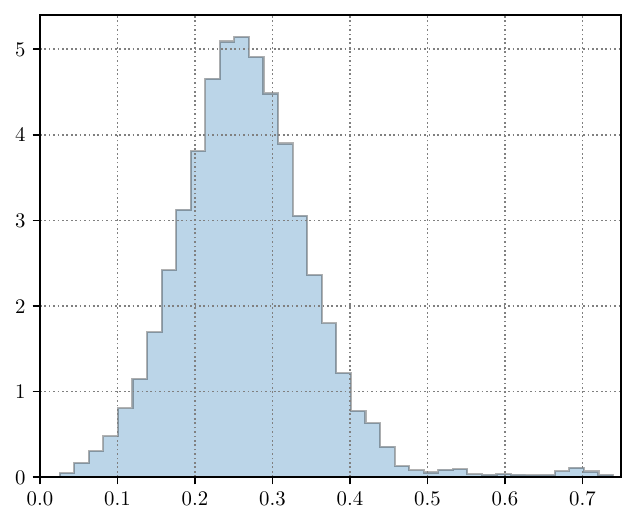}}
\\
\caption{\small
    \textbf{Measurement of the degree of self-reference of the $a$RGB encoder.}
    The sample image is brought from Urban100 dataset \citep{cv:data:huang15-urban100}.
    For a perfect autoencoder with no structure encoding capability, the values of $\mA \partial f / \partial \xx$ should be 1 for every pixel.
    However, in spite of our high reconstruction accuracy ($62.738 \mathrm{dB}$ PSNR for this patch, which corresponds to an average color deviation of a pixel of $0.069 / 255$), the value of $\mA \partial f / \partial \xx$ vary from 0.0066 to 0.7582, with average of 0.2654.
    This is unexpectedly low regarding the high accuracy, indicating that the reconstruction of our autoencoder heavily relies on the pixel's neighboring structure.
    Note that $\mA \in \R^{C \times C'}$ and $f / \partial \xx \in \R^{C' \times H \times W}\,$, where each element is obtained mutually independently.
    The heatmap and the histogram is obtained by taking the root-mean-square of the values over the three color channels.
}
\label{fig:appx_math_imperfect}
\end{figure}

%% file: sections/B_impl_detail.tex
This section presents additional details for training the main $a$RGB autoencoder used throughout the experiments in Section~\ref{sec:4_experiment}.

\subsection{Training the $a$RGB autoencoder}
\label{sec:b_impl_detail:autoencoder_training}

\paragraph{Architecture.}
The $a$RGB autoencoder consists of three models: a convolutional router $f_{\text{r}}\,$, $K=20$ convolutional experts $f_{1}, \ldots, f_{K}\,$, and a linear decoder $g\,$.
The architecture is drawn in Table~\ref{tab:appx_architecture:router}, \ref{tab:appx_architecture:expert}, and \ref{tab:appx_architecture:decoder}, where C3 and C1 denote convolutions with kernel size 3 and 1, respectively, and L0.2 is a leaky ReLU with negative slope 0.2.
Also, BN is a batch normalization with the same channel width as the output channel of the convolution in the same line.
For $3 \times 3$ convolutions, we use zero padding of size 1.
The router $f_{\text{r}}$ consists of three $3 \times 3$ and two $1 \times 1$ convolutions, with batch normalization \citep{cv:layer:ioffe15-batchnorm} and leaky ReLU with negative slope 0.2 between each pair of convolutions, except for after the first convolution, where we put only a leaky ReLU.
Each expert $f_{k}$ for all $k \in \{1, \ldots, K\}$ has identical architecture, with four $3 \times 3$ convolutions and leaky ReLUs with negative slope of 0.2 in between.
We do not use normalization layers in the experts.
All the layers are initialized by the default PyTorch initializer \citep{sw:pytorch}, meaning that each expert is initialized with different weights.
Overall, the router $f_{\text{r}}$ has a receptive field of size $7 \times 7\,$, and each expert $f_{k}$ has a receptive field of size $9 \times 9\,$, hence $R = 9$ in Section~\ref{sec:3_method:training}.
For the linear decoder, we use three $1 \times 1$ convolutions without activations and normalization layers, with each layer gradually reducing the channel dimension.
Only the last convolution of the decoder has a bias term.
We have empirically found that this three linear layer architecture leads to a slightly better convergence.
We can easily compute the effective weight of the decoder by multiplying all the internal weights with standard matrix multiplication.
The decoder has receptive field of a single pixel and is also linear.

\input{sections/tables/appx_architecture}

\paragraph{Data preparation.}
Our training dataset consists of a mix of image patches obtained from several sources, including DIV2K~\citep{cv:data:agustsson17-div2k}, Flickr2K~\citep{cv:data:timofte17-flickr2k}, DIV8K~\citep{cv:data:gu19-div8k}, and ImageNet-1k~\citep{cv:data:russakovsky15-imagenet1k} datasets.
The DIV2K, Flickr2K, and DIV8K datasets are high quality high resolution image datasets designed for training image super-resolution models.
Those are selected in order to provide our autoencoder with rich structural information in clean, visually pleasing natural images.
The DIV2K training set contains 800 high quality images, the Flickr2K set has 2,650 images, and the DIV8K dataset consists of 1,500 very high quality images up to 8K resolution.
We also added the ImageNet-1k dataset to the training data to increase the diversity of image structures for supervision.
Since the DIV2K, Flickr2K, and DIV8K datasets have much higher image size than one typically used for training a network, we have preprocessed these three dataset by cropping into patches of size $480 \times 480$ with stride $240\,$.
The results are 32,592 patches for the DIV2K, 107,403 patches for the Flickr2K, and 551,136 patches for the DIV8K dataset.
The patch datasets are then concatenated with 1,281,167 images from the ImageNet-1k training data to build up the training data with 1,972,298 images in total.
For each image fetched for training, a random crop of $256 \times 256$ is applied first, and then random horizontal and vertical flips, followed by a random 90 degrees rotation are applied consecutively.

\paragraph{Finding the optimal hyperparameters.}
We searched for the optimal architecture and other training hyperparameters by ablation studies mentioned in Section~\ref{sec:5_discussion:ablation}.
For the ease of comparison, we unify the training process for the ablation studies in a simpler setting.
Only using DIV2K dataset \citep{cv:data:agustsson17-div2k} except for the second batch of experiments in Table~\ref{tab:ablation}, we vary the hyperparameters for training.
Because the training dataset and the task become simple, we reduce the patch size to $192 \times 192$ and train our autoencoder for 300k iterations.
Furthermore, since the per-pixel router $f_{\text{r}}$ of our $a$RGB encoder distributes each pixel to a single expert, the number of iterations required for each expert to be trained using the same amount of data scales linearly to the number of experts.
To compensate the effect, we changed the number of training iterations for the autoencoder accordingly.
All the results in Section~\ref{sec:5_discussion:ablation} is quantified by the accuracy of $\times 4$ image super-resolution task using RRDBNets \citep{cv:sr:wang18-esrgan} trained to minimize the $L_{1}$ loss defined over the $a$RGB space.
All the RRDBNet models are trained only with DIV2K dataset for 300k iterations from scratch.

\paragraph{Training hyperparameters.}
The weight of the load-balancing loss is selected to be $\lambda = 10^{-2}\,$ based on empirical observations.
We found this to be the minimum value that ensures uniform expert assignment throughout the training process.
This was chosen from a parameter sweep in the ranges from $10^{0}$ to $10^{-5}$ in powers of 10\,.
The network is trained with a batch size of 16\,.
An Adam \citep{ml:opt:kingma15-adam} optimizer is used with its default hyperparameters and an initial learning rate of 5e-4 is used.
We use cosine learning rate schedule \citep{ml:opt:Loshchilov17-cosinelr}, starting from a period of 1k iteration, increased doubly up to 256k iterations.
The training ends at 511k iterations.

\paragraph{Autoencoding accuracy.}
For completeness, we provide the autoencoding accuracy for the validation datasets used throughout this work in Table~\ref{tab:appx:autoencode_acc}.
Because the output of the $a$RGB autoencoder deviates from the original image only less than a single quantization step (1/255) on average, we regard the $a$RGB autoencoder as almost lossless in our analysis except for Appendix~\ref{sec:a_math}, where we find that the nonideal reconstruction capability helps the autoencoder to learn structural prior of images.

\input{sections/tables/appx_autoencode_acc}

\subsection{Notes on training image restoration models}
\label{sec:b_impl_detail:gradscale}

\paragraph{Change in optimal training hyperparameters}
As emphasized throughout our manuscript, we have not changed the hyperparameters except for the very loss function in every experiment in Section~\ref{sec:4_experiment}.
However, we also note that our demonstration does not necessarily mean that those setups are optimal.
For instance, as mentioned in Section~\ref{sec:5_discussion:metricspace}, altering the representation space leads to dramatic change in the scale and the shape of the gradients fed into the image restoration model during its training.
Under stochastic gradient descent algorithms, this increase in the size of gradients of more than a hundredfold leads to significant changes in the training dynamics of those models.
It is, therefore, less likely that the original set of hyperparameters is still optimal in the new representation space.
Likewise, replacing the representation space may also change the optimal architecture for the image restoration task defined over the new space.
Although we strongly believe that searching through the new possibilities allowed by our $a$RGB representation should be a fascinating research topic, this is beyond the scope of our paper.
One example close to this direction is our demonstration of the NAFNet trained for $L_{1, \text{$a$RGB}}$ loss, reported in Section~\ref{sec:4_experiment}.
In this experiment, because the replacement of metric causes the change in scale and shape of its gradients, we have doubled the weight of the metric for better convergence.
As a result, this new setup has led us to better denoisers than both the original one and the one trained with only the representation space being altered.
We leave further exploration of this topic for future work.

\paragraph{Recommendation regarding gradient clipping}
As a final remark, it is highly recommended to remove gradient clipping to maximize the advantage of using $a$RGB-based losses.
Section~\ref{sec:3_method:training} attributes the performance gain caused by the additive noise to the sensitivity increase in the $a$RGB encoder $f\,$.
In practice, the effect can be observed as an increment of two orders of magnitude in the size of gradients of the image restoration models being trained.
The same scale of optimizer's step size can only be achieved by increasing the learning rate a hundredfold, which quickly leads to training instability.
We may safely conclude that the per-pixel distance losses in our $a$RGB space helps training of image restoration models by stably increasing the internal gradients.
However, in recent image restoration techniques \citep{cv:deblur:zamir21-mprnet,cv:res:chen22-nafnet}, especially for the models with attention layers \citep{nlp:archi:vaswani17-transformer}, gradient clipping is a common practice to stabilize the training of the model.
To take advantage of our method, in Section~\ref{sec:4_experiment}, we changed gradient clipping mechanism to clamp at a inf norm of 20 for every experiment.
This value is barely touched throughout the training process.

%% file: sections/tables/appx_architecture.tex
\begin{table*}[t]
\newcommand{\figwidth}{.32\linewidth}
\begin{minipage}{\figwidth}
\begin{table}[H]
\caption{\small
\textbf{Architecture of $f_{\text{r}}\,$.}
}
\label{tab:appx_architecture:router}
\vspace{.5em}
\centering
\resizebox{\linewidth}{!}{%
\begin{tabular}{cc}
    \toprule
    C3 $\rightarrow$ L0.2 & $3ch \rightarrow 64ch$ \\
    \midrule
    C3 $\rightarrow$ BN $\rightarrow$ L0.2 & $64ch \rightarrow 128ch$ \\
    \midrule
    C3 $\rightarrow$ BN $\rightarrow$ L0.2 & $128ch \rightarrow 256ch$ \\
    \midrule
    C1 $\rightarrow$ BN $\rightarrow$ L0.2 & $256ch \rightarrow 512ch$ \\
    \midrule
    C1 $\rightarrow$ Softmax & $512ch \rightarrow 20ch$ \\
    \bottomrule
\end{tabular}%
}%
\end{table}
\end{minipage}
\hfill
\begin{minipage}{\figwidth}
\begin{table}[H]
\caption{\small
\textbf{Architecture of $f_{k}\,$.}
}
\label{tab:appx_architecture:expert}
\vspace{.5em}
\centering
\resizebox{\linewidth}{!}{%
\begin{tabular}{cc}
    \toprule
    C3 $\rightarrow$ L0.2 & $3ch \rightarrow 32ch$ \\
    \midrule
    C3 $\rightarrow$ L0.2 & $32ch \rightarrow 64ch$ \\
    \midrule
    C3 $\rightarrow$ L0.2 & $64ch \rightarrow 128ch$ \\
    \midrule
    C3 & $128ch \rightarrow 128ch$ \\
    \bottomrule
\end{tabular}%
}%
\end{table}
\end{minipage}
\hfill
\begin{minipage}{\figwidth}
\begin{table}[H]
\caption{\small
\textbf{Architecture of $g\,$.}
}
\label{tab:appx_architecture:decoder}
\vspace{.5em}
\centering
\resizebox{\linewidth}{!}{%
\begin{tabular}{cc}
    \toprule
    C1 (no bias) & $128ch \rightarrow 64ch$ \\
    \midrule
    C1 (no bias) & $64ch \rightarrow 32ch$ \\
    \midrule
    C1 & $32ch \rightarrow 3ch$ \\
    \bottomrule
\end{tabular}%
}%
\end{table}
\vfill
\end{minipage}
\vspace{-.7em}
\end{table*}

%% file: sections/tables/appx_autoencode_acc.tex
\begin{table*}[t]
\caption{\small
\textbf{Reconstruction accuracy of the $a$RGB autoencoder on out-of-distribution datasets.}
We measured the PSNR scores \emph{wihtout} quantization to 255 scale.
Average RGB difference in the second row stands for the average absolute difference in the pixel's color value out of the maximum range of 255.
All the values are significantly below the quantization gap of $0.5\,$, indicating almost perfect reconstruction of the input.
}
\label{tab:appx:autoencode_acc}
\vspace{.5em}
\centering
\resizebox{.75\linewidth}{!}{%
\begin{tabular}{lcccccc}
    \toprule
    & Set5 & Set14 & Urban100 & DIV2K-Val & SIDD-Val & GoPro-Val \\
    \midrule
    PSNR [$\mathrm{dB}$] & 67.206 & 64.531 & 65.556 & 70.812 & 72.007 & 72.853 \\
    Avg. RGB diff. & 0.0477 & 0.0602 & 0.0669 & 0.0418 & 0.0301 & 0.0266 \\
    \bottomrule
\end{tabular}%
}%
\vspace{-0.5em}
\end{table*}

%% file: sections/C_esrgan.tex
Table~\ref{tab:appx:result_gan} extends Section~\ref{sec:4_experiment} to provide full evaluation results on the $4 \times$ perceptual super-resolution task using ESRGAN \citep{cv:sr:wang18-esrgan}.
As mentioned in Section~\ref{sec:4_experiment}, our training process exactly follows the official implementation \citep{cv:sr:wang18-esrgan}.
We first pre-train the network, RRDBNet, with DIV2K \citep{cv:data:agustsson17-div2k} and Flickr2K \citep{cv:data:timofte17-flickr2k} combined, for 1M iterations using RGB $L_{1}$ loss.
Then, the weights are fine-tuned with the loss written in the first column for 400k iterations using DIV2K, Flickr2K, and OSTv2 \citep{cv:sr:wang18-sftgan} datasets combined.
Preparation of the datasets are done with the same code the authors has provided.
In this task, the $a$RGB loss is a simple $L_{1}$ loss over the range of the $a$RGB encoder.
We leave all the other training hyperparameters untouched.

In addition to LPIPS \citep{cv:obj:zhang18-lpips}, NIQE \citep{cv:obj:mittal13-niqe}, and FID \citep{cv:gan:heusel17-tturfid} metrics, we added DISTS \citep{cv:obj:Ding20-dists}, a pairwise perceptual metric for image restoration.
Moreover, four common benchmarks for the percpeptual super-resolution, \ie, Set14 \citep{cv:data:zeyde10-set14}, BSD100 \citep{cv:data:martin01-bsd300}, Manga109 \citep{cv:data:Matsui17-manga109}, and OutdoorSceneTest300 \citep{cv:sr:wang18-sftgan} datasets are added for comparison.
The updated results are shown in Table~\ref{tab:appx:result_gan}.
The reported values are calculated in the RGB domain, cropping 4 pixels from the outer sides for PSNR, SSIM, and NIQE \citep{cv:obj:mittal13-niqe} scores.
For DISTS \citep{cv:obj:Ding20-dists} and LPIPS \citep{cv:obj:zhang18-lpips}, we use the code from official repo, and for FID \citep{cv:gan:heusel17-tturfid}, we report scores from the PyTorch reimplementation \citep{cv:metric:Seitzer20-pytorchfid} of the original Tensorflow code.

The data shows that $L_{1, \text{$a$RGB}}$ loss without the VGG perceptual loss yields comparable results to the $L_{1}$ loss without the perceptual loss in adversarial training.
However, when equipped with VGG perceptual loss, the results show a general increase in performance over the original ESRGAN in all the distoration-based metrics (PSNR, SSIM), the pairwise perceptual metrics (LPIPS, DISTS), and the unpaired quality assessment metrics (NIQE, FID).

%% file: sections/D_more_result.tex
This section provides more visual results from the main experiments.
Figure~\ref{fig:appx:result_gan:p1} through \ref{fig:appx:result_gan:p6} show results from the ESRGAN models \citep{cv:sr:wang18-esrgan} trained with and without $a$RGB representation.
Empirically, the supervision given by our $a$RGB encoder helps the model avoid generating visual artifacts and color inconsistency induced by adversarial training.
Figure~\ref{fig:appx:result_denoise} shows additional results from the real image denoising task solved with NAFNet \citep{cv:res:chen22-nafnet}.
Lastly, image deblurring using MPRNet \citep{cv:deblur:zamir21-mprnet} trained from our $a$RGB representation is further demontrated in Figure~\ref{fig:appx:result_deblur:p1} and \ref{fig:appx:result_deblur:p2}.
As visualized in Appendix~\ref{sec:f_decomposition}, the additional information embodied in the extra dimensions of the $a$RGB representation resembles edgeness information of an image.
We conjecture that this allows a pairwise per-pixel distance defined over our $a$RGB space to provide image restoration models with stronger supervision leading to the reconstruction of sharper edges.
The results reveal that this effect is realized as suppression of artifacts in the perceptual image super-resolution task, sharper produced images in the image denoising task, and better reconstruction of edges and more accurate alignments in the image deblurring task.

\input{sections/tables/appx_result_gan}

\FloatBarrier
\input{figures/appx_result_esrgan/main1}
\input{figures/appx_result_esrgan/main5}
\FloatBarrier

\FloatBarrier
\input{figures/appx_result_esrgan/main3}
\input{figures/appx_result_esrgan/main4}
\FloatBarrier

\FloatBarrier
\input{figures/appx_result_esrgan/main2}
\FloatBarrier

\FloatBarrier
\input{figures/appx_result_esrgan/main6}
\FloatBarrier

\FloatBarrier
\input{figures/appx_result_denoise/main}
\FloatBarrier

\FloatBarrier
\input{figures/appx_result_deblur/main1}
\FloatBarrier

\FloatBarrier
\input{figures/appx_result_deblur/main2}
\FloatBarrier

%% file: sections/tables/appx_result_gan.tex
\begin{table*}
\caption{\small
\textbf{Complete quantitative results for training ESRGAN $\times 4$ in the $a$RGB space.}
Improved results are highlighted in \textbf{boldface} characters.
}
\label{tab:appx:result_gan}
\vspace{.5em}
\resizebox{\linewidth}{!}{%
\begin{tabular}{lcccccccccccc}
    \toprule
    & \multicolumn{6}{c}{{Set14}} & \multicolumn{6}{c}{{B100}} \\
    \cmidrule(lr){2-7} \cmidrule(lr){8-13}
    Training objective & PSNR$\uparrow$ & SSIM$\uparrow$ & LPIPS$\downarrow$ & DISTS$\downarrow$ & NIQE$\downarrow$ & FID$\downarrow$ & PSNR$\uparrow$ & SSIM$\uparrow$ & LPIPS$\downarrow$ & DISTS$\downarrow$ & NIQE$\downarrow$ & FID$\downarrow$ \\
    \midrule
    $0.01 L_{1} \qquad \hspace{.07em} + L_{\text{VGG}} + 0.005 L_{\text{Adv}}$\textsuperscript{$\dagger$}
        & 24.494 & 0.6543 & 0.1341 & 0.06374 & 3.8774 & 56.700 & 23.909 & 0.6205 & 0.1617 & 0.09603 & 3.6636 & 51.521 \\
    ${\color{BrickRed} 0.01 L_{1, \text{$a$RGB}}} + L_{\text{VGG}} + 0.005 L_{\text{Adv}}$
        & \textbf{24.796} & \textbf{0.6623} & \textbf{0.1281} & \textbf{0.06052} & \textbf{3.7230} & \textbf{52.797} & \textbf{24.138} & \textbf{0.6306} & \textbf{0.1595} & \textbf{0.09578} & \textbf{3.4185} & \textbf{47.665} \\ 
    \midrule
    & \multicolumn{6}{c}{{Manga109}} & \multicolumn{6}{c}{{Urban100}} \\
    \cmidrule(lr){2-7} \cmidrule(lr){8-13}
    & PSNR$\uparrow$ & SSIM$\uparrow$ & LPIPS$\downarrow$ & DISTS$\downarrow$ & NIQE$\downarrow$ & FID$\downarrow$ & PSNR$\uparrow$ & SSIM$\uparrow$ & LPIPS$\downarrow$ & DISTS$\downarrow$ & NIQE$\downarrow$ & FID$\downarrow$ \\
    \midrule
    $0.01 L_{1} \qquad \hspace{.07em} + L_{\text{VGG}} + 0.005 L_{\text{Adv}}$\textsuperscript{$\dagger$}
        & 26.441 & 0.8170 & 0.0646 & 0.01036 & 3.5758 & 11.282 & 22.776 & 0.7033 & 0.1232 & 0.02432 & 4.2067 & 20.616 \\
    ${\color{BrickRed} 0.01 L_{1, \text{$a$RGB}}} + L_{\text{VGG}} + 0.005 L_{\text{Adv}}$
        & \textbf{26.651} & \textbf{0.8186} & \textbf{0.0630} & \textbf{0.00863} & \textbf{3.4245} & \textbf{10.907} & \textbf{23.270} & \textbf{0.7196} & \textbf{0.1183} & \textbf{0.02050} & \textbf{3.8982} & \textbf{17.739} \\ 
    \midrule
    & \multicolumn{6}{c}{{DIV2K-Val}} & \multicolumn{6}{c}{{OST300}} \\
    \cmidrule(lr){2-7} \cmidrule(lr){8-13}
    & PSNR$\uparrow$ & SSIM$\uparrow$ & LPIPS$\downarrow$ & DISTS$\downarrow$ & NIQE$\downarrow$ & FID$\downarrow$ & PSNR$\uparrow$ & SSIM$\uparrow$ & LPIPS$\downarrow$ & DISTS$\downarrow$ & NIQE$\downarrow$ & FID$\downarrow$ \\
    \midrule
    $0.01 L_{1} \qquad \hspace{.07em} + L_{\text{VGG}} + 0.005 L_{\text{Adv}}$\textsuperscript{$\dagger$}
        & 26.627 & 0.7033 & 0.1154 & 0.00545 & 3.0913 & 13.557 & 23.249 & 0.6166 & 0.1688 & 0.05734 & \textbf{3.3834} & 21.851 \\
    ${\color{BrickRed} 0.01 L_{1, \text{$a$RGB}}} + L_{\text{VGG}} + 0.005 L_{\text{Adv}}$
        & \textbf{26.845} & \textbf{0.7500} & \textbf{0.1110} & \textbf{0.00351} & \textbf{2.9615} & \textbf{12.799} & \textbf{23.649} & \textbf{0.6338} & \textbf{0.1662} & \textbf{0.05484} & 3.5247 & \textbf{19.952} \\
    \bottomrule
    \multicolumn{13}{l}{\textsuperscript{$\dagger$}\footnotesize{The official ESRGAN models \citep{cv:sr:wang18-esrgan}.}}
\end{tabular}%
}%
\end{table*}

%% file: figures/appx_result_esrgan/main1.tex
\begin{figure*}[t]
\newcommand{\figwidth}{.199\linewidth}
\newcommand{\varA}{LR}%
\newcommand{\varB}{RRDBNet}%
\newcommand{\varC}{L1+Adv}%
\newcommand{\varD}{L1aRGB+Adv}%
\newcommand{\varE}{ESRGAN}%
\newcommand{\varF}{L1aRGB+VGG+Adv}%
\newcommand{\varG}{GT}%
\centering
\newcommand{\seed}{DIV2K100_0830_2493}%
    \subfloat{\includegraphics[width=\figwidth]{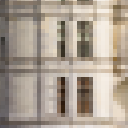}}
    \hfill
    \subfloat{\includegraphics[width=\figwidth]{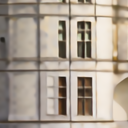}}
    \hfill
    \subfloat{\includegraphics[width=\figwidth]{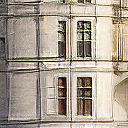}}
    \hfill
    \subfloat{\includegraphics[width=\figwidth]{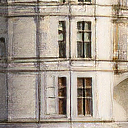}}
    \hfill
    \subfloat{\includegraphics[width=\figwidth]{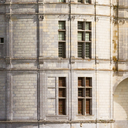}}
    \\[-0.05em]
\renewcommand{\seed}{DIV2K100_0869_c82d}
    \subfloat{\includegraphics[width=\figwidth]{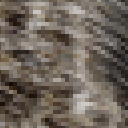}}
    \hfill
    \subfloat{\includegraphics[width=\figwidth]{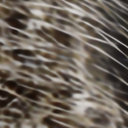}}
    \hfill
    \subfloat{\includegraphics[width=\figwidth]{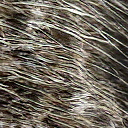}}
    \hfill
    \subfloat{\includegraphics[width=\figwidth]{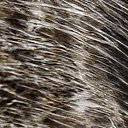}}
    \hfill
    \subfloat{\includegraphics[width=\figwidth]{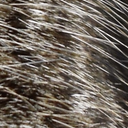}}
    \\[-0.05em]
\renewcommand{\seed}{DIV2K100_0883_2351}
    \subfloat{\includegraphics[width=\figwidth]{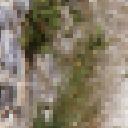}}
    \hfill
    \subfloat{\includegraphics[width=\figwidth]{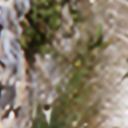}}
    \hfill
    \subfloat{\includegraphics[width=\figwidth]{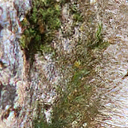}}
    \hfill
    \subfloat{\includegraphics[width=\figwidth]{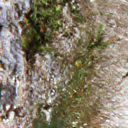}}
    \hfill
    \subfloat{\includegraphics[width=\figwidth]{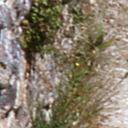}}
    \\[-0.05em]
\renewcommand{\seed}{DIV2K100_0807_660c}
    \addtocounter{subfigure}{-15}
    \subfloat[LR]{\includegraphics[width=\figwidth]{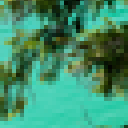}}
    \hfill
    \subfloat[RRDBNet]{\includegraphics[width=\figwidth]{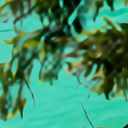}}
    \hfill
    \subfloat[ESRGAN]{\includegraphics[width=\figwidth]{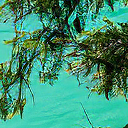}}
    \hfill
    \subfloat[\resizebox{7em}{!}{\scriptsize\color{BrickRed} $L_{1, \text{$a$RGB}} + L_{\text{V}} + L_{\text{A}}$}]{\includegraphics[width=\figwidth]{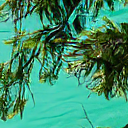}}
    \hfill
    \subfloat[HR]{\includegraphics[width=\figwidth]{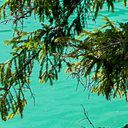}}
    \\[-.7em]
\caption{\small
\textbf{Qualitative comparison of ESRGAN models trained with different loss functions on DIV2K-Val \citep{cv:data:agustsson17-div2k} benchmark.}
}
\label{fig:appx:result_gan:p1}
\end{figure*}

%% file: figures/appx_result_esrgan/main5.tex
\begin{figure*}[t]
\newcommand{\figwidth}{.199\linewidth}
\newcommand{\varA}{LR}%
\newcommand{\varB}{RRDBNet}%
\newcommand{\varC}{L1+Adv}%
\newcommand{\varD}{L1aRGB+Adv}%
\newcommand{\varE}{ESRGAN}%
\newcommand{\varF}{L1aRGB+VGG+Adv}%
\newcommand{\varG}{GT}%
\centering
\newcommand{\seed}{B100_223061_7669}%
    \subfloat{\includegraphics[width=\figwidth]{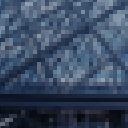}}
    \hfill
    \subfloat{\includegraphics[width=\figwidth]{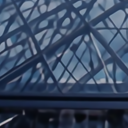}}
    \hfill
    \subfloat{\includegraphics[width=\figwidth]{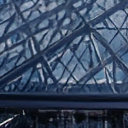}}
    \hfill
    \subfloat{\includegraphics[width=\figwidth]{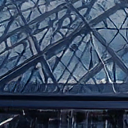}}
    \hfill
    \subfloat{\includegraphics[width=\figwidth]{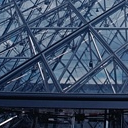}}
    \\[-0.05em]
\renewcommand{\seed}{B100_304034_a170}%
    \subfloat{\includegraphics[width=\figwidth]{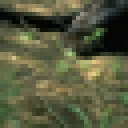}}
    \hfill
    \subfloat{\includegraphics[width=\figwidth]{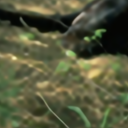}}
    \hfill
    \subfloat{\includegraphics[width=\figwidth]{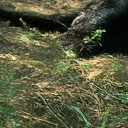}}
    \hfill
    \subfloat{\includegraphics[width=\figwidth]{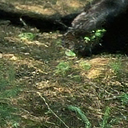}}
    \hfill
    \subfloat{\includegraphics[width=\figwidth]{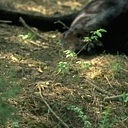}}
    \\[-0.05em]
\renewcommand{\seed}{B100_105025_9ca3}%
    \addtocounter{subfigure}{-10}
    \subfloat[LR]{\includegraphics[width=\figwidth]{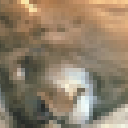}}
    \hfill
    \subfloat[RRDBNet]{\includegraphics[width=\figwidth]{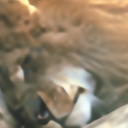}}
    \hfill
    \subfloat[ESRGAN]{\includegraphics[width=\figwidth]{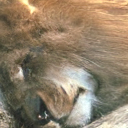}}
    \hfill
    \subfloat[\resizebox{7em}{!}{\scriptsize\color{BrickRed} $L_{1, \text{$a$RGB}} + L_{\text{V}} + L_{\text{A}}$}]{\includegraphics[width=\figwidth]{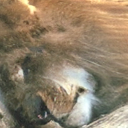}}
    \hfill
    \subfloat[HR]{\includegraphics[width=\figwidth]{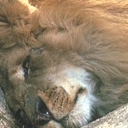}}
    \\[-.7em]
\caption{\small
\textbf{Qualitative comparison of ESRGAN models trained with different loss functions on B100 \citep{cv:data:martin01-bsd300} benchmark.}
}
\label{fig:appx:result_gan:p5}
\end{figure*}

%% file: figures/appx_result_esrgan/main3.tex
\begin{figure*}[t]
\newcommand{\figwidth}{.199\linewidth}
\newcommand{\varA}{LR}%
\newcommand{\varB}{RRDBNet}%
\newcommand{\varC}{L1+Adv}%
\newcommand{\varD}{L1aRGB+Adv}%
\newcommand{\varE}{ESRGAN}%
\newcommand{\varF}{L1aRGB+VGG+Adv}%
\newcommand{\varG}{GT}%
\centering
\newcommand{\seed}{Set14_baboon_1d1a}%
    \subfloat{\includegraphics[width=\figwidth]{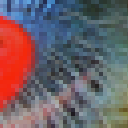}}
    \hfill
    \subfloat{\includegraphics[width=\figwidth]{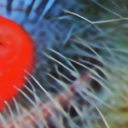}}
    \hfill
    \subfloat{\includegraphics[width=\figwidth]{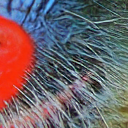}}
    \hfill
    \subfloat{\includegraphics[width=\figwidth]{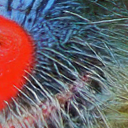}}
    \hfill
    \subfloat{\includegraphics[width=\figwidth]{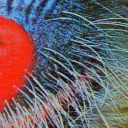}}
    \\[-0.05em]
\renewcommand{\seed}{Set14_barbara_cba9}%
    \subfloat{\includegraphics[width=\figwidth]{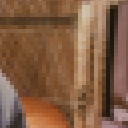}}
    \hfill
    \subfloat{\includegraphics[width=\figwidth]{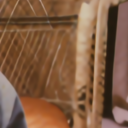}}
    \hfill
    \subfloat{\includegraphics[width=\figwidth]{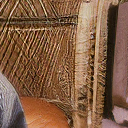}}
    \hfill
    \subfloat{\includegraphics[width=\figwidth]{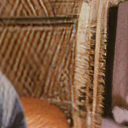}}
    \hfill
    \subfloat{\includegraphics[width=\figwidth]{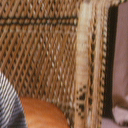}}
    \\[-0.05em]
\renewcommand{\seed}{Set14_man_1dad}%
    \addtocounter{subfigure}{-10}
    \subfloat[LR]{\includegraphics[width=\figwidth]{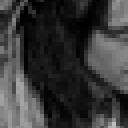}}
    \hfill
    \subfloat[RRDBNet]{\includegraphics[width=\figwidth]{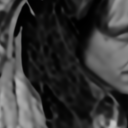}}
    \hfill
    \subfloat[ESRGAN]{\includegraphics[width=\figwidth]{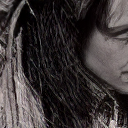}}
    \hfill
    \subfloat[\resizebox{7em}{!}{\scriptsize\color{BrickRed} $L_{1, \text{$a$RGB}} + L_{\text{V}} + L_{\text{A}}$}]{\includegraphics[width=\figwidth]{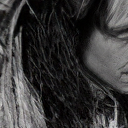}}
    \hfill
    \subfloat[HR]{\includegraphics[width=\figwidth]{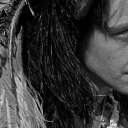}}
    \\[-.7em]
\caption{\small
\textbf{Qualitative comparison of ESRGAN models trained with different loss functions on Set14 \citep{cv:data:zeyde10-set14} benchmark.}
}
\label{fig:appx:result_gan:p3}
\end{figure*}

%% file: figures/appx_result_esrgan/main4.tex
\begin{figure*}[t]
\newcommand{\figwidth}{.199\linewidth}
\newcommand{\varA}{LR}%
\newcommand{\varB}{RRDBNet}%
\newcommand{\varC}{L1+Adv}%
\newcommand{\varD}{L1aRGB+Adv}%
\newcommand{\varE}{ESRGAN}%
\newcommand{\varF}{L1aRGB+VGG+Adv}%
\newcommand{\varG}{GT}%
\centering
\newcommand{\seed}{Manga109_MagicianLoad_97e5}%
    \subfloat{\includegraphics[width=\figwidth]{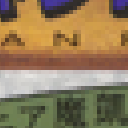}}
    \hfill
    \subfloat{\includegraphics[width=\figwidth]{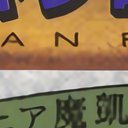}}
    \hfill
    \subfloat{\includegraphics[width=\figwidth]{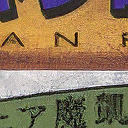}}
    \hfill
    \subfloat{\includegraphics[width=\figwidth]{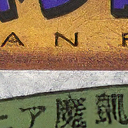}}
    \hfill
    \subfloat{\includegraphics[width=\figwidth]{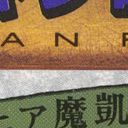}}
    \\[-0.05em]
\renewcommand{\seed}{Manga109_WarewareHaOniDearu_4e39}%
    \subfloat{\includegraphics[width=\figwidth]{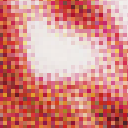}}
    \hfill
    \subfloat{\includegraphics[width=\figwidth]{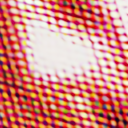}}
    \hfill
    \subfloat{\includegraphics[width=\figwidth]{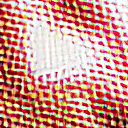}}
    \hfill
    \subfloat{\includegraphics[width=\figwidth]{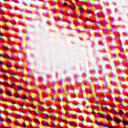}}
    \hfill
    \subfloat{\includegraphics[width=\figwidth]{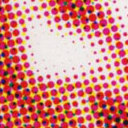}}
    \\[-0.05em]
\renewcommand{\seed}{Manga109_KarappoHighschool_6aa2}%
    \subfloat{\includegraphics[width=\figwidth]{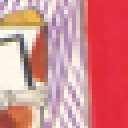}}
    \hfill
    \subfloat{\includegraphics[width=\figwidth]{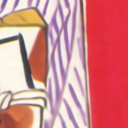}}
    \hfill
    \subfloat{\includegraphics[width=\figwidth]{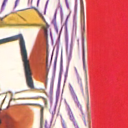}}
    \hfill
    \subfloat{\includegraphics[width=\figwidth]{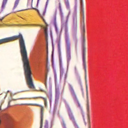}}
    \hfill
    \subfloat{\includegraphics[width=\figwidth]{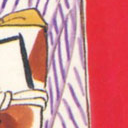}}
    \\[-0.05em]
\renewcommand{\seed}{Manga109_YumeiroCooking_bc45}%
    \addtocounter{subfigure}{-15}
    \subfloat[LR]{\includegraphics[width=\figwidth]{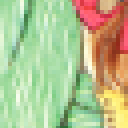}}
    \hfill
    \subfloat[RRDBNet]{\includegraphics[width=\figwidth]{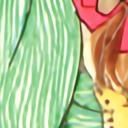}}
    \hfill
    \subfloat[ESRGAN]{\includegraphics[width=\figwidth]{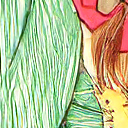}}
    \hfill
    \subfloat[\resizebox{7em}{!}{\scriptsize\color{BrickRed} $L_{1, \text{$a$RGB}} + L_{\text{V}} + L_{\text{A}}$}]{\includegraphics[width=\figwidth]{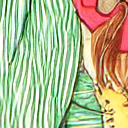}}
    \hfill
    \subfloat[HR]{\includegraphics[width=\figwidth]{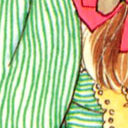}}
    \\[-.7em]
\caption{\small
\textbf{Qualitative comparison of ESRGAN models trained with different loss functions on Manga109 \citep{cv:data:Matsui17-manga109} benchmark.}
}
\label{fig:appx:result_gan:p4}
\end{figure*}

%% file: figures/appx_result_esrgan/main2.tex
\begin{figure*}[t]
\newcommand{\figwidth}{.199\linewidth}
\newcommand{\varA}{LR}%
\newcommand{\varB}{RRDBNet}%
\newcommand{\varC}{L1+Adv}%
\newcommand{\varD}{L1aRGB+Adv}%
\newcommand{\varE}{ESRGAN}%
\newcommand{\varF}{L1aRGB+VGG+Adv}%
\newcommand{\varG}{GT}%
\centering
\newcommand{\seed}{DIV2K100_img004_12f8}%
    \subfloat{\includegraphics[width=\figwidth]{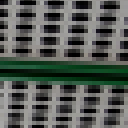}}
    \hfill
    \subfloat{\includegraphics[width=\figwidth]{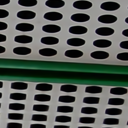}}
    \hfill
    \subfloat{\includegraphics[width=\figwidth]{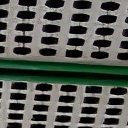}}
    \hfill
    \subfloat{\includegraphics[width=\figwidth]{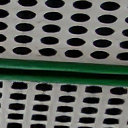}}
    \hfill
    \subfloat{\includegraphics[width=\figwidth]{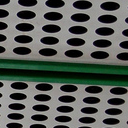}}
    \\[-0.05em]
\renewcommand{\seed}{DIV2K100_img030_f2fa}%
    \subfloat{\includegraphics[width=\figwidth]{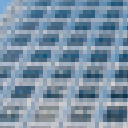}}
    \hfill
    \subfloat{\includegraphics[width=\figwidth]{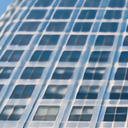}}
    \hfill
    \subfloat{\includegraphics[width=\figwidth]{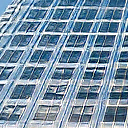}}
    \hfill
    \subfloat{\includegraphics[width=\figwidth]{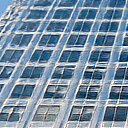}}
    \hfill
    \subfloat{\includegraphics[width=\figwidth]{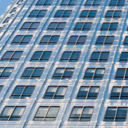}}
    \\[-0.05em]
\renewcommand{\seed}{DIV2K100_img034_68ae}%
    \subfloat{\includegraphics[width=\figwidth]{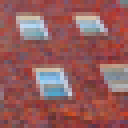}}
    \hfill
    \subfloat{\includegraphics[width=\figwidth]{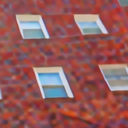}}
    \hfill
    \subfloat{\includegraphics[width=\figwidth]{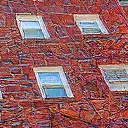}}
    \hfill
    \subfloat{\includegraphics[width=\figwidth]{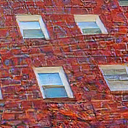}}
    \hfill
    \subfloat{\includegraphics[width=\figwidth]{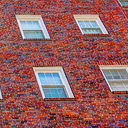}}
    \\[-0.05em]
\renewcommand{\seed}{DIV2K100_img065_ca56}%
    \subfloat{\includegraphics[width=\figwidth]{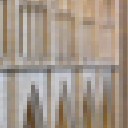}}
    \hfill
    \subfloat{\includegraphics[width=\figwidth]{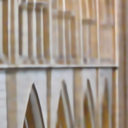}}
    \hfill
    \subfloat{\includegraphics[width=\figwidth]{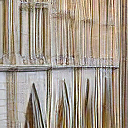}}
    \hfill
    \subfloat{\includegraphics[width=\figwidth]{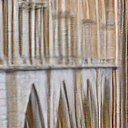}}
    \hfill
    \subfloat{\includegraphics[width=\figwidth]{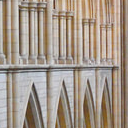}}
    \\[-0.05em]
\renewcommand{\seed}{DIV2K100_img074_63ce}%
    \subfloat{\includegraphics[width=\figwidth]{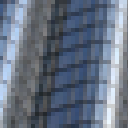}}
    \hfill
    \subfloat{\includegraphics[width=\figwidth]{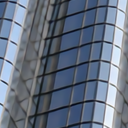}}
    \hfill
    \subfloat{\includegraphics[width=\figwidth]{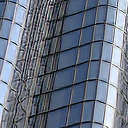}}
    \hfill
    \subfloat{\includegraphics[width=\figwidth]{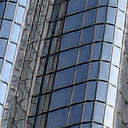}}
    \hfill
    \subfloat{\includegraphics[width=\figwidth]{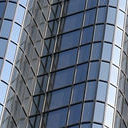}}
    \\[-0.05em]
\renewcommand{\seed}{DIV2K100_img088_d4a7}%
    \addtocounter{subfigure}{-25}
    \subfloat[LR]{\includegraphics[width=\figwidth]{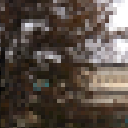}}
    \hfill
    \subfloat[RRDBNet]{\includegraphics[width=\figwidth]{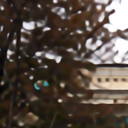}}
    \hfill
    \subfloat[ESRGAN]{\includegraphics[width=\figwidth]{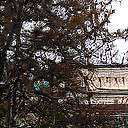}}
    \hfill
    \subfloat[\resizebox{7em}{!}{\scriptsize\color{BrickRed} $L_{1, \text{$a$RGB}} + L_{\text{V}} + L_{\text{A}}$}]{\includegraphics[width=\figwidth]{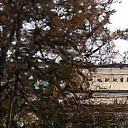}}
    \hfill
    \subfloat[HR]{\includegraphics[width=\figwidth]{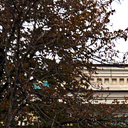}}
    \\[-.7em]
\caption{\small
\textbf{Qualitative comparison of ESRGAN models trained with different loss functions on Urban100 \citep{cv:data:huang15-urban100} benchmark.}
}
\label{fig:appx:result_gan:p2}
\end{figure*}

%% file: figures/appx_result_esrgan/main6.tex
\begin{figure*}[t]
\newcommand{\figwidth}{.199\linewidth}
\newcommand{\varA}{LR}%
\newcommand{\varB}{RRDBNet}%
\newcommand{\varC}{L1+Adv}%
\newcommand{\varD}{L1aRGB+Adv}%
\newcommand{\varE}{ESRGAN}%
\newcommand{\varF}{L1aRGB+VGG+Adv}%
\newcommand{\varG}{GT}%
\centering
\newcommand{\seed}{OST300_OST_023_cf05}%
    \subfloat{\includegraphics[width=\figwidth]{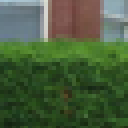}}
    \hfill
    \subfloat{\includegraphics[width=\figwidth]{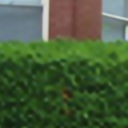}}
    \hfill
    \subfloat{\includegraphics[width=\figwidth]{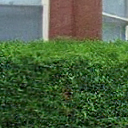}}
    \hfill
    \subfloat{\includegraphics[width=\figwidth]{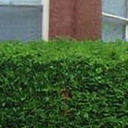}}
    \hfill
    \subfloat{\includegraphics[width=\figwidth]{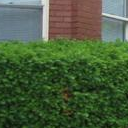}}
    \\[-0.05em]
\renewcommand{\seed}{OST300_OST_130_6e70}%
    \subfloat{\includegraphics[width=\figwidth]{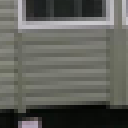}}
    \hfill
    \subfloat{\includegraphics[width=\figwidth]{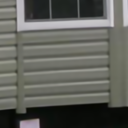}}
    \hfill
    \subfloat{\includegraphics[width=\figwidth]{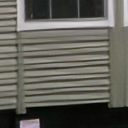}}
    \hfill
    \subfloat{\includegraphics[width=\figwidth]{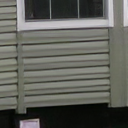}}
    \hfill
    \subfloat{\includegraphics[width=\figwidth]{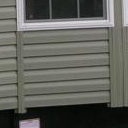}}
    \\[-0.05em]
\renewcommand{\seed}{OST300_OST_206_61fd}%
    \subfloat{\includegraphics[width=\figwidth]{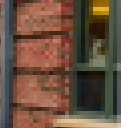}}
    \hfill
    \subfloat{\includegraphics[width=\figwidth]{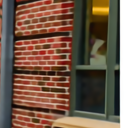}}
    \hfill
    \subfloat{\includegraphics[width=\figwidth]{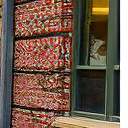}}
    \hfill
    \subfloat{\includegraphics[width=\figwidth]{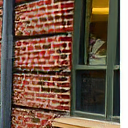}}
    \hfill
    \subfloat{\includegraphics[width=\figwidth]{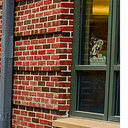}}
    \\[-0.05em]
\renewcommand{\seed}{OST300_OST_223_f304}%
    \subfloat{\includegraphics[width=\figwidth]{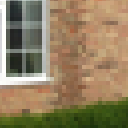}}
    \hfill
    \subfloat{\includegraphics[width=\figwidth]{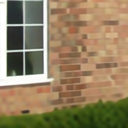}}
    \hfill
    \subfloat{\includegraphics[width=\figwidth]{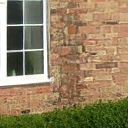}}
    \hfill
    \subfloat{\includegraphics[width=\figwidth]{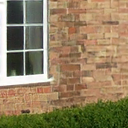}}
    \hfill
    \subfloat{\includegraphics[width=\figwidth]{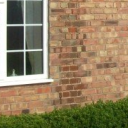}}
    \\[-0.05em]
\renewcommand{\seed}{OST300_OST_276_de2e}%
    \subfloat{\includegraphics[width=\figwidth]{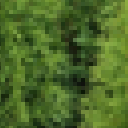}}
    \hfill
    \subfloat{\includegraphics[width=\figwidth]{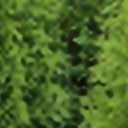}}
    \hfill
    \subfloat{\includegraphics[width=\figwidth]{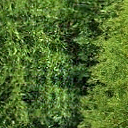}}
    \hfill
    \subfloat{\includegraphics[width=\figwidth]{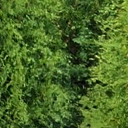}}
    \hfill
    \subfloat{\includegraphics[width=\figwidth]{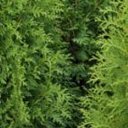}}
    \\[-0.05em]
\renewcommand{\seed}{OST300_OST_047_b694}%
    \addtocounter{subfigure}{-25}
    \subfloat[LR]{\includegraphics[width=\figwidth]{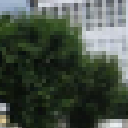}}
    \hfill
    \subfloat[RRDBNet]{\includegraphics[width=\figwidth]{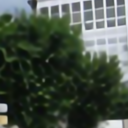}}
    \hfill
    \subfloat[ESRGAN]{\includegraphics[width=\figwidth]{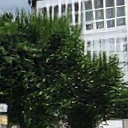}}
    \hfill
    \subfloat[\resizebox{5em}{!}{\scriptsize\color{BrickRed} $L_{1, \text{$a$RGB}} + L_{\text{V}} + L_{\text{A}}$}]{\includegraphics[width=\figwidth]{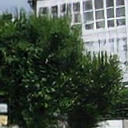}}
    \hfill
    \subfloat[HR]{\includegraphics[width=\figwidth]{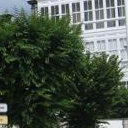}}
    \\[-.7em]
\caption{\small
\textbf{Qualitative comparison of ESRGAN models trained with different loss functions on OutdoorSceneTest300 \citep{cv:sr:wang18-sftgan} benchmark.}
}
\label{fig:appx:result_gan:p6}
\end{figure*}

%% file: figures/appx_result_denoise/main.tex
\begin{figure*}
\newcommand{\figwidth}{0.1418\linewidth}
\newcommand{\varA}{LQ}%
\newcommand{\varB}{N32}%
\newcommand{\varC}{N32+LPaRGB}%
\newcommand{\varD}{N32+L1aRGB}%
\newcommand{\varE}{N64}%
\newcommand{\varF}{N64+L1aRGB}%
\newcommand{\varG}{GT}%
\centering
\begin{subfigure}[b]{.9\linewidth}
\newcommand{\seed}{0349}%
    \subfloat{\includegraphics[width=\figwidth]{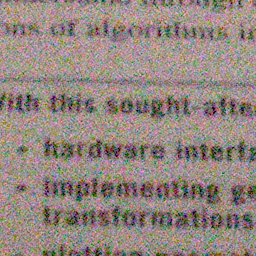}}
    \hfill
    \subfloat{\includegraphics[width=\figwidth]{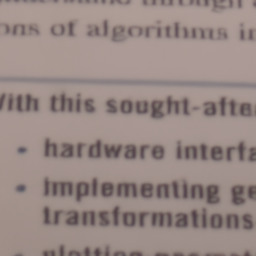}}
    \hfill
    \subfloat{\includegraphics[width=\figwidth]{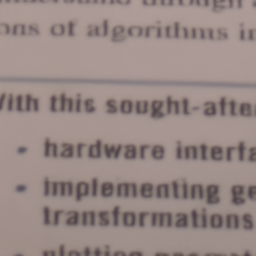}}
    \hfill
    \subfloat{\includegraphics[width=\figwidth]{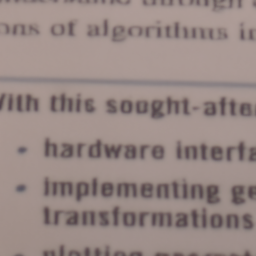}}
    \hfill
    \subfloat{\includegraphics[width=\figwidth]{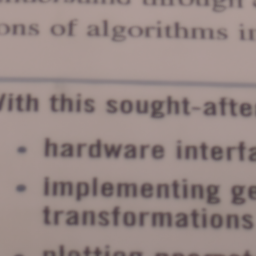}}
    \hfill
    \subfloat{\includegraphics[width=\figwidth]{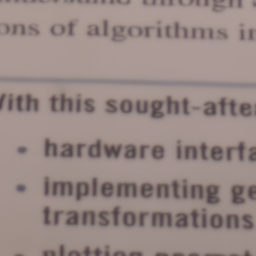}}
    \hfill
    \subfloat{\includegraphics[width=\figwidth]{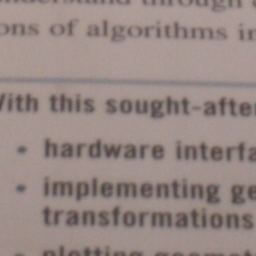}}
    \\[-0.05em]
    \addtocounter{subfigure}{-7}
    \subfloat[Noisy\\ \hspace*{1.3em}\resizebox{3.5em}{!}{19.7557 $\mathrm{dB}$}]{\includegraphics[width=\figwidth]{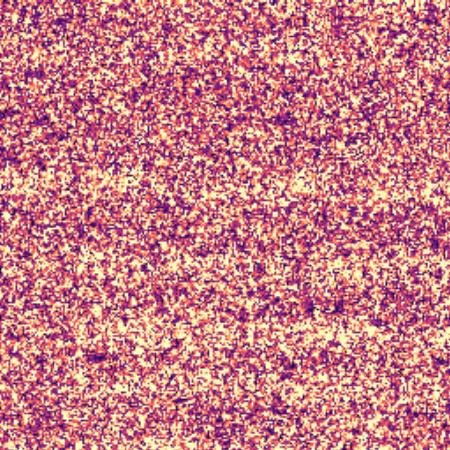}}
    \hfill
    \subfloat[N32 $L_{\text{PSNR}}$\\ \hspace*{1.3em}\resizebox{3.5em}{!}{37.0814 $\mathrm{dB}$}]{\includegraphics[width=\figwidth]{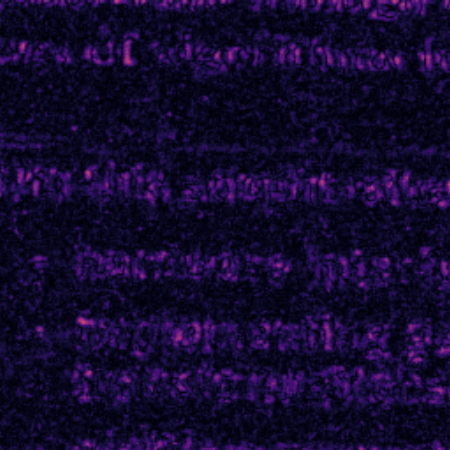}}
    \hfill
    \subfloat[{\scriptsize N32 {\color{BrickRed}$L_{\text{PSNR}, \text{$a$RGB}}$}}\\ \hspace*{1.3em}\resizebox{3.5em}{!}{\textbf{37.2545} $\mathrm{dB}$}]{\includegraphics[width=\figwidth]{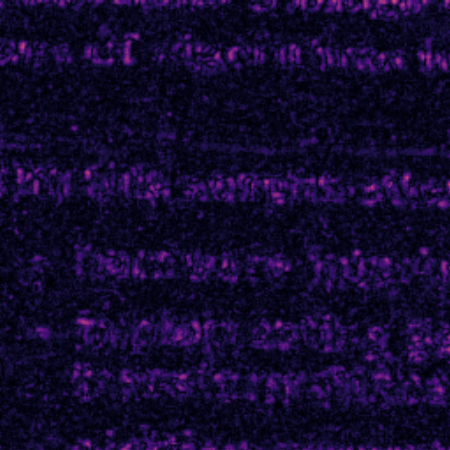}}
    \hfill
    \subfloat[\resizebox{4.9em}{!}{N32 {\color{BrickRed}$L_{1, \text{$a$RGB}}$}}\\ \hspace*{1.3em}\resizebox{3.5em}{!}{\textbf{37.0479} $\mathrm{dB}$}]{\includegraphics[width=\figwidth]{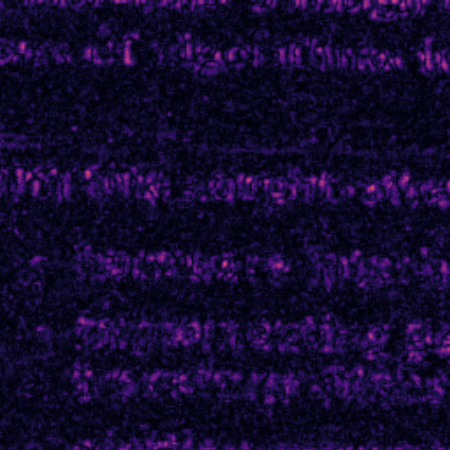}}
    \hfill
    \subfloat[N64 $L_{\text{PSNR}}$\\ \hspace*{1.3em}\resizebox{3.5em}{!}{37.5211 $\mathrm{dB}$}]{\includegraphics[width=\figwidth]{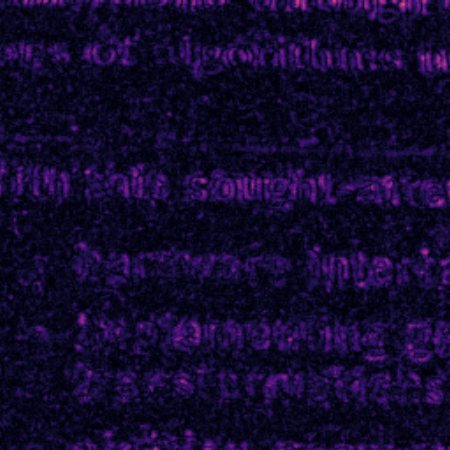}}
    \hfill
    \subfloat[\resizebox{4.9em}{!}{N64 {\color{BrickRed}$L_{1, \text{$a$RGB}}$}}\\ \hspace*{1.3em}\resizebox{3.5em}{!}{\textbf{38.2478} $\mathrm{dB}$}]{\includegraphics[width=\figwidth]{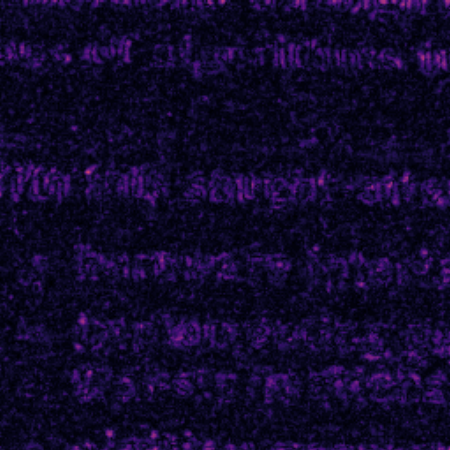}}
    \hfill
    \subfloat[Clean GT\\ \hspace*{1.4em}\resizebox{3.5em}{!}{PSNR ($\mathrm{dB}$)}]{\vspace*{-1.12mm}\includegraphics[width=\figwidth]{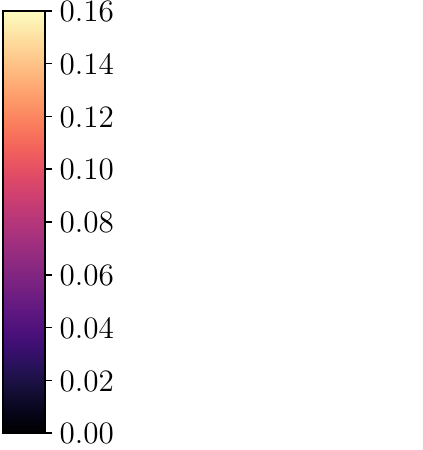}}
    \\[0.2em]
\renewcommand{\seed}{0515}%
    \subfloat{\includegraphics[width=\figwidth]{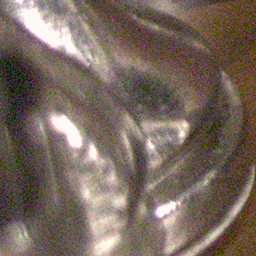}}
    \hfill
    \subfloat{\includegraphics[width=\figwidth]{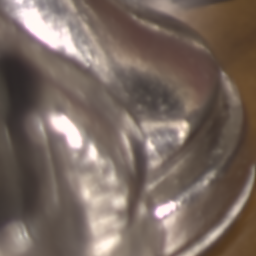}}
    \hfill
    \subfloat{\includegraphics[width=\figwidth]{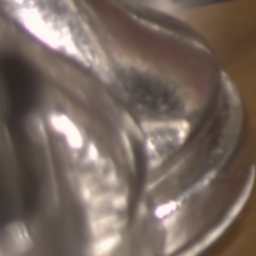}}
    \hfill
    \subfloat{\includegraphics[width=\figwidth]{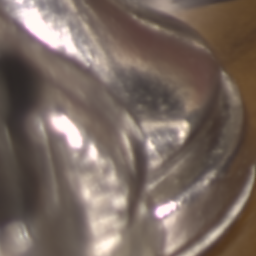}}
    \hfill
    \subfloat{\includegraphics[width=\figwidth]{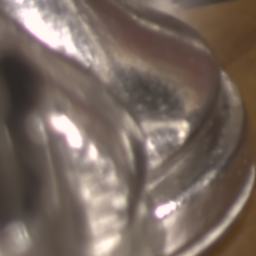}}
    \hfill
    \subfloat{\includegraphics[width=\figwidth]{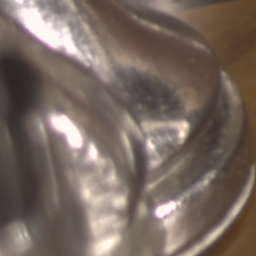}}
    \hfill
    \subfloat{\includegraphics[width=\figwidth]{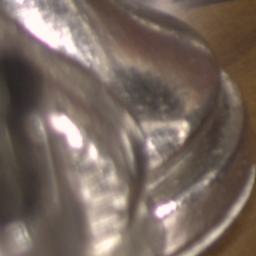}}
    \\[-0.05em]
    \addtocounter{subfigure}{-14}
    \subfloat[Noisy\\ \hspace*{1.3em}\resizebox{3.5em}{!}{25.9609 $\mathrm{dB}$}]{\includegraphics[width=\figwidth]{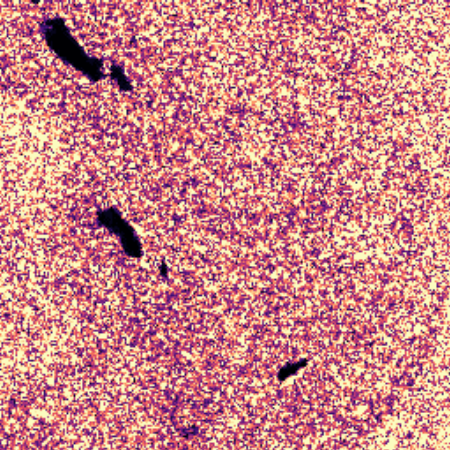}}
    \hfill
    \subfloat[N32 $L_{\text{PSNR}}$\\ \hspace*{1.3em}\resizebox{3.5em}{!}{38.7499 $\mathrm{dB}$}]{\includegraphics[width=\figwidth]{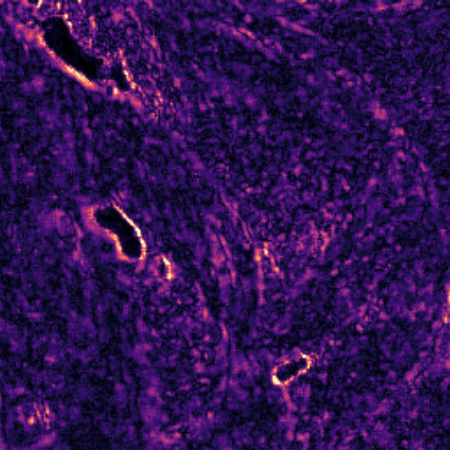}}
    \hfill
    \subfloat[{\scriptsize N32 {\color{BrickRed}$L_{\text{PSNR}, \text{$a$RGB}}$}}\\ \hspace*{1.3em}\resizebox{3.5em}{!}{\textbf{40.1597} $\mathrm{dB}$}]{\includegraphics[width=\figwidth]{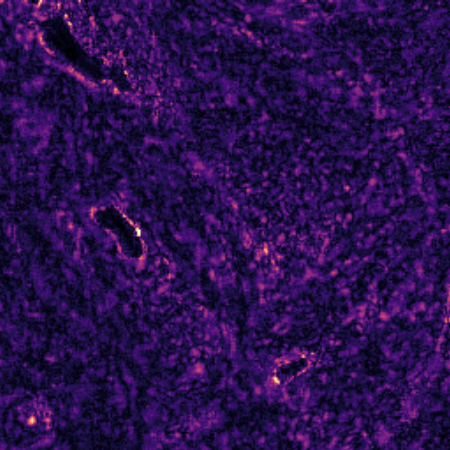}}
    \hfill
    \subfloat[\resizebox{4.9em}{!}{N32 {\color{BrickRed}$L_{1, \text{$a$RGB}}$}}\\ \hspace*{1.3em}\resizebox{3.5em}{!}{\textbf{40.2695} $\mathrm{dB}$}]{\includegraphics[width=\figwidth]{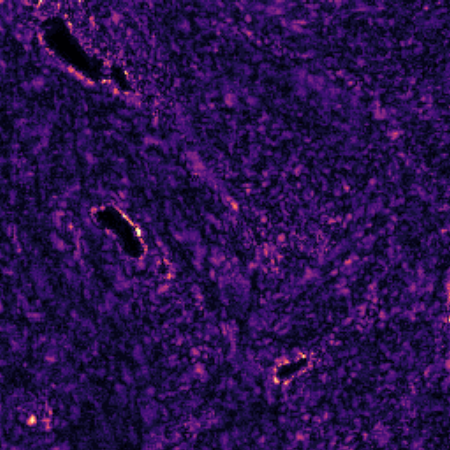}}
    \hfill
    \subfloat[N64 $L_{\text{PSNR}}$\\ \hspace*{1.3em}\resizebox{3.5em}{!}{39.6910 $\mathrm{dB}$}]{\includegraphics[width=\figwidth]{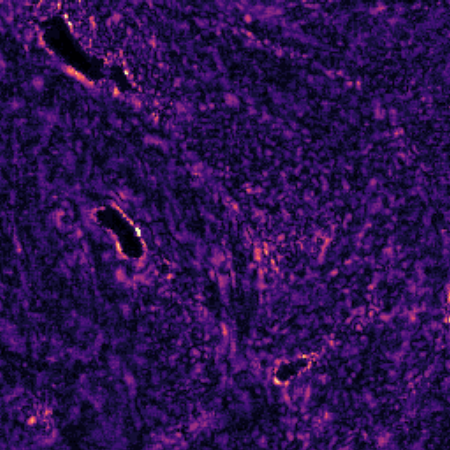}}
    \hfill
    \subfloat[\resizebox{4.9em}{!}{N64 {\color{BrickRed}$L_{1, \text{$a$RGB}}$}}\\ \hspace*{1.3em}\resizebox{3.5em}{!}{\textbf{40.3953} $\mathrm{dB}$}]{\includegraphics[width=\figwidth]{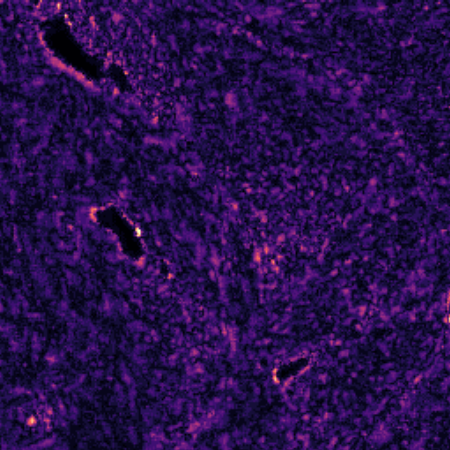}}
    \hfill
    \subfloat[Clean GT\\ \hspace*{1.4em}\resizebox{3.5em}{!}{PSNR ($\mathrm{dB}$)}]{\vspace*{-0.72mm}\includegraphics[width=\figwidth]{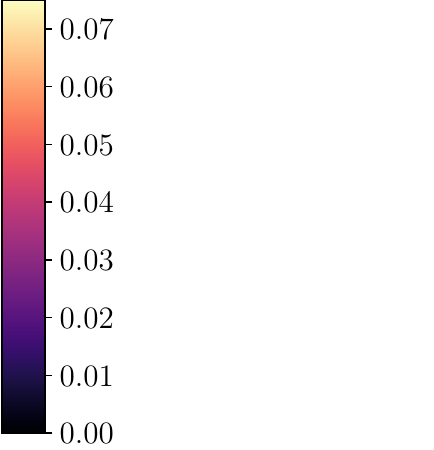}}
    \\[0.2em]
\renewcommand{\seed}{0379}%
    \subfloat{\includegraphics[width=\figwidth]{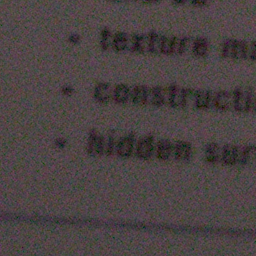}}
    \hfill
    \subfloat{\includegraphics[width=\figwidth]{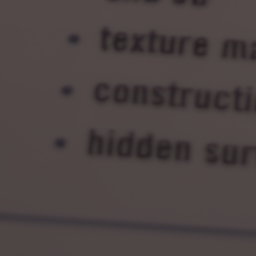}}
    \hfill
    \subfloat{\includegraphics[width=\figwidth]{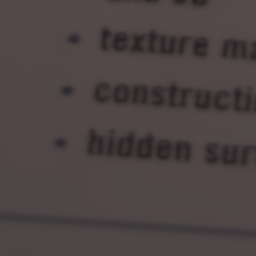}}
    \hfill
    \subfloat{\includegraphics[width=\figwidth]{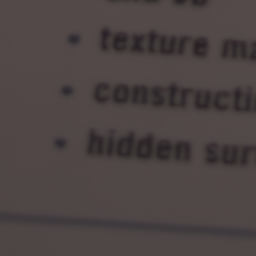}}
    \hfill
    \subfloat{\includegraphics[width=\figwidth]{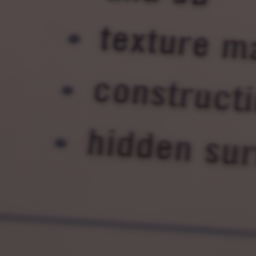}}
    \hfill
    \subfloat{\includegraphics[width=\figwidth]{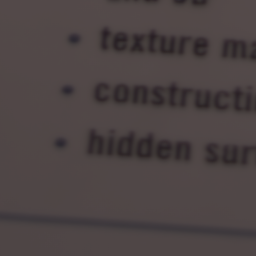}}
    \hfill
    \subfloat{\includegraphics[width=\figwidth]{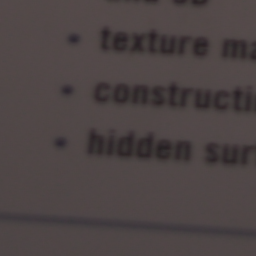}}
    \\[-0.05em]
    \addtocounter{subfigure}{-14}
    \subfloat[Noisy\\ \hspace*{1.3em}\resizebox{3.5em}{!}{29.4107 $\mathrm{dB}$}]{\includegraphics[width=\figwidth]{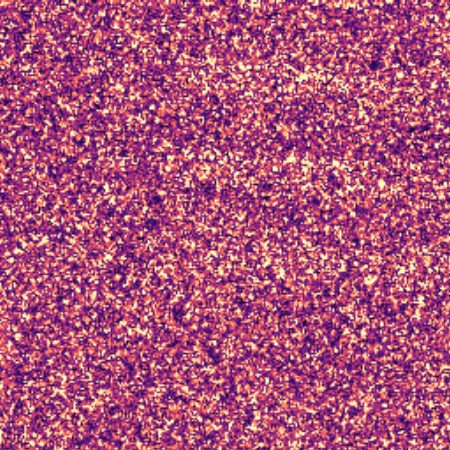}}
    \hfill
    \subfloat[N32 $L_{\text{PSNR}}$\\ \hspace*{1.3em}\resizebox{3.5em}{!}{47.1689 $\mathrm{dB}$}]{\includegraphics[width=\figwidth]{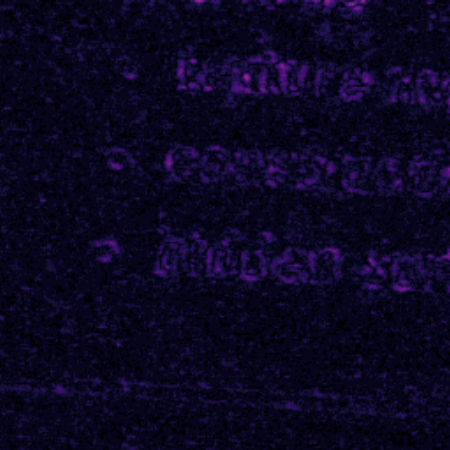}}
    \hfill
    \subfloat[{\scriptsize N32 {\color{BrickRed}$L_{\text{PSNR}, \text{$a$RGB}}$}}\\ \hspace*{1.3em}\resizebox{3.5em}{!}{\textbf{47.5380} $\mathrm{dB}$}]{\includegraphics[width=\figwidth]{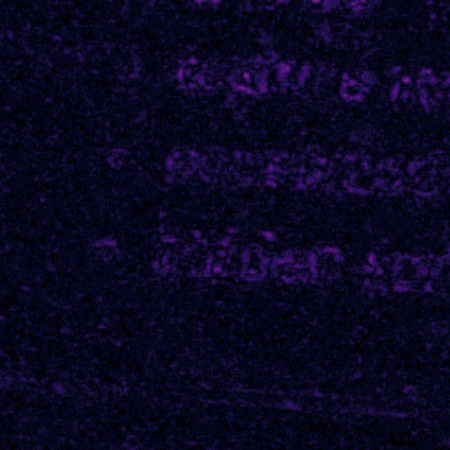}}
    \hfill
    \subfloat[\resizebox{4.9em}{!}{N32 {\color{BrickRed}$L_{1, \text{$a$RGB}}$}}\\ \hspace*{1.3em}\resizebox{3.5em}{!}{\textbf{47.4457} $\mathrm{dB}$}]{\includegraphics[width=\figwidth]{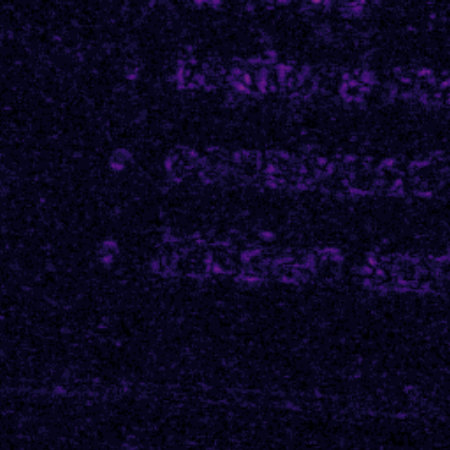}}
    \hfill
    \subfloat[N64 $L_{\text{PSNR}}$\\ \hspace*{1.3em}\resizebox{3.5em}{!}{46.9483 $\mathrm{dB}$}]{\includegraphics[width=\figwidth]{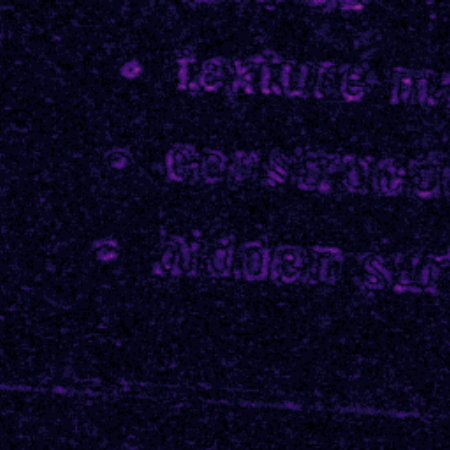}}
    \hfill
    \subfloat[\resizebox{4.9em}{!}{N64 {\color{BrickRed}$L_{1, \text{$a$RGB}}$}}\\ \hspace*{1.3em}\resizebox{3.5em}{!}{\textbf{47.8430} $\mathrm{dB}$}]{\includegraphics[width=\figwidth]{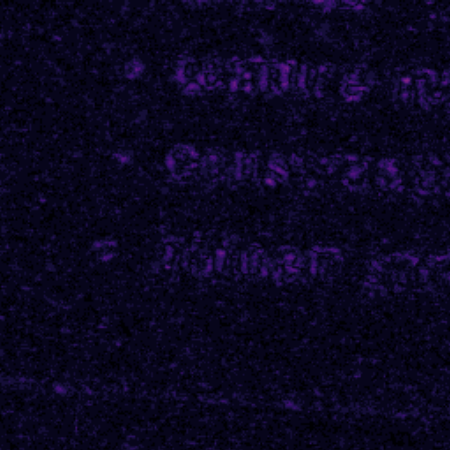}}
    \hfill
    \subfloat[Clean GT\\ \hspace*{1.4em}\resizebox{3.5em}{!}{PSNR ($\mathrm{dB}$)}]{\vspace*{-0.72mm}\includegraphics[width=\figwidth]{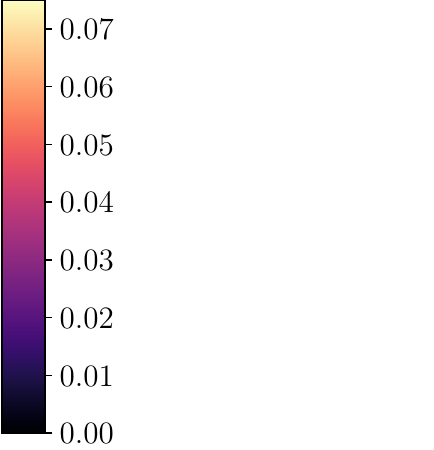}}
    \\[0.2em]
\renewcommand{\seed}{1156}%
    \subfloat{\includegraphics[width=\figwidth]{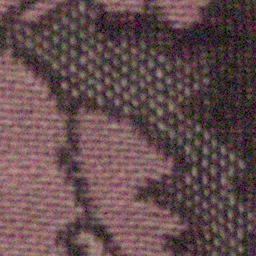}}
    \hfill
    \subfloat{\includegraphics[width=\figwidth]{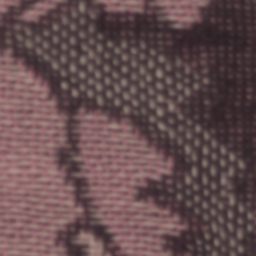}}
    \hfill
    \subfloat{\includegraphics[width=\figwidth]{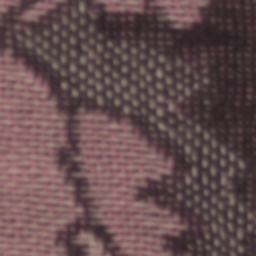}}
    \hfill
    \subfloat{\includegraphics[width=\figwidth]{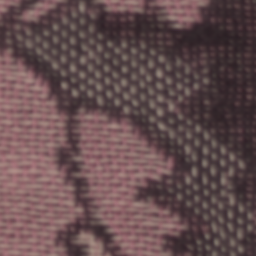}}
    \hfill
    \subfloat{\includegraphics[width=\figwidth]{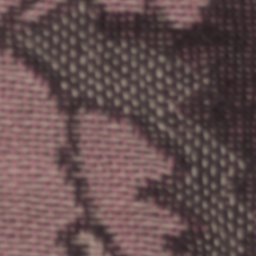}}
    \hfill
    \subfloat{\includegraphics[width=\figwidth]{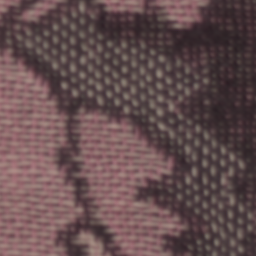}}
    \hfill
    \subfloat{\includegraphics[width=\figwidth]{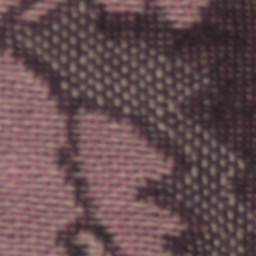}}
    \\[-0.05em]
    \addtocounter{subfigure}{-14}
    \subfloat[Noisy\\ \hspace*{1.3em}\resizebox{3.5em}{!}{25.5477 $\mathrm{dB}$}]{\includegraphics[width=\figwidth]{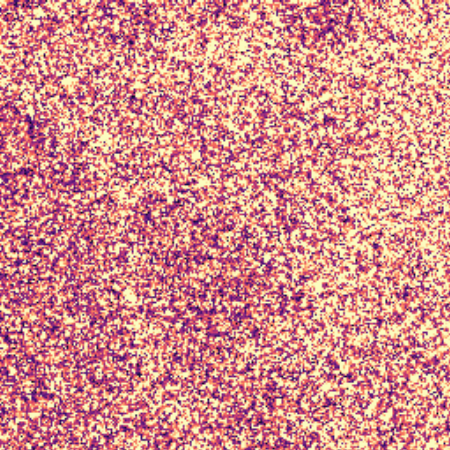}}
    \hfill
    \subfloat[N32 $L_{\text{PSNR}}$\\ \hspace*{1.3em}\resizebox{3.5em}{!}{39.4739 $\mathrm{dB}$}]{\includegraphics[width=\figwidth]{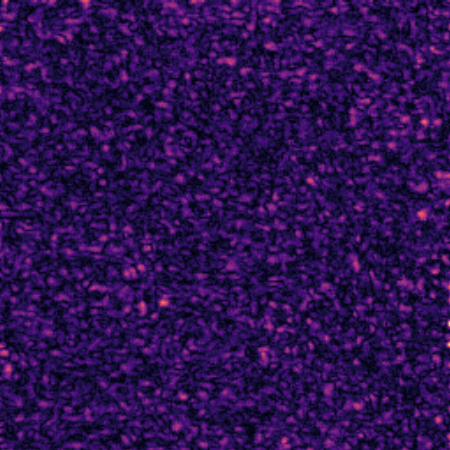}}
    \hfill
    \subfloat[{\scriptsize N32 {\color{BrickRed}$L_{\text{PSNR}, \text{$a$RGB}}$}}\\ \hspace*{1.3em}\resizebox{3.5em}{!}{\textbf{39.4866} $\mathrm{dB}$}]{\includegraphics[width=\figwidth]{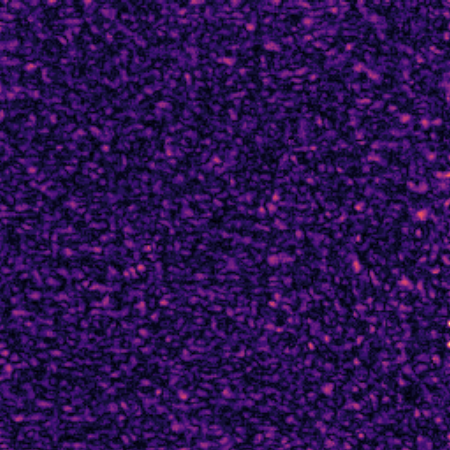}}
    \hfill
    \subfloat[\resizebox{4.9em}{!}{N32 {\color{BrickRed}$L_{1, \text{$a$RGB}}$}}\\ \hspace*{1.3em}\resizebox{3.5em}{!}{\textbf{39.5938} $\mathrm{dB}$}]{\includegraphics[width=\figwidth]{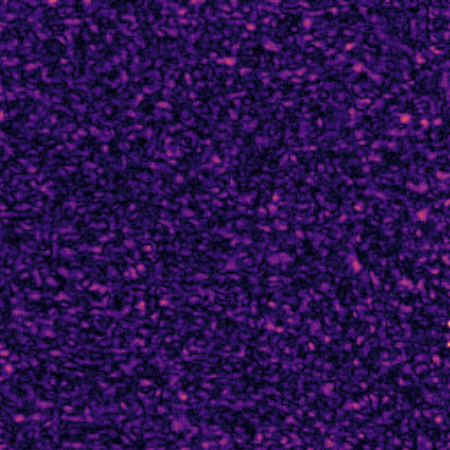}}
    \hfill
    \subfloat[N64 $L_{\text{PSNR}}$\\ \hspace*{1.3em}\resizebox{3.5em}{!}{39.8979 $\mathrm{dB}$}]{\includegraphics[width=\figwidth]{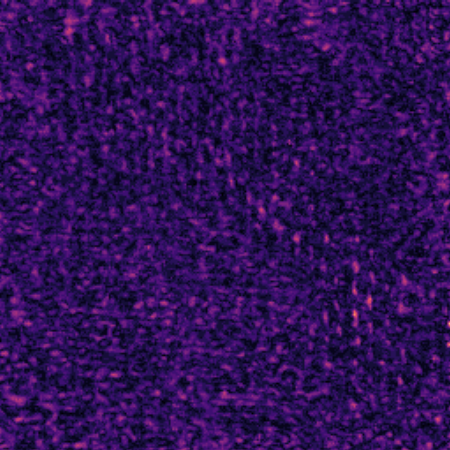}}
    \hfill
    \subfloat[\resizebox{4.9em}{!}{N64 {\color{BrickRed}$L_{1, \text{$a$RGB}}$}}\\ \hspace*{1.3em}\resizebox{3.5em}{!}{\textbf{40.6133} $\mathrm{dB}$}]{\includegraphics[width=\figwidth]{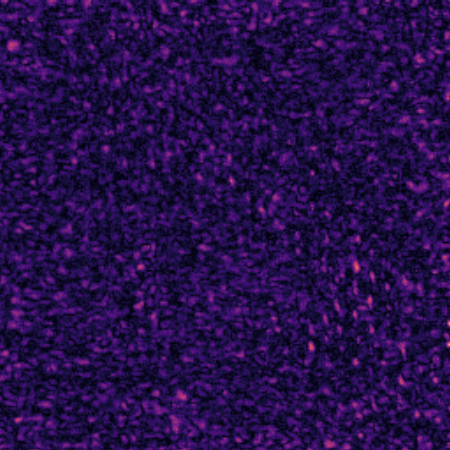}}
    \hfill
    \subfloat[Clean GT\\ \hspace*{1.4em}\resizebox{3.5em}{!}{PSNR ($\mathrm{dB}$)}]{\vspace*{-0.72mm}\includegraphics[width=\figwidth]{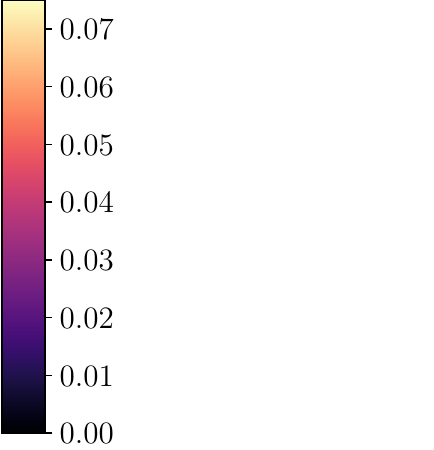}}
    \\[-.7em]
\end{subfigure}
\caption{\small
    \textbf{Qualitative comparison of real image denoising models on SIDD benchmark \citep{cv:data:abdelhamed18-sidd}.}
    Each column corresponds to each row in Table~\ref{tab:result_denoise}.
    N32 corresponds to NAFNet-width32 and N64 corresponds to NAFNet-width64.
    The bottom rows show the maximum absolute difference in color with a range of $[0, 1]\,$.
}
\label{fig:appx:result_denoise}
\end{figure*}

%% file: figures/appx_result_deblur/main1.tex
\begin{figure*}
\newcommand{\figwidth}{0.165\linewidth}
\newcommand{\patchsize}{192_256}
\newcommand{\varA}{LQ}%
\newcommand{\varB}{Lchar}%
\newcommand{\varC}{LcharaRGB}%
\newcommand{\varD}{Lchar+TLC}%
\newcommand{\varE}{LcharaRGB+TLC}%
\newcommand{\varF}{GT}%
\centering
\newcommand{\seed}{GoPro_GOPR0410_11_00-000159_86_940}%
    \subfloat{\includegraphics[width=\figwidth]{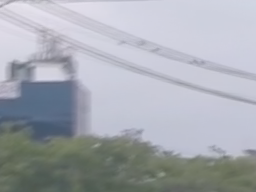}}
    \hfill
    \subfloat{\includegraphics[width=\figwidth]{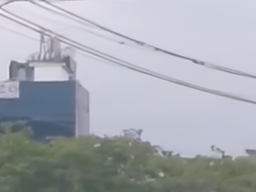}}
    \hfill
    \subfloat{\includegraphics[width=\figwidth]{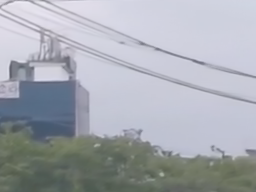}}
    \hfill
    \subfloat{\includegraphics[width=\figwidth]{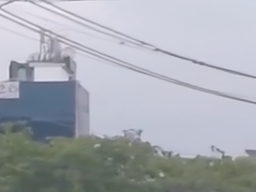}}
    \hfill
    \subfloat{\includegraphics[width=\figwidth]{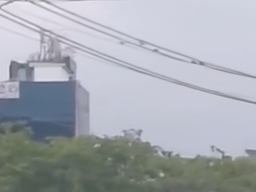}}
    \hfill
    \subfloat{\includegraphics[width=\figwidth]{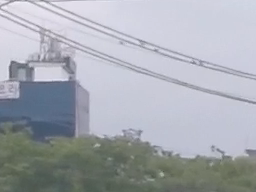}}
    \\[-0.14em]
    \addtocounter{subfigure}{-6}
    \subfloat[Blurry\\ \hspace*{1.3em}\resizebox{3.5em}{!}{30.3011 $\mathrm{dB}$}]{\includegraphics[width=\figwidth]{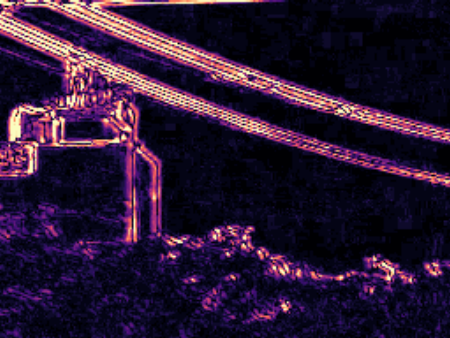}}
    \hfill
    \subfloat[$L_{\text{Char}}$\\ \hspace*{1.3em}\resizebox{3.5em}{!}{31.1395 $\mathrm{dB}$}]{\includegraphics[width=\figwidth]{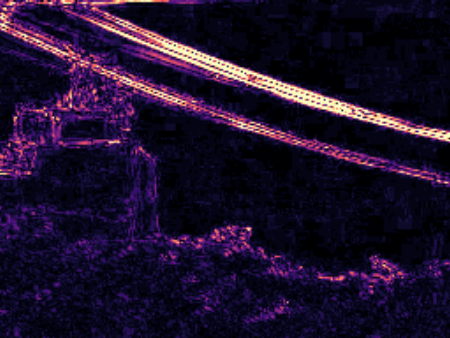}}
    \hfill
    \subfloat[{\color{BrickRed}$L_{\text{Char}, \text{$a$RGB}}$}\\ \hspace*{1.3em}\resizebox{3.5em}{!}{\textbf{32.0593} $\mathrm{dB}$}]{\includegraphics[width=\figwidth]{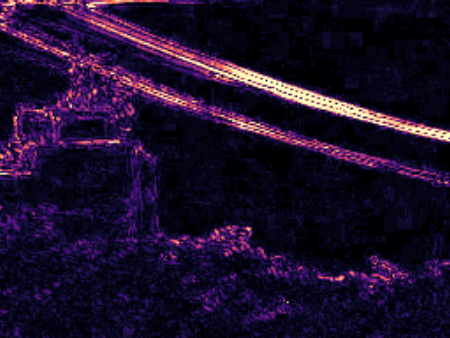}}
    \hfill
    \subfloat[$L_{\text{Char}}\,$,TLC\\ \hspace*{1.3em}\resizebox{3.5em}{!}{32.5316 $\mathrm{dB}$}]{\includegraphics[width=\figwidth]{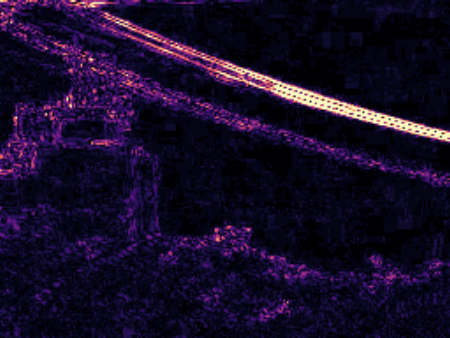}}
    \hfill
    \subfloat[\resizebox{6em}{!}{{\color{BrickRed}$L_{\text{Char}, \text{$a$RGB}}\,$},TLC}\\ \hspace*{1.3em}\resizebox{3.5em}{!}{\textbf{36.9283} $\mathrm{dB}$}]{\includegraphics[width=\figwidth]{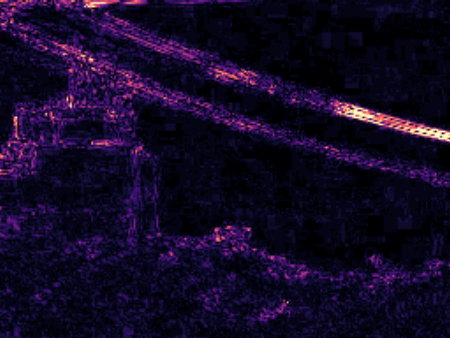}}
    \hfill
    \subfloat[Clean GT\\ \hspace*{1.4em}\resizebox{3.5em}{!}{PSNR ($\mathrm{dB}$)}]{\vspace*{-1.02mm}\includegraphics[width=\figwidth]{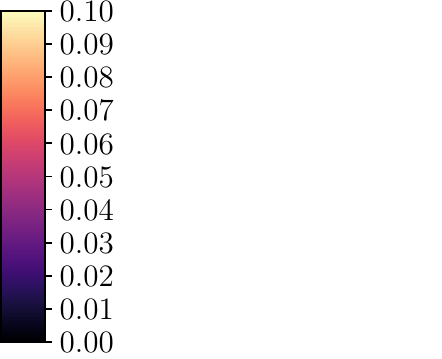}}
\\[0.2em]
\renewcommand{\seed}{GoPro_GOPR0410_11_00-000231_472_23}%
    \subfloat{\includegraphics[width=\figwidth]{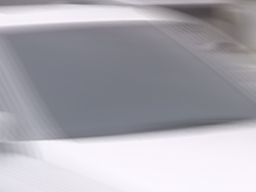}}
    \hfill
    \subfloat{\includegraphics[width=\figwidth]{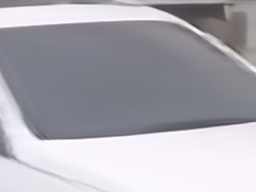}}
    \hfill
    \subfloat{\includegraphics[width=\figwidth]{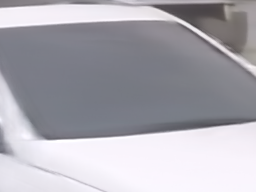}}
    \hfill
    \subfloat{\includegraphics[width=\figwidth]{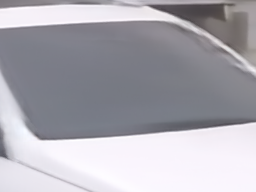}}
    \hfill
    \subfloat{\includegraphics[width=\figwidth]{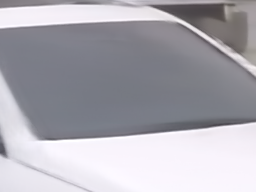}}
    \hfill
    \subfloat{\includegraphics[width=\figwidth]{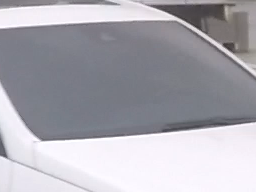}}
    \\[-0.14em]
    \addtocounter{subfigure}{-12}
    \subfloat[Blurry\\ \hspace*{1.3em}\resizebox{3.5em}{!}{22.6186 $\mathrm{dB}$}]{\includegraphics[width=\figwidth]{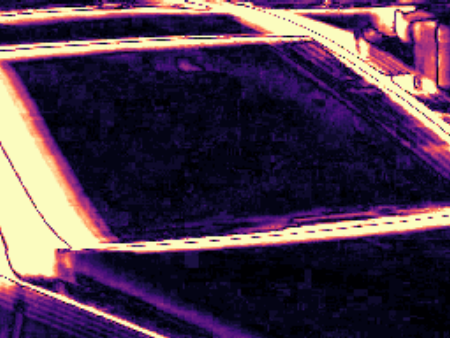}}
    \hfill
    \subfloat[$L_{\text{Char}}$\\ \hspace*{1.3em}\resizebox{3.5em}{!}{28.3096 $\mathrm{dB}$}]{\includegraphics[width=\figwidth]{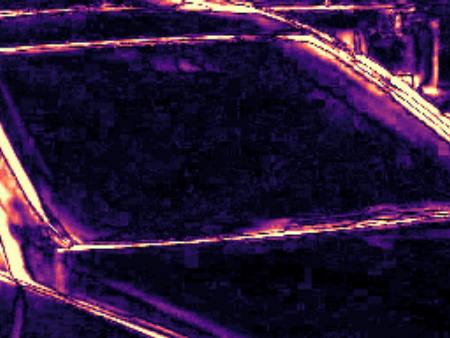}}
    \hfill
    \subfloat[{\color{BrickRed}$L_{\text{Char}, \text{$a$RGB}}$}\\ \hspace*{1.3em}\resizebox{3.5em}{!}{{28.2597} $\mathrm{dB}$}]{\includegraphics[width=\figwidth]{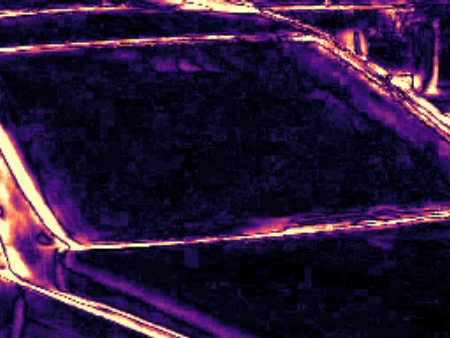}}
    \hfill
    \subfloat[$L_{\text{Char}}\,$,TLC\\ \hspace*{1.3em}\resizebox{3.5em}{!}{27.2944 $\mathrm{dB}$}]{\includegraphics[width=\figwidth]{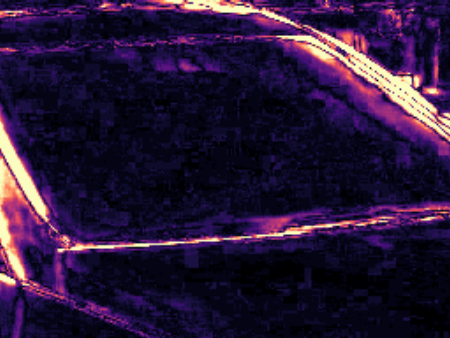}}
    \hfill
    \subfloat[\resizebox{6em}{!}{{\color{BrickRed}$L_{\text{Char}, \text{$a$RGB}}\,$},TLC}\\ \hspace*{1.3em}\resizebox{3.5em}{!}{\textbf{29.3727} $\mathrm{dB}$}]{\includegraphics[width=\figwidth]{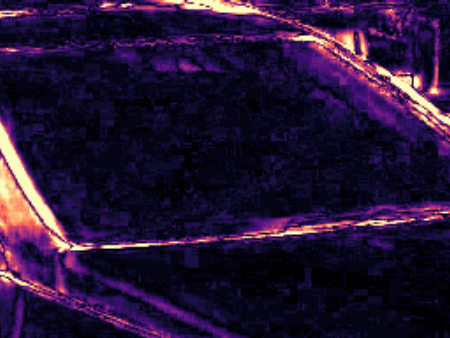}}
    \hfill
    \subfloat[Clean GT\\ \hspace*{1.4em}\resizebox{3.5em}{!}{PSNR ($\mathrm{dB}$)}]{\vspace*{-1.02mm}\includegraphics[width=\figwidth]{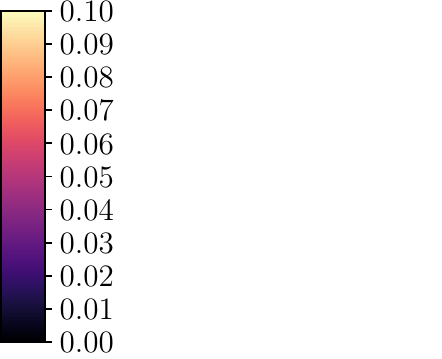}}
\\[0.2em]
\renewcommand{\seed}{GoPro_GOPR0384_11_00-000073_481_824}%
    \subfloat{\includegraphics[width=\figwidth]{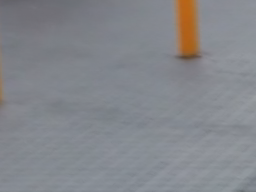}}
    \hfill
    \subfloat{\includegraphics[width=\figwidth]{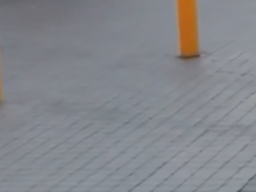}}
    \hfill
    \subfloat{\includegraphics[width=\figwidth]{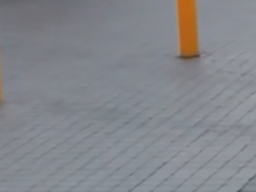}}
    \hfill
    \subfloat{\includegraphics[width=\figwidth]{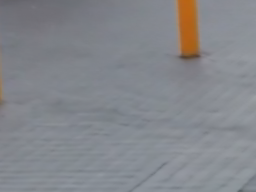}}
    \hfill
    \subfloat{\includegraphics[width=\figwidth]{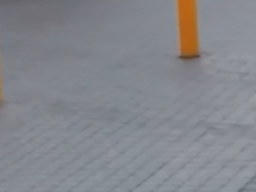}}
    \hfill
    \subfloat{\includegraphics[width=\figwidth]{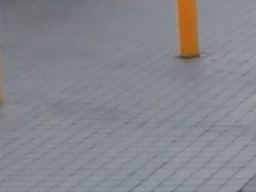}}
    \\[-0.14em]
    \addtocounter{subfigure}{-12}
    \subfloat[Blurry\\ \hspace*{1.3em}\resizebox{3.5em}{!}{38.1823 $\mathrm{dB}$}]{\includegraphics[width=\figwidth]{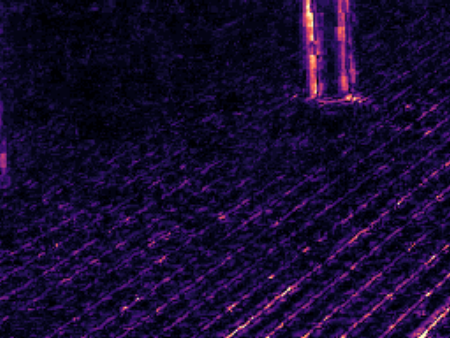}}
    \hfill
    \subfloat[$L_{\text{Char}}$\\ \hspace*{1.3em}\resizebox{3.5em}{!}{41.8038 $\mathrm{dB}$}]{\includegraphics[width=\figwidth]{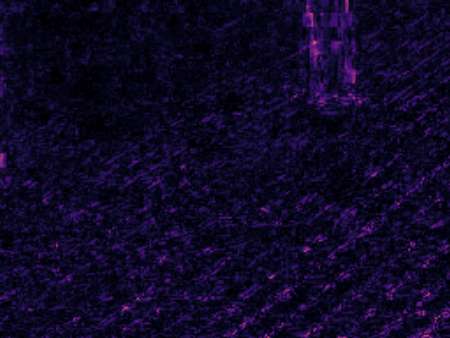}}
    \hfill
    \subfloat[{\color{BrickRed}$L_{\text{Char}, \text{$a$RGB}}$}\\ \hspace*{1.3em}\resizebox{3.5em}{!}{\textbf{42.0144} $\mathrm{dB}$}]{\includegraphics[width=\figwidth]{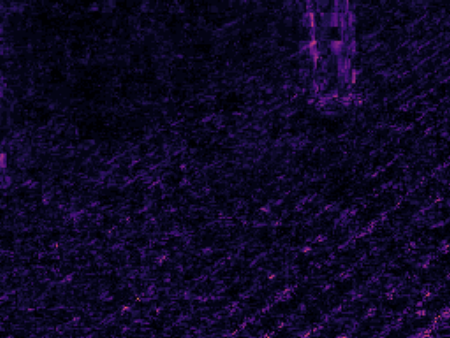}}
    \hfill
    \subfloat[$L_{\text{Char}}\,$,TLC\\ \hspace*{1.3em}\resizebox{3.5em}{!}{37.5406 $\mathrm{dB}$}]{\includegraphics[width=\figwidth]{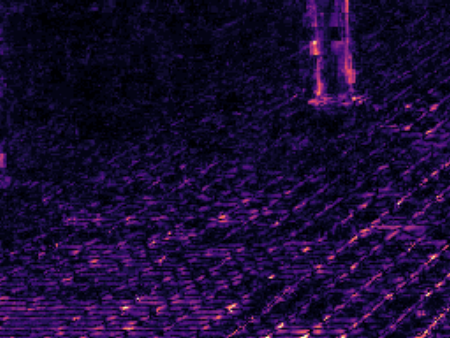}}
    \hfill
    \subfloat[\resizebox{6em}{!}{{\color{BrickRed}$L_{\text{Char}, \text{$a$RGB}}\,$},TLC}\\ \hspace*{1.3em}\resizebox{3.5em}{!}{\textbf{39.7551} $\mathrm{dB}$}]{\includegraphics[width=\figwidth]{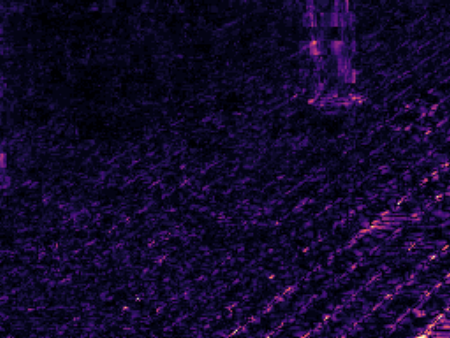}}
    \hfill
    \subfloat[Clean GT\\ \hspace*{1.4em}\resizebox{3.5em}{!}{PSNR ($\mathrm{dB}$)}]{\vspace*{-1.02mm}\includegraphics[width=\figwidth]{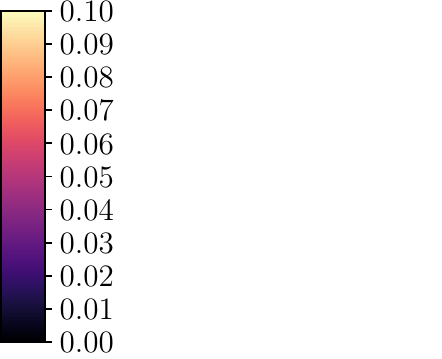}}
\\[0.2em]
\renewcommand{\seed}{GoPro_GOPR0384_11_00-000017_138_440}%
    \subfloat{\includegraphics[width=\figwidth]{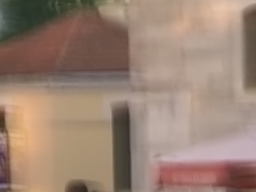}}
    \hfill
    \subfloat{\includegraphics[width=\figwidth]{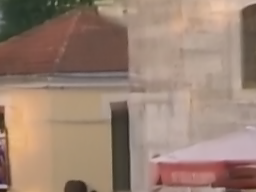}}
    \hfill
    \subfloat{\includegraphics[width=\figwidth]{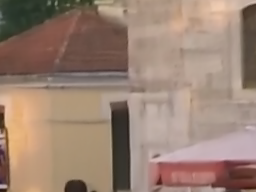}}
    \hfill
    \subfloat{\includegraphics[width=\figwidth]{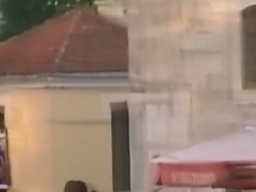}}
    \hfill
    \subfloat{\includegraphics[width=\figwidth]{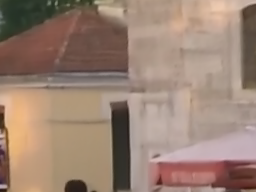}}
    \hfill
    \subfloat{\includegraphics[width=\figwidth]{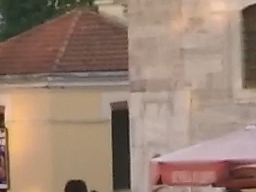}}
    \\[-0.14em]
    \addtocounter{subfigure}{-12}
    \subfloat[Blurry\\ \hspace*{1.3em}\resizebox{3.5em}{!}{24.9808 $\mathrm{dB}$}]{\includegraphics[width=\figwidth]{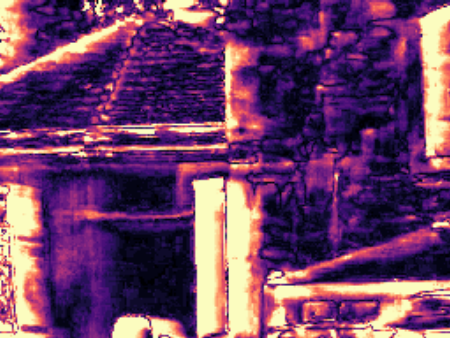}}
    \hfill
    \subfloat[$L_{\text{Char}}$\\ \hspace*{1.3em}\resizebox{3.5em}{!}{30.5654 $\mathrm{dB}$}]{\includegraphics[width=\figwidth]{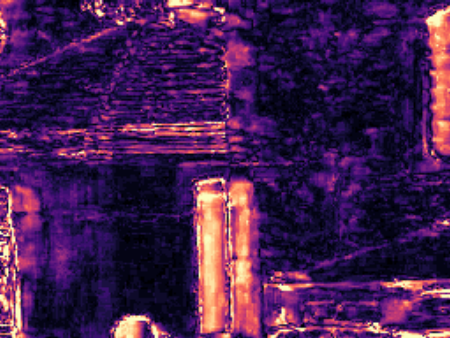}}
    \hfill
    \subfloat[{\color{BrickRed}$L_{\text{Char}, \text{$a$RGB}}$}\\ \hspace*{1.3em}\resizebox{3.5em}{!}{\textbf{33.5537} $\mathrm{dB}$}]{\includegraphics[width=\figwidth]{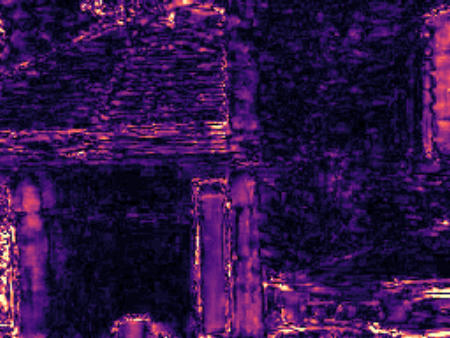}}
    \hfill
    \subfloat[$L_{\text{Char}}\,$,TLC\\ \hspace*{1.3em}\resizebox{3.5em}{!}{28.8951 $\mathrm{dB}$}]{\includegraphics[width=\figwidth]{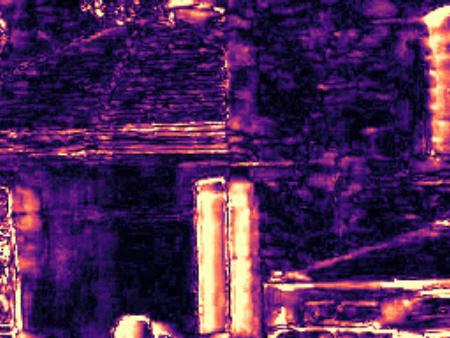}}
    \hfill
    \subfloat[\resizebox{6em}{!}{{\color{BrickRed}$L_{\text{Char}, \text{$a$RGB}}\,$},TLC}\\ \hspace*{1.3em}\resizebox{3.5em}{!}{\textbf{33.2253} $\mathrm{dB}$}]{\includegraphics[width=\figwidth]{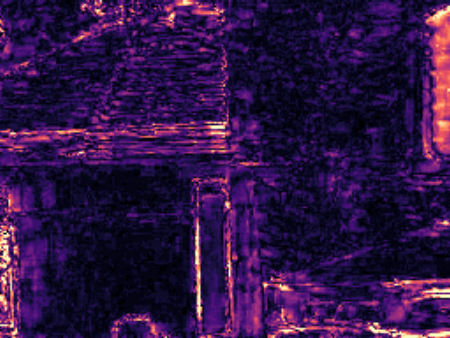}}
    \hfill
    \subfloat[Clean GT\\ \hspace*{1.4em}\resizebox{3.5em}{!}{PSNR ($\mathrm{dB}$)}]{\vspace*{-1.02mm}\includegraphics[width=\figwidth]{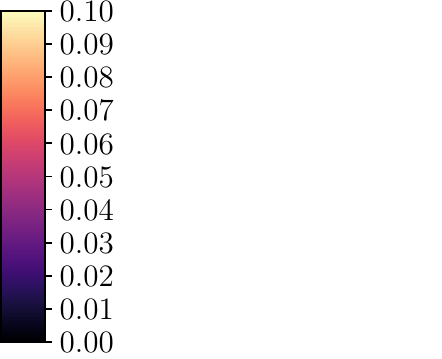}}
\\[-0.7em]
\caption{\small
    \textbf{Qualitative comparison of motion blur deblurring models in GoPro benchmark \citep{cv:deblur:nah17-deepdeblur}.}
    Each column corresponds to each row in Table~\ref{tab:result_deblur}.
    The bottom rows show the maximum absolute difference in color with a range of $[0, 1]\,$.
}
\label{fig:appx:result_deblur:p1}
\end{figure*}

%% file: figures/appx_result_deblur/main2.tex
\begin{figure*}
\newcommand{\figwidth}{0.165\linewidth}
\newcommand{\patchsize}{192_256}
\newcommand{\varA}{LQ}%
\newcommand{\varB}{Lchar}%
\newcommand{\varC}{LcharaRGB}%
\newcommand{\varD}{Lchar+TLC}%
\newcommand{\varE}{LcharaRGB+TLC}%
\newcommand{\varF}{GT}%
\centering
\newcommand{\seed}{HIDE_8fromGOPR1040.MP4_240_184}%
    \subfloat{\includegraphics[width=\figwidth]{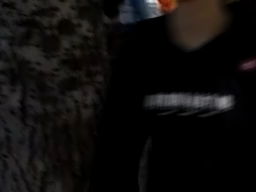}}
    \hfill
    \subfloat{\includegraphics[width=\figwidth]{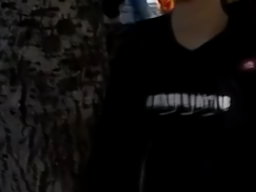}}
    \hfill
    \subfloat{\includegraphics[width=\figwidth]{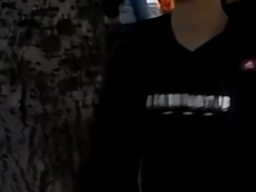}}
    \hfill
    \subfloat{\includegraphics[width=\figwidth]{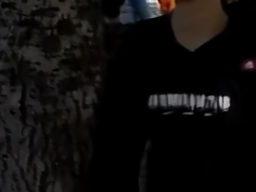}}
    \hfill
    \subfloat{\includegraphics[width=\figwidth]{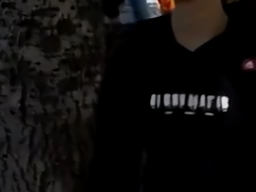}}
    \hfill
    \subfloat{\includegraphics[width=\figwidth]{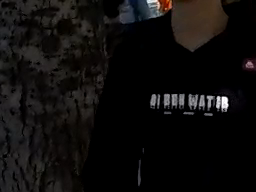}}
    \\[-0.14em]
    \addtocounter{subfigure}{-6}
    \subfloat[Blurry\\ \hspace*{1.3em}\resizebox{3.5em}{!}{27.2154 $\mathrm{dB}$}]{\includegraphics[width=\figwidth]{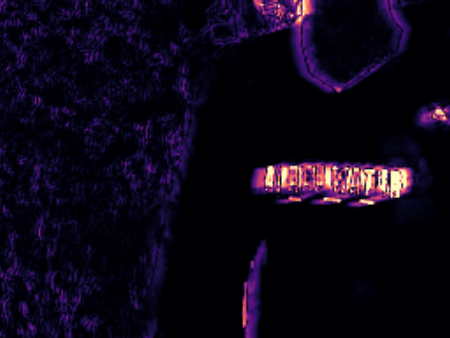}}
    \hfill
    \subfloat[$L_{\text{Char}}$\\ \hspace*{1.3em}\resizebox{3.5em}{!}{27.0476 $\mathrm{dB}$}]{\includegraphics[width=\figwidth]{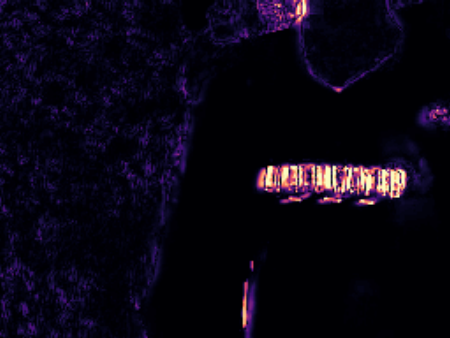}}
    \hfill
    \subfloat[{\color{BrickRed}$L_{\text{Char}, \text{$a$RGB}}$}\\ \hspace*{1.3em}\resizebox{3.5em}{!}{\textbf{27.5802} $\mathrm{dB}$}]{\includegraphics[width=\figwidth]{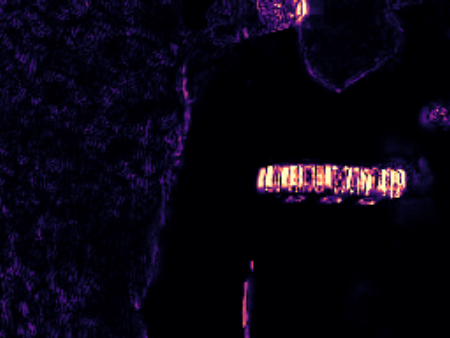}}
    \hfill
    \subfloat[$L_{\text{Char}}\,$,TLC\\ \hspace*{1.3em}\resizebox{3.5em}{!}{27.2900 $\mathrm{dB}$}]{\includegraphics[width=\figwidth]{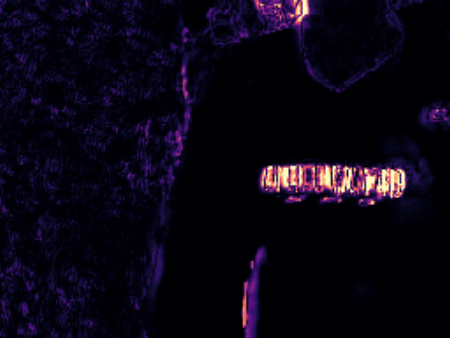}}
    \hfill
    \subfloat[\resizebox{6em}{!}{{\color{BrickRed}$L_{\text{Char}, \text{$a$RGB}}\,$},TLC}\\ \hspace*{1.3em}\resizebox{3.5em}{!}{\textbf{31.4381} $\mathrm{dB}$}]{\includegraphics[width=\figwidth]{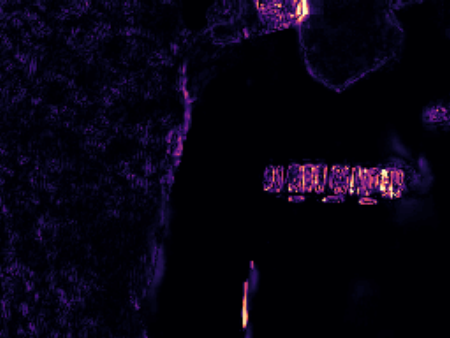}}
    \hfill
    \subfloat[Clean GT\\ \hspace*{1.4em}\resizebox{3.5em}{!}{PSNR ($\mathrm{dB}$)}]{\vspace*{-1.02mm}\includegraphics[width=\figwidth]{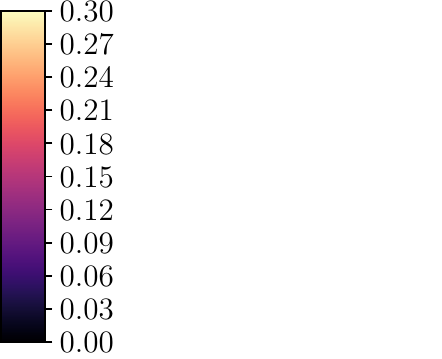}}
\\[0.2em]
\renewcommand{\seed}{HIDE_6fromGOPR0950_433_567}%
    \subfloat{\includegraphics[width=\figwidth]{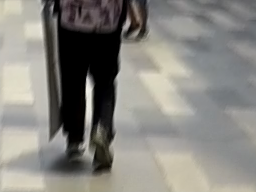}}
    \hfill
    \subfloat{\includegraphics[width=\figwidth]{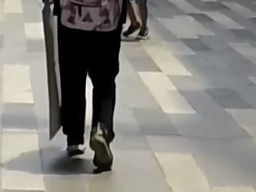}}
    \hfill
    \subfloat{\includegraphics[width=\figwidth]{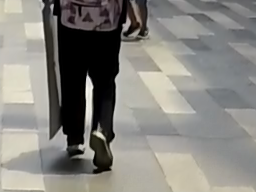}}
    \hfill
    \subfloat{\includegraphics[width=\figwidth]{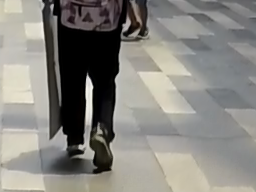}}
    \hfill
    \subfloat{\includegraphics[width=\figwidth]{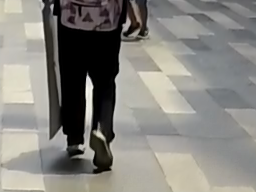}}
    \hfill
    \subfloat{\includegraphics[width=\figwidth]{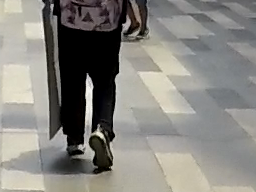}}
    \\[-0.14em]
    \addtocounter{subfigure}{-12}
    \subfloat[Blurry\\ \hspace*{1.3em}\resizebox{3.5em}{!}{26.3822 $\mathrm{dB}$}]{\includegraphics[width=\figwidth]{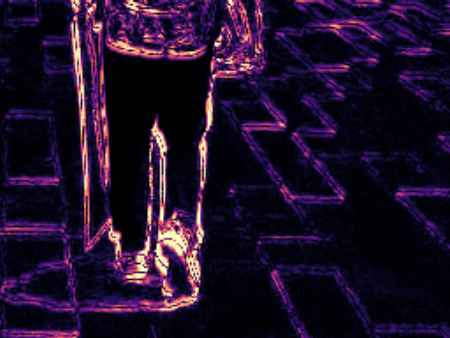}}
    \hfill
    \subfloat[$L_{\text{Char}}$\\ \hspace*{1.3em}\resizebox{3.5em}{!}{30.8486 $\mathrm{dB}$}]{\includegraphics[width=\figwidth]{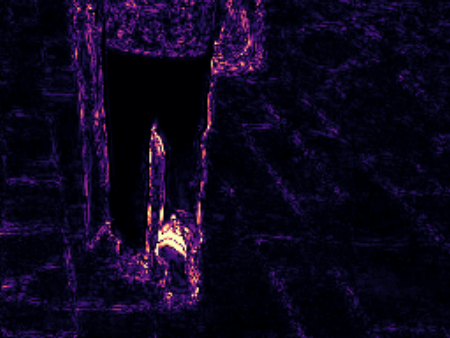}}
    \hfill
    \subfloat[{\color{BrickRed}$L_{\text{Char}, \text{$a$RGB}}$}\\ \hspace*{1.3em}\resizebox{3.5em}{!}{\textbf{32.9989} $\mathrm{dB}$}]{\includegraphics[width=\figwidth]{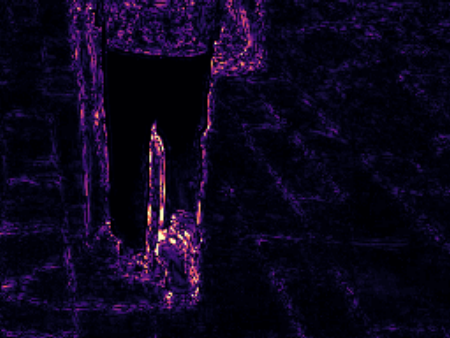}}
    \hfill
    \subfloat[$L_{\text{Char}}\,$,TLC\\ \hspace*{1.3em}\resizebox{3.5em}{!}{30.6344 $\mathrm{dB}$}]{\includegraphics[width=\figwidth]{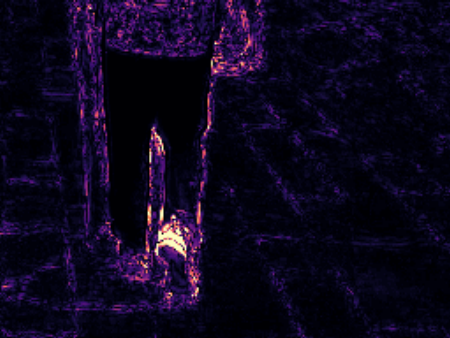}}
    \hfill
    \subfloat[\resizebox{6em}{!}{{\color{BrickRed}$L_{\text{Char}, \text{$a$RGB}}\,$},TLC}\\ \hspace*{1.3em}\resizebox{3.5em}{!}{\textbf{32.6934} $\mathrm{dB}$}]{\includegraphics[width=\figwidth]{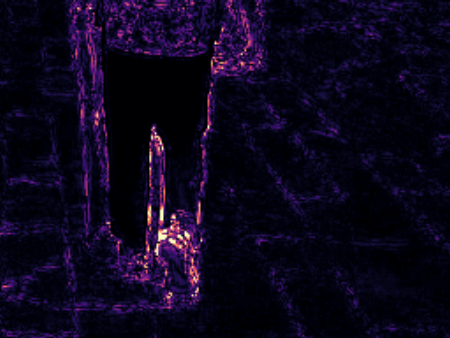}}
    \hfill
    \subfloat[Clean GT\\ \hspace*{1.4em}\resizebox{3.5em}{!}{PSNR ($\mathrm{dB}$)}]{\vspace*{-1.02mm}\includegraphics[width=\figwidth]{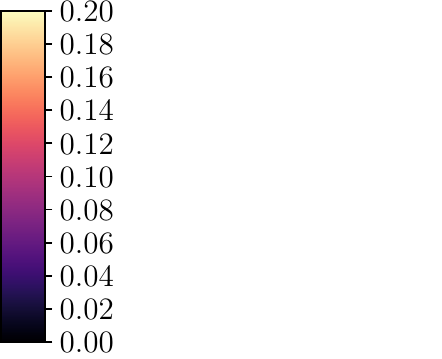}}
\\[0.2em]
\renewcommand{\seed}{HIDE_269fromGOPR1089.MP4_427_801}%
    \subfloat{\includegraphics[width=\figwidth]{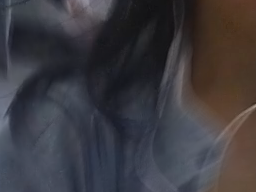}}
    \hfill
    \subfloat{\includegraphics[width=\figwidth]{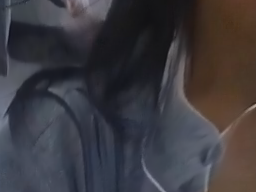}}
    \hfill
    \subfloat{\includegraphics[width=\figwidth]{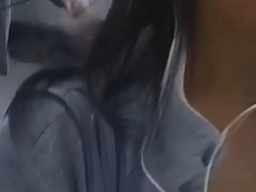}}
    \hfill
    \subfloat{\includegraphics[width=\figwidth]{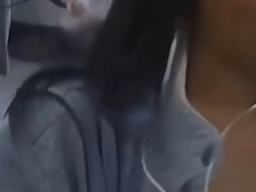}}
    \hfill
    \subfloat{\includegraphics[width=\figwidth]{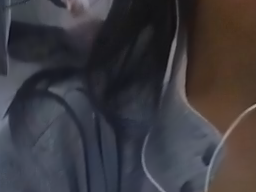}}
    \hfill
    \subfloat{\includegraphics[width=\figwidth]{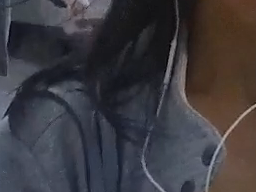}}
    \\[-0.14em]
    \addtocounter{subfigure}{-12}
    \subfloat[Blurry\\ \hspace*{1.3em}\resizebox{3.5em}{!}{28.0323 $\mathrm{dB}$}]{\includegraphics[width=\figwidth]{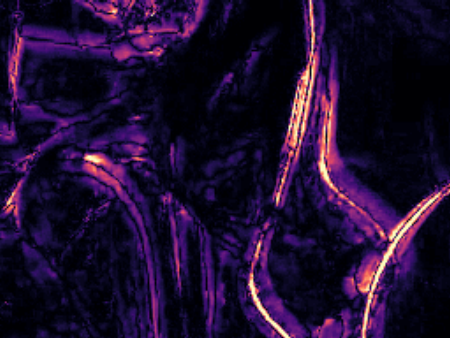}}
    \hfill
    \subfloat[$L_{\text{Char}}$\\ \hspace*{1.3em}\resizebox{3.5em}{!}{29.3776 $\mathrm{dB}$}]{\includegraphics[width=\figwidth]{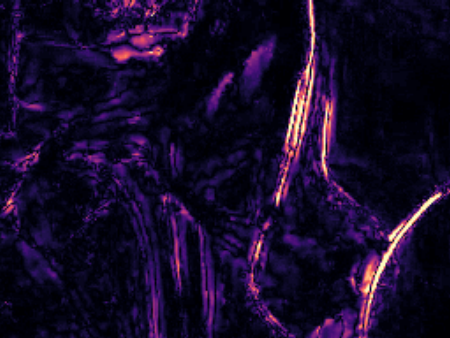}}
    \hfill
    \subfloat[{\color{BrickRed}$L_{\text{Char}, \text{$a$RGB}}$}\\ \hspace*{1.3em}\resizebox{3.5em}{!}{\textbf{30.0996} $\mathrm{dB}$}]{\includegraphics[width=\figwidth]{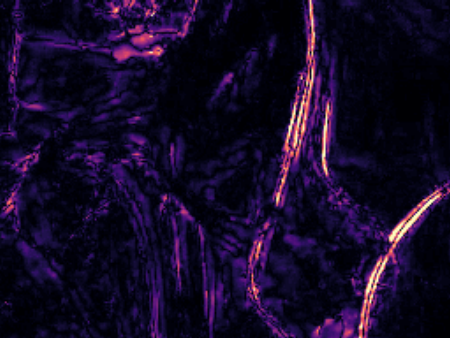}}
    \hfill
    \subfloat[$L_{\text{Char}}\,$,TLC\\ \hspace*{1.3em}\resizebox{3.5em}{!}{30.0415 $\mathrm{dB}$}]{\includegraphics[width=\figwidth]{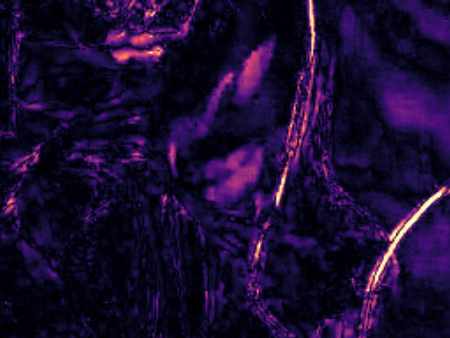}}
    \hfill
    \subfloat[\resizebox{6em}{!}{{\color{BrickRed}$L_{\text{Char}, \text{$a$RGB}}\,$},TLC}\\ \hspace*{1.3em}\resizebox{3.5em}{!}{\textbf{32.9874} $\mathrm{dB}$}]{\includegraphics[width=\figwidth]{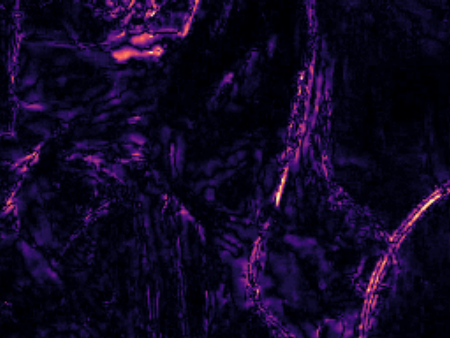}}
    \hfill
    \subfloat[Clean GT\\ \hspace*{1.4em}\resizebox{3.5em}{!}{PSNR ($\mathrm{dB}$)}]{\vspace*{-1.02mm}\includegraphics[width=\figwidth]{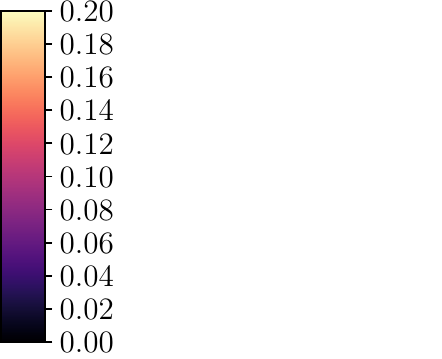}}
\\[0.2em]
\renewcommand{\seed}{HIDE_77fromGOPR1087.MP4_209_635}%
    \subfloat{\includegraphics[width=\figwidth]{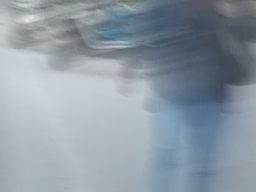}}
    \hfill
    \subfloat{\includegraphics[width=\figwidth]{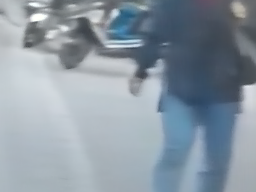}}
    \hfill
    \subfloat{\includegraphics[width=\figwidth]{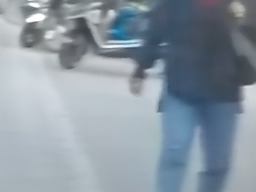}}
    \hfill
    \subfloat{\includegraphics[width=\figwidth]{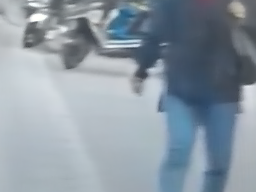}}
    \hfill
    \subfloat{\includegraphics[width=\figwidth]{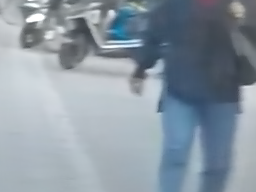}}
    \hfill
    \subfloat{\includegraphics[width=\figwidth]{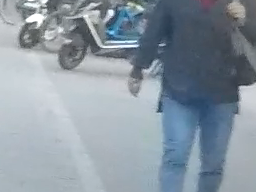}}
    \\[-0.14em]
    \addtocounter{subfigure}{-12}
    \subfloat[Blurry\\ \hspace*{1.3em}\resizebox{3.5em}{!}{21.5565 $\mathrm{dB}$}]{\includegraphics[width=\figwidth]{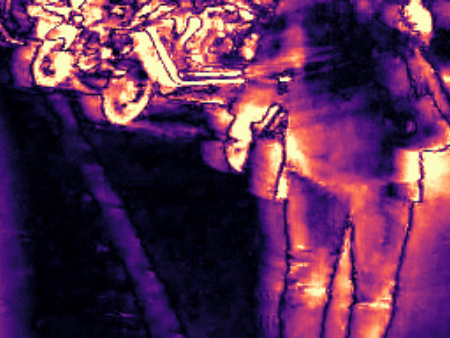}}
    \hfill
    \subfloat[$L_{\text{Char}}$\\ \hspace*{1.3em}\resizebox{3.5em}{!}{23.6871 $\mathrm{dB}$}]{\includegraphics[width=\figwidth]{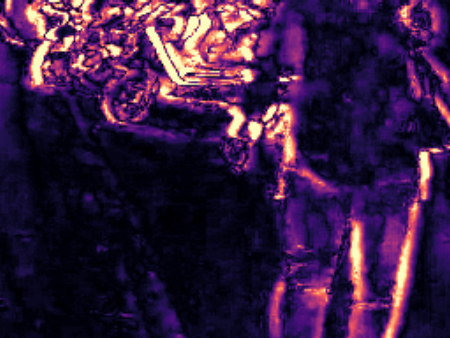}}
    \hfill
    \subfloat[{\color{BrickRed}$L_{\text{Char}, \text{$a$RGB}}$}\\ \hspace*{1.3em}\resizebox{3.5em}{!}{\textbf{25.9344} $\mathrm{dB}$}]{\includegraphics[width=\figwidth]{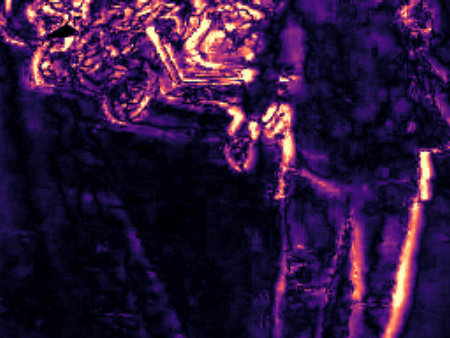}}
    \hfill
    \subfloat[$L_{\text{Char}}\,$,TLC\\ \hspace*{1.3em}\resizebox{3.5em}{!}{23.7183 $\mathrm{dB}$}]{\includegraphics[width=\figwidth]{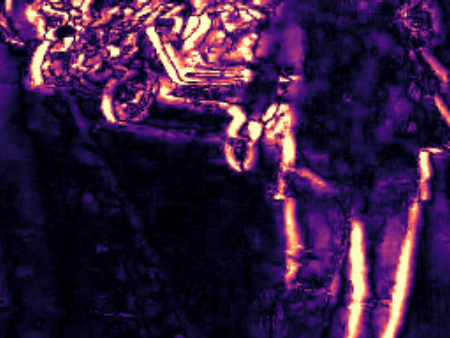}}
    \hfill
    \subfloat[\resizebox{6em}{!}{{\color{BrickRed}$L_{\text{Char}, \text{$a$RGB}}\,$},TLC}\\ \hspace*{1.3em}\resizebox{3.5em}{!}{\textbf{26.4518} $\mathrm{dB}$}]{\includegraphics[width=\figwidth]{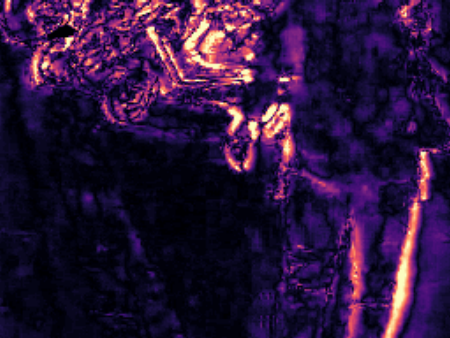}}
    \hfill
    \subfloat[Clean GT\\ \hspace*{1.4em}\resizebox{3.5em}{!}{PSNR ($\mathrm{dB}$)}]{\vspace*{-1.02mm}\includegraphics[width=\figwidth]{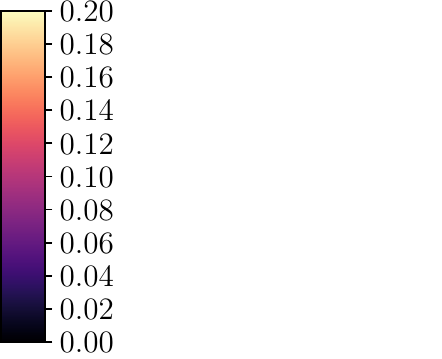}}
\\[-0.7em]
\caption{\small
    \textbf{Qualitative comparison of motion blur deblurring models in HIDE benchmark \citep{cv:deblur:shen19-hide-dataset}.}
    Each column corresponds to each row in Table~\ref{tab:result_deblur}.
    The bottom rows show the maximum absolute difference in color with a range of $[0, 1]\,$.
}
\label{fig:appx:result_deblur:p2}
\end{figure*}

%% file: sections/E_decomposition.tex
This section is divided into three parts:
In the first part of this section, we discuss further on the embedding decomposition test conducted in Section~\ref{sec:5_discussion:nullspace}.
Mixing different types of images in the same way as we did in Section~\ref{sec:5_discussion:nullspace} reveals how our $a$RGB representation encodes additional information in the extra dimensions.
Next, we show more examples of the t-SNE \citep{ml:vis:vdmaaten08-tsne} visualization of our learned embeddings in addition to Figure~\ref{fig:discussion:tsne}.
Unlike segmentation maps typically generated and assessed for semantic tasks, the ones produced by our $a$RGB encoder are extremely fine-grained and complicated, yet we find some common structural features in the results.
In the last part, we conduct another experiment to visualize how each expert in the $a$RGB encoder $f$ learns to specialize.

\subsection{Decomposition of $a$RGB representation}
\label{sec:f_decomposition:decomposition}
In Section~\ref{sec:5_discussion:nullspace}, we have discussed how the linearity of our decoder $g$ helps understanding the learned representation space.
In particular, given its linear weight $\mA$ and bias $\vb\,$, we can decompose any given embedding in the $a$RGB representation $\xxi \in \R^{C}$ into a sum of two orthogonal vectors:
\begin{equation}
\label{eq:ortho_decomposition_restate}
\xxi = \mA^{\dagger} \mA \xxi + (\mI - \mA^{\dagger} \mA) \xxi\,,
\end{equation}
where we denote the matrix $\mA^{\dagger}$ as the Moore-Penrose pseudoinverse of $\mA\,$.
We can regard the multiplicand of the second term $\mI - \mA^{\dagger} \mA$ as a linear projection operator onto the nullspace of $\mA\,$, and the multiplicand of the first term $\mA^{\dagger} \mA$ as a linear projection onto the orthogonal complement of the nullspace of $\mA\,$.
Mathematically, and also empirically, it is easy to see that summation of any vector projected onto the nullspace of $\mA$ leads to no change in the decoded image $\xx' = g(\xxi)\,$.
That is,
\begin{align}
\label{eq:nullspace_summation_nullified}
g(\xxi + (\mI - \mA^{\dagger} \mA) \bm{\zeta}) &= \mA (\xxi + (\mI - \mA^{\dagger} \mA) \bm{\zeta}) + \vb \\
&= \mA \xxi + \vb + \mA (\mI - \mA^{\dagger} \mA) \bm{\zeta} \\
&= \mA \xxi + \vb + (\mA - \mA \mA^{\dagger} \mA) \bm{\zeta} \\
&= \mA \xxi + \vb + (\mA - \mA) \bm{\zeta} \\
&= \mA \xxi + \vb \\
&= g(\xxi)\,.
\end{align}
The identity in the fourth row is from the equality $\mA = \mA \mA^{\dagger} \mA$ of Moore-Penrose pseudoinverse.
Therefore, the subspace to which the projection operator $\mI - \mA^{\dagger} \mA$ is mapped is the allowed degree of freedom that additional information can be embedded.

Inversion of mixed embedding uncovers what information lies in this particular subspace.
In addition to the results in Figure~\ref{fig:discussion:inversion}, Figure~\ref{fig:appx:decomposition} and \ref{fig:appx:decomposition:laplace} show visual results of $a$RGB embedding inversion.
The flat image source is manually synthesized by a single color gradient.
The real image source is a patch brought from DIV2K validation dataset \citep{cv:data:agustsson17-div2k}, and the Gaussian noise source is sampled from $\mathcal{N}(0.5, 0.5)\,$, where the RGB range is $[0, 1]\,$.
Each patch has an equal size of $256 \times 256\,$.
First, we obtain the two $a$RGB embeddings from both sources.
Then, we take the parallel part $f_{\parallel}$ from the first source and the perpendicular part $f_{\perp}$ from the second source.
An image is optimized to produce the synthesized embedding as its $a$RGB representation.
Due to the simple training setup, all the optimization quickly converges after 50 iterations of SGD with learning rate 0.1\,, as shown in Figure~\ref{fig:appx:decomposition:losscurve}.
We show the result of the optimization in Figure~\ref{fig:appx:decomposition}.
The edge structure is extracted and highlighted by applying the Laplacian operator on the final image and displayed in Figure~\ref{fig:appx:decomposition:laplace}.

From the results, we can conclude the followings:
Firstly, the parallel part of the $a$RGB embedding dominates the color information.
This is predictable since this parallel component is the information that is decoded back to the RGB image with our linear decoder $g\,$.
Secondly, the perpendicular part of the $a$RGB embedding conveys edge structure from the source image.
As a consequence, we can retrieve this information from the $a$RGB representation following the same process used to train the inversion in this section.
Moreover, this edgeness is the additional information brought with our $a$RGB representation to serve as a performance boost for training the image restoration models.
As our main results in Section~\ref{sec:4_experiment} suggests, the reduction of visual artifacts in perceptual image super-resolution and the enhancement of edge structures in image denoising and deblurring are attributed to this particular information carried in the perpendicular embeddings.
Lastly, we can clearly observe that this additional information is diminished away if the underlying image is highly noisy, far from the manifold of clean, natural images.
From this observation, we can conclude that the information learned by the $a$RGB autoencoder from its training dataset gives the model a structural prior in order to process images similar to its original training data.
This argument aligns with our finding in the ablation studies in Section~\ref{sec:5_discussion:ablation}, where a network trained from different dataset generally produce poorer restoration models.

\input{figures/appx_disc_decomposition/main}

\subsection{Topology of the learned $a$RGB space}
\label{sec:f_decomposition:tsne}
We have conjectured in Section~\ref{sec:3_method:autoencoder} that the underlying distribution of small image patches are disconnected, exhibiting very complex structures.
We can indirectly check the validity of our argument by visualizing the learned embeddings of an image using dimension reduction techniques.
Specifically, we use t-SNE \citep{ml:vis:vdmaaten08-tsne} algorithm to embed $H \times W = 65,536$ vectors of size $C = 128$ into a two dimensional plane to visualize the structure of our learned embedding for a particular image patch.

In this small experiment, we expect two outcomes:
First, we expect perceptually distinct groups of embedding vectors to be appeared in the t-SNE results.
Although the geometric information of the 2D projection carries little meaning in depicting the exact structure of the underlying feature space, visually distinguishable clustering in this region will signify that the structure of underlying representation is not connected into a single manifold, but consists of multiple disconnected regions with distinct characteristic values.
Second, we expect that the experts are well specialized, requiring that clusters of $a$RGB \emph{pixels} each assigned to a single expert to be appeared in the embeddings.
Figure~\ref{fig:appx:tsne} shows three more examples in addition to Figure~\ref{fig:discussion:segm} and \ref{fig:discussion:tsne} from our main manuscript.
The sample images are picked from different datasets, \ie, DIV2K \citep{cv:data:agustsson17-div2k}, Urban100 \citep{cv:data:huang15-urban100}, and Manga109 \citep{cv:data:Matsui17-manga109}, having different color distributions, contents, and styles.
However, as the results show that the embeddings generated from this set of images have several commonalities:
First, the embeddings are clustered in a well-separated regions.
We can observe two distinct types of clusters: \emph{common} groups where multiple experts are involved in generating similar embeddings and \emph{expert-specific} groups where a single expert dominates in encoding the information of pixels.
Existence of the first type of groups indicates the existence of common subspace between the feature spaces of each specialized expert.
The latter type of groups show a clear evidence of expert specialization in our $a$RGB encoder.
From the observations, we conclude that our initial design philosophy of the network serves its original purpose.

\input{figures/appx_disc_tsne/main}

\subsection{Visualization of the learned features of the $a$RGB encoder}
\label{sec:f_decomposition:feature_viz}
In this last section of our paper, we provide another visualization results to facilitate understanding of the behavior of our $a$RGB encoder.
In order to visualized the learned features for individual expert of our encoder, we simply maximize the activation of a single last channel of one of the experts.
Starting from a random image of size $32 \times 32\,$, which is more than three times larger than the receptive field of our network, we run a simple maximization of the average activation of each channels to produce a single feature image for each channel of an expert.
Figure~\ref{sec:f_decomposition:feature_viz} shows some of the results.
First, we observe that channels of the same index at each of the experts are maximally activated at the similar color distribution.
This similarity comes from the shared linear decoder of our autoencoder.
However, we also notice that the same set of filters are maximally stimulated at different \emph{patterns}.
The results uncover another evidence of expert specialization in our $a$RGB autoencoder.

\input{figures/appx_filter_visualization/main}

%% file: figures/appx_disc_decomposition/main.tex
\begin{figure*}[t]
\newcommand{\figurewidth}{.6165\linewidth}
\newcommand{\h}{25mm}
\newcommand{\hh}{2.5mm}
\renewcommand{\vv}{\vspace*{-1.15mm}}
\vspace{.7em}
\centering
\begin{minipage}[c]{\figurewidth}
\raisebox{27mm}[0mm][0mm]{\makebox[\hh][c]{\hspace*{132mm}%
\makebox[\h][c]{\hspace{-0.\linewidth}\normalsize{Flat image}}%
\makebox[\h][c]{\hspace{-0.\linewidth}\normalsize{Natural image}}%
\makebox[\h][c]{\hspace{-0.\linewidth}\normalsize{Gaussian noise}}%
}}\hfill%
\makebox[\h][c]{\textbf{\normalsize{\raisebox{1mm}[0mm][0mm]{\small \ \ Source $f_{\parallel} (\xx_{1})$}}}}\hfill%
\rotatebox[origin=l]{90}{\makebox[0mm][l]{\hspace*{0.05\linewidth}\textbf{\normalsize{\raisebox{1mm}[0mm][0mm]{\small Source $f_{\perp} (\xx_{2})$}}}}}\hfill%
\newcommand{\extt}{png}
\newcommand{\varA}{B_flat}\includegraphics[height=\h]{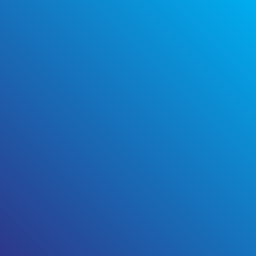}%
\newcommand{\varB}{B_image}\includegraphics[height=\h]{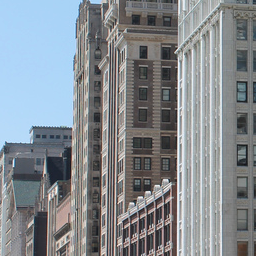}%
\newcommand{\varC}{B_noise}\includegraphics[height=\h]{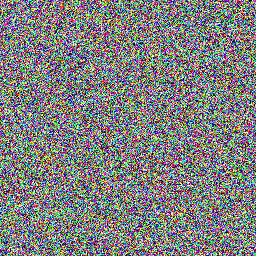}\vspace*{-3.1mm}\\
\begin{tikzpicture}\draw (0,0) -- (\linewidth,0);\end{tikzpicture}\vspace*{-0.5mm}\\%
\makebox[\hh]{\rotatebox[origin=l]{90}{\makebox[\h][c]{\hspace{-0.\linewidth}\normalsize{Flat image}}}}\hspace{0.5mm}%
\newcommand{\seed}{A_flat}
\includegraphics[height=\h]{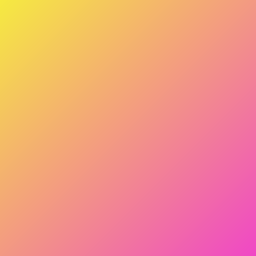}\hfill%
\includegraphics[height=\h]{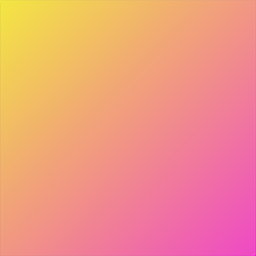}%
\includegraphics[height=\h]{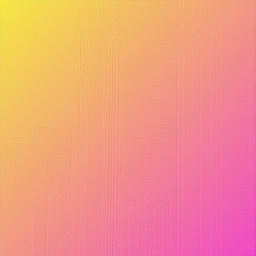}%
\includegraphics[height=\h]{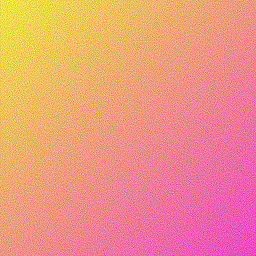}\vv\\
\makebox[\hh]{\rotatebox[origin=l]{90}{\makebox[\h][c]{\hspace{-0.\linewidth}\normalsize{Natural image}}}}\hspace{0.5mm}%
\renewcommand{\seed}{A_image}
\includegraphics[height=\h]{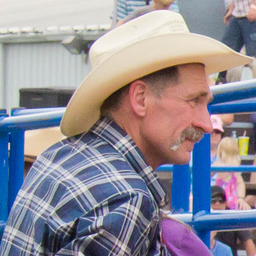}\hfill%
\includegraphics[height=\h]{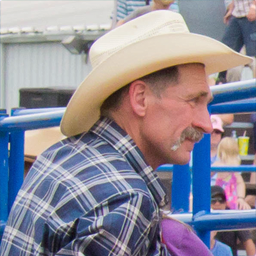}%
\includegraphics[height=\h]{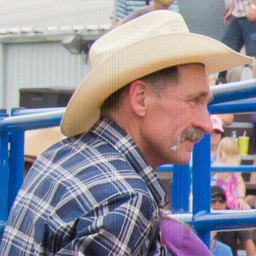}%
\includegraphics[height=\h]{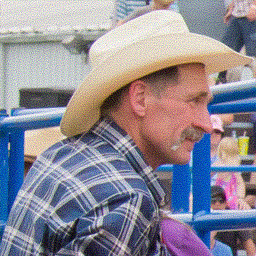}\vv\\
\makebox[\hh]{\rotatebox[origin=l]{90}{\makebox[\h][c]{\hspace{-0.\linewidth}\normalsize{Gaussian noise}}}}\hspace{0.5mm}%
\renewcommand{\seed}{A_noise}
\includegraphics[height=\h]{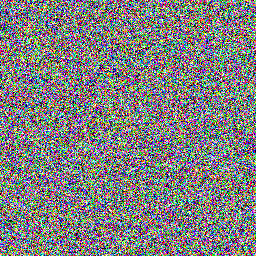}\hfill%
\raisebox{0mm}[0mm][0mm]{\makebox[0mm]{\hspace*{-1.5mm}\begin{tikzpicture}\draw (0,0mm) -- (0,-106.2mm);\end{tikzpicture}}}%
\includegraphics[height=\h]{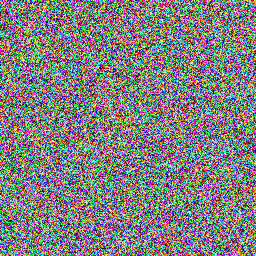}%
\includegraphics[height=\h]{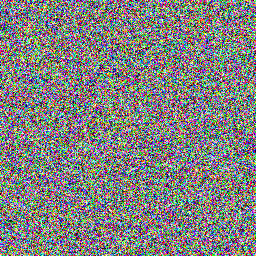}%
\includegraphics[height=\h]{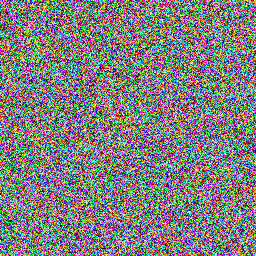}\vspace*{-3.5mm}
\\
\end{minipage}
\\[-.4em]
\caption{\small
    \textbf{Decomposition of the $a$RGB representation space.}
    The $a$RGB embeddings of the images on the sides are decomposed into orthogonal components and mixed $\xxi_{\text{mix}} = f_{\parallel}(\xx_{1}) + f_{\perp}(\xx_{2})\,$.
    Each cell is the image corresponding to the mixed embedding $f^{-1}(\xxi_{\text{mix}})\,$.
    \\[1.5em]
}
\label{fig:appx:decomposition}%
\begin{minipage}[c]{\figurewidth}
\raisebox{27mm}[0mm][0mm]{\makebox[\hh][c]{\hspace*{132mm}%
\makebox[\h][c]{\hspace{-0.\linewidth}\normalsize{Flat image}}%
\makebox[\h][c]{\hspace{-0.\linewidth}\normalsize{Natural image}}%
\makebox[\h][c]{\hspace{-0.\linewidth}\normalsize{Gaussian noise}}%
}}\hfill%
\makebox[\h][c]{\textbf{\normalsize{\raisebox{1mm}[0mm][0mm]{\small \ \ Source $f_{\parallel} (\xx_{1})$}}}}\hfill%
\rotatebox[origin=l]{90}{\makebox[0mm][l]{\hspace*{0.05\linewidth}\textbf{\normalsize{\raisebox{1mm}[0mm][0mm]{\small Source $f_{\perp} (\xx_{2})$}}}}}\hfill%
\newcommand{\extt}{png}
\newcommand{\varA}{B_flat}\includegraphics[height=\h]{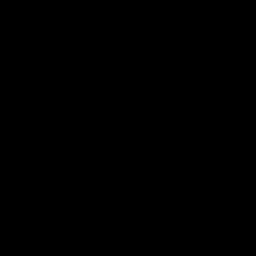}%
\newcommand{\varB}{B_image}\includegraphics[height=\h]{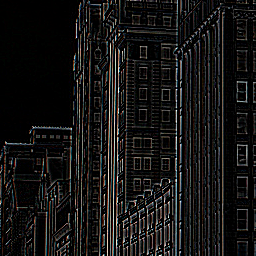}%
\newcommand{\varC}{B_noise}\includegraphics[height=\h]{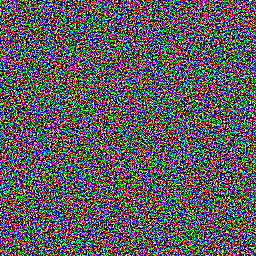}\vspace*{-3.1mm}\\
\begin{tikzpicture}\draw (0,0) -- (\linewidth,0);\end{tikzpicture}\vspace*{-0.5mm}\\%
\makebox[\hh]{\rotatebox[origin=l]{90}{\makebox[\h][c]{\hspace{-0.\linewidth}\normalsize{Flat image}}}}\hspace{0.5mm}%
\newcommand{\seed}{A_flat}
\includegraphics[height=\h]{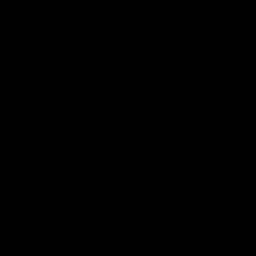}\hfill%
\includegraphics[height=\h]{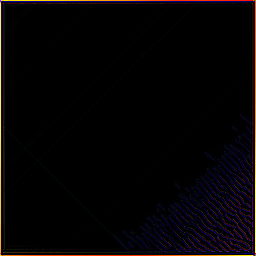}%
\includegraphics[height=\h]{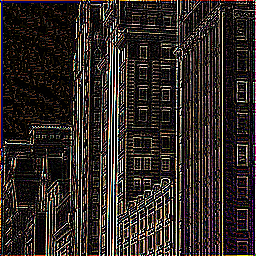}%
\includegraphics[height=\h]{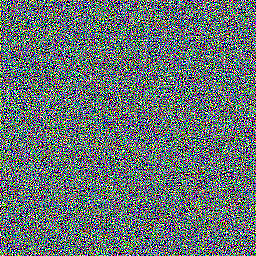}\vv\\
\makebox[\hh]{\rotatebox[origin=l]{90}{\makebox[\h][c]{\hspace{-0.\linewidth}\normalsize{Natural image}}}}\hspace{0.5mm}%
\renewcommand{\seed}{A_image}
\includegraphics[height=\h]{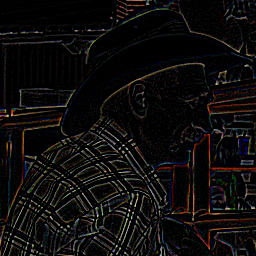}\hfill%
\includegraphics[height=\h]{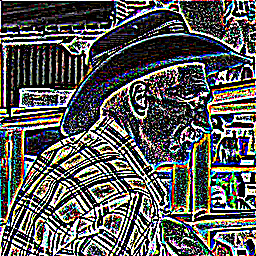}%
\includegraphics[height=\h]{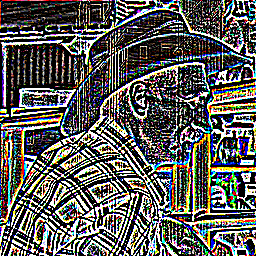}%
\includegraphics[height=\h]{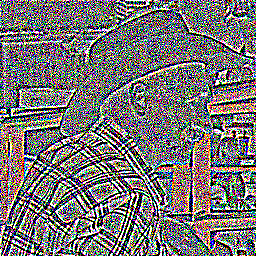}\vv\\
\makebox[\hh]{\rotatebox[origin=l]{90}{\makebox[\h][c]{\hspace{-0.\linewidth}\normalsize{Gaussian noise}}}}\hspace{0.5mm}%
\renewcommand{\seed}{A_noise}
\includegraphics[height=\h]{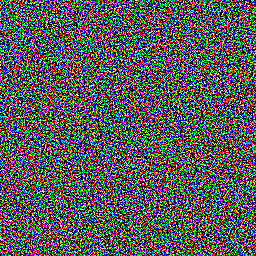}\hfill%
\raisebox{0mm}[0mm][0mm]{\makebox[0mm]{\hspace*{-1.5mm}\begin{tikzpicture}\draw (0,0mm) -- (0,-106.2mm);\end{tikzpicture}}}%
\includegraphics[height=\h]{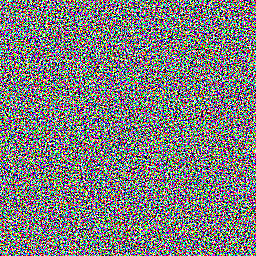}%
\includegraphics[height=\h]{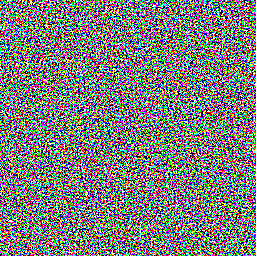}%
\includegraphics[height=\h]{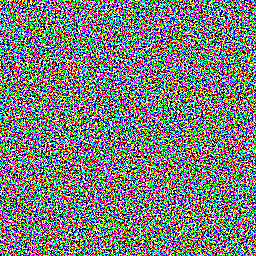}\vspace*{-3.5mm}
\\
\end{minipage}
\\[-.4em]
\caption{\small
    \textbf{Edge-enhanced inversion results of Figure~\ref{fig:appx:decomposition}.}
    A discrete Laplacian operator is applied to the same images in Figure~\ref{fig:appx:decomposition} to enhance the high-frequency structures for clearer understanding.
    The results reveal that the perpendicular component of the $a$RGB embedding $f_{\perp}$ contributes to high-frequency structures.
}
\label{fig:appx:decomposition:laplace}
\end{figure*}

\begin{figure*}[t]
\centering
\includegraphics[width=\linewidth]{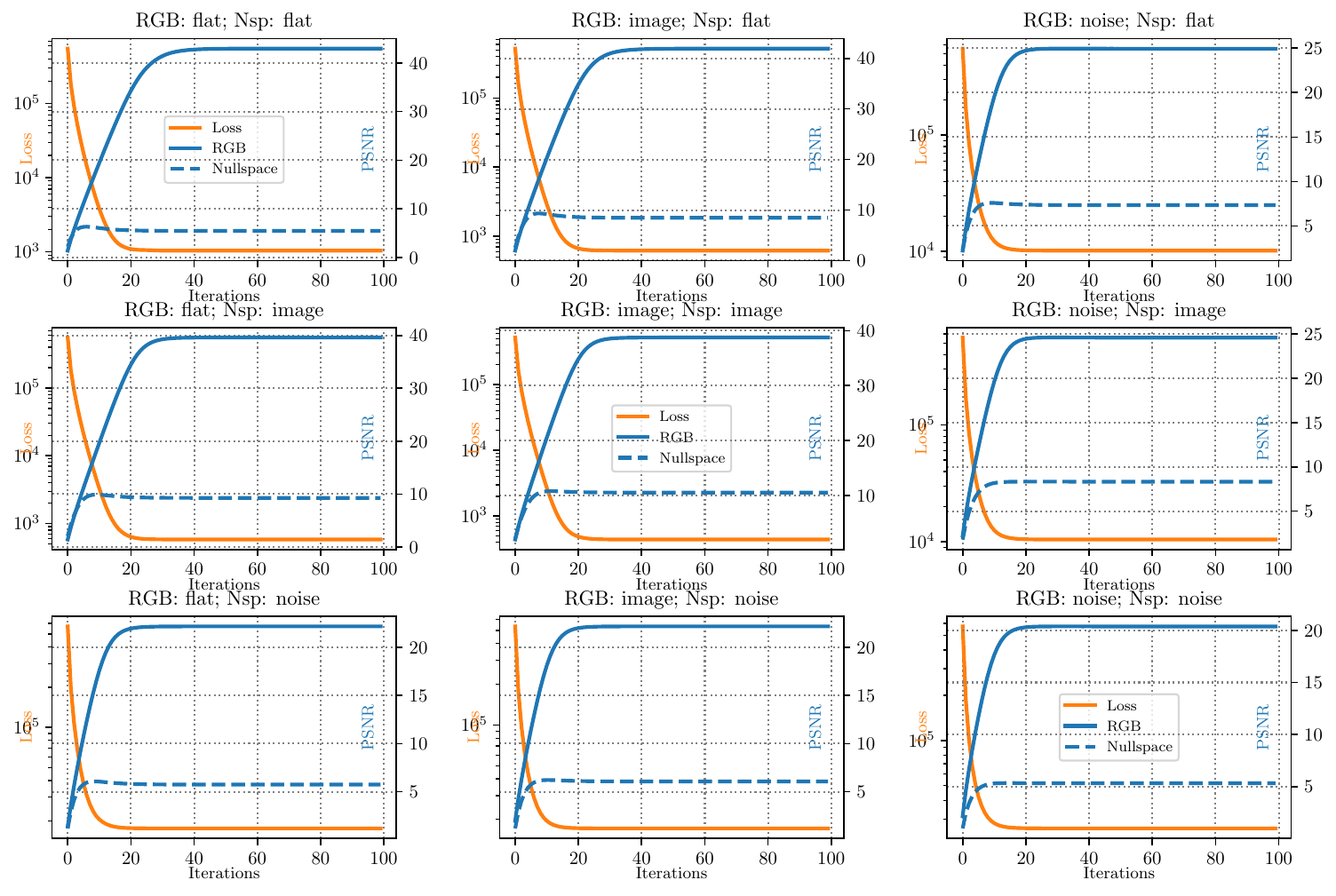}
\\
\caption{\small
    \textbf{Training curve for the decomposition test.}
    All the embedding inversion test quickly converge after 50 iterations.
    \textit{RGB} corresponds to the source image $\xx_{1}$ used for the parallel component $\xxi_{\parallel} = \mA^{\dagger} \mA f(\xx_{1}) = \mA^{\dagger} \mA \xxi_{\text{mix}}\,$ of the target $a$RGB embedding $\xxi_{\text{mix}}\,$, and \textit{Nullspace} corresponds to the source image $\xx_{w}$ used for the perpendicular component $\xxi_{\perp} = (\mI - \mA^{\dagger} \mA) f(\xx_{2}) = (\mI - \mA^{\dagger} \mA) \xxi_{\text{mix}}\,$, where $mA$ is the weight of the linear decoder $g\,$.
    As shown in Figure~\ref{fig:appx:decomposition}, The low-frequency color distribution of the resulting inversion follows that of the parallel component's source $\xx_{1}\,$, resulting in high PSNR scores.
    Although the PSNR scores between the inversions $f^{-1}(\xxi_{\text{mix}})$ and the corresponding source images $\xx_{2}$ of the perpendicular component $f_{\perp}(\xx_{2})$ are low, Figure~\ref{fig:appx:decomposition:laplace} reveals that the perpendicular components encode high frequency information of the image.
}
\label{fig:appx:decomposition:losscurve}
\end{figure*}

%% file: figures/appx_disc_tsne/main.tex
\begin{figure*}[t]
\newcommand{\width}{0.329\linewidth}
\centering
\includegraphics[width=\width]{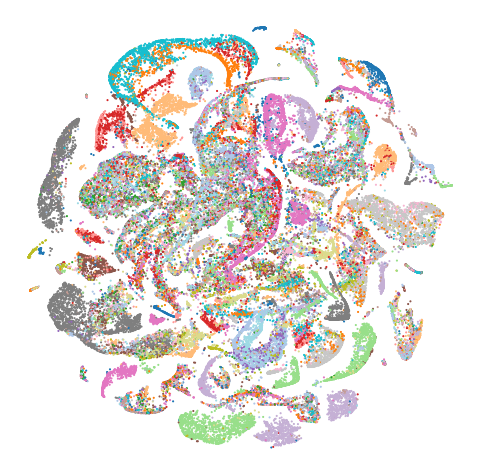}
\hfill
\includegraphics[width=\width]{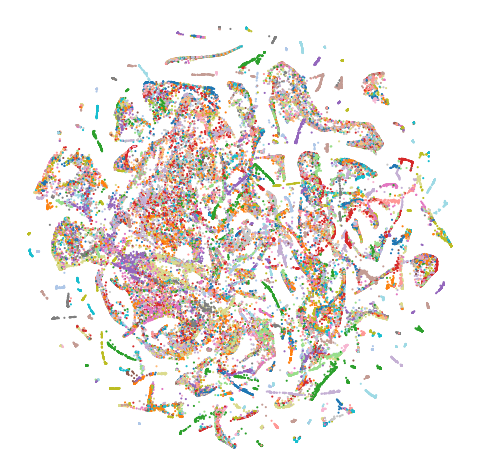}
\hfill
\includegraphics[width=\width]{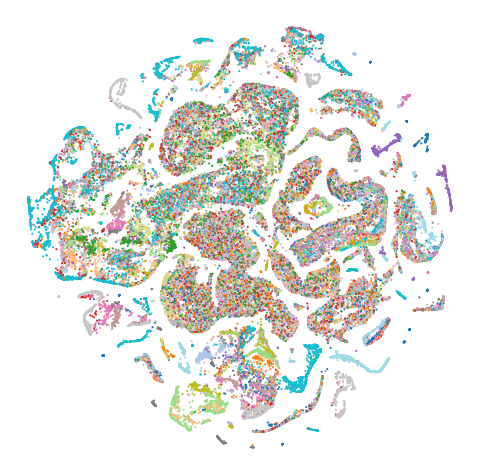}
\\
\subfloat[\small 0899 from DIV2K-Val. \label{fig:appx:tsne:img1}]{\includegraphics[width=\width]{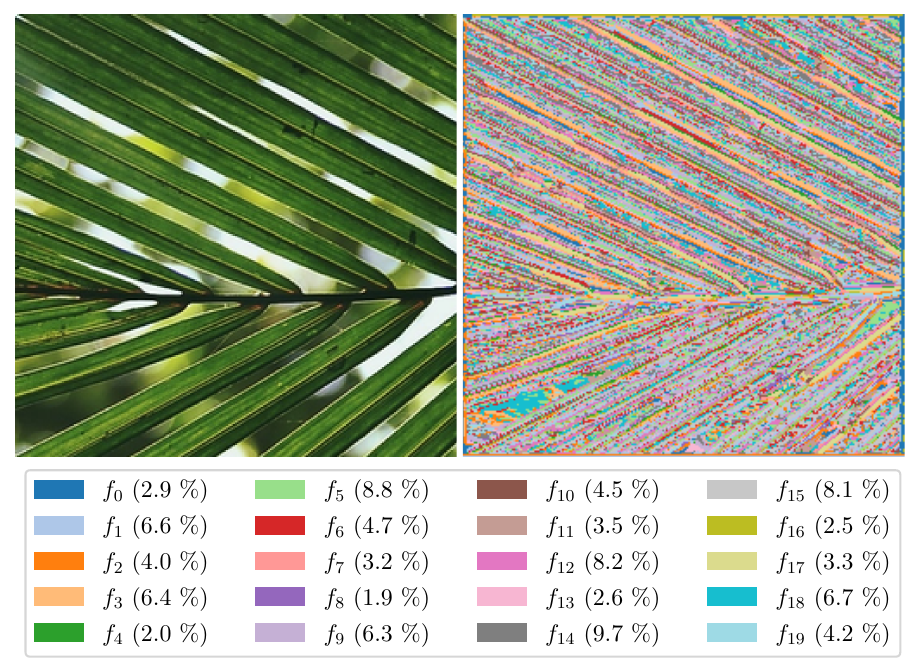}}
\hfill
\subfloat[\small img009 from Urban100. \label{fig:appx:tsne:img2}]{\includegraphics[width=\width]{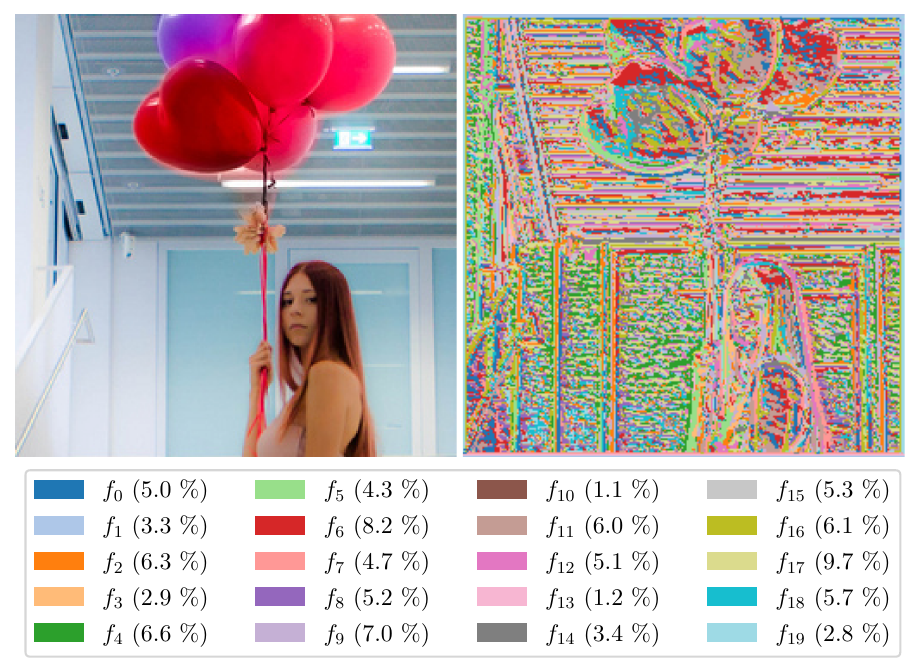}}
\hfill
\subfloat[\small Raphael from Manga109. \label{fig:appx:tsne:img3}]{\includegraphics[width=\width]{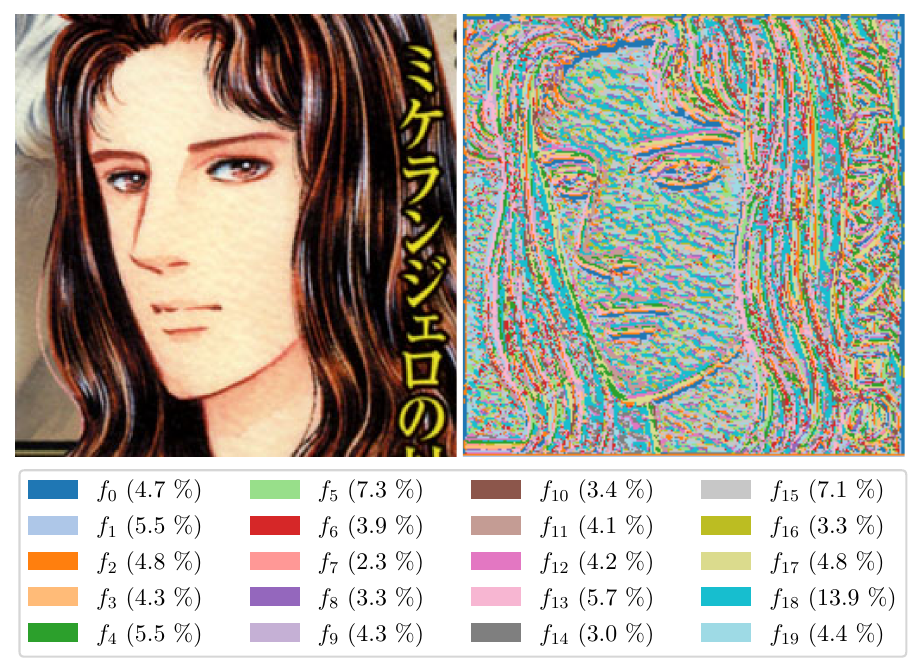}}
\\
\caption{\small
    \textbf{More examples on expert specialization using t-SNE and segmentation map.}
    Sample images are brought from three well-used super-resolution benchmark datasets, \ie,  DIV2K \citep{cv:data:agustsson17-div2k}, Urban100 \citep{cv:data:huang15-urban100}, and Manga109 \citep{cv:data:Matsui17-manga109}.
    Although the content and the style of each patches are widely different, the distribution of the learned $a$RGB embeddings in these patches exhibit similar pattern: the distributions are decomposed into \emph{common} groups, where multiple experts are involved in the encoding, and \emph{expert-specific} groups.
}
\label{fig:appx:tsne}
\end{figure*}

%% file: figures/appx_filter_visualization/main.tex
\begin{figure*}[t]
\newcommand{\width}{0.245\linewidth}
\centering
\subfloat[\small Expert 5\,. \label{fig:appx:visualize:expert5}]{\includegraphics[width=\width]{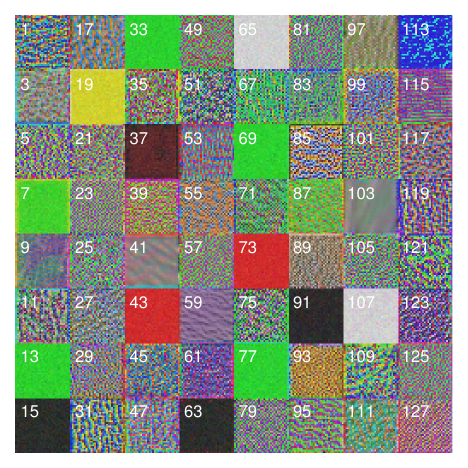}}
\hfill
\subfloat[\small Expert 10\,. \label{fig:appx:visualize:expert10}]{\includegraphics[width=\width]{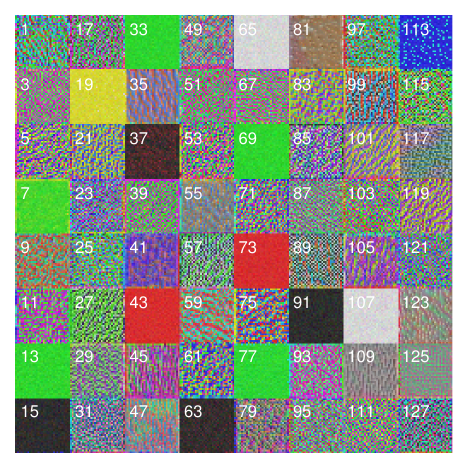}}
\hfill
\subfloat[\small Expert 15\,. \label{fig:appx:visualize:expert15}]{\includegraphics[width=\width]{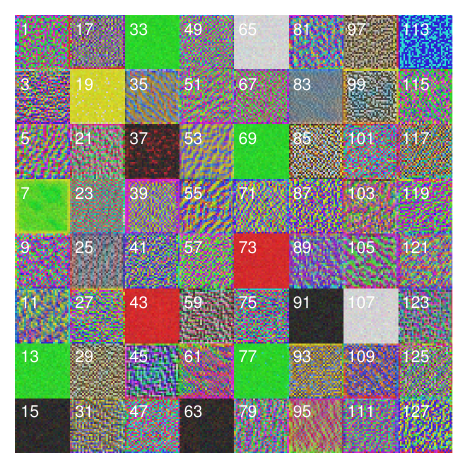}}
\hfill
\subfloat[\small Expert 20\,. \label{fig:appx:visualize:expert20}]{\includegraphics[width=\width]{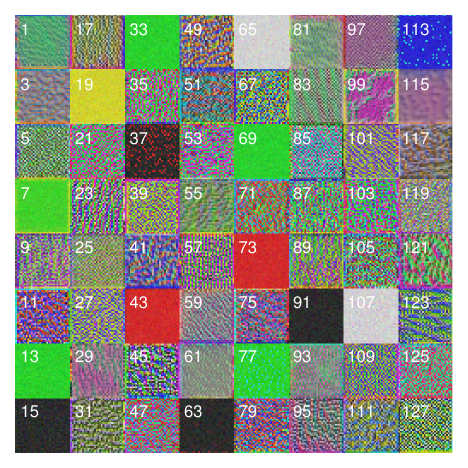}}
\\
\caption{\small
    \textbf{Visualization of the output filters of the experts.}
    Randomly initialized $32 \times 32$ images are trained to maximize a specific filter at the last convolutional layer of the selected expert.
    The ID of each filter is annotated with white numbers.
    Note that the $a$RGB representation space has a dimension of 128\,, the same as the number of filters in the last layer of each experts.
    The results show that while filters of different experts encoding the same channel is maximally activated at a similar average color, the high-frequency patterns each filter maximally attends to vary significantly.
}
\label{fig:appx:visualize:expert}
\end{figure*}